\documentclass[
  a4paper,reprint,aps,prd,
  groupedaddress,superscriptaddress,
  nofootinbib,
  dvipsnames
]{revtex4-2}
\usepackage{amsmath,amssymb,mathtools}
\allowdisplaybreaks

\usepackage{graphicx}
\usepackage{dcolumn}
\usepackage{bm}
\usepackage{booktabs}
\usepackage{tabularx}
\usepackage{multirow}
\usepackage{rotating}
\usepackage{adjustbox}

\usepackage{csquotes}
\usepackage{orcidlink}

\usepackage{xcolor}
\usepackage{hyperref}
\hypersetup{
  colorlinks=true,
  linkcolor=blue,
  citecolor=MidnightBlue,
  urlcolor=Blue
}
\usepackage[capitalise]{cleveref}

\newcommand{\dd}{\mathrm{d}}

\begin{document}

\title{Background-level reconstruction of scalar-field potentials from dark-energy histories and comparison with analytic potential families}
  
\author{Shahnawaz A. Adil\,\orcidlink{0000-0003-4999-7801}}
\email{shahnawaz@icf.unam.mx}
\affiliation{Instituto de Ciencias F\'isicas, Universidad Nacional Aut\'onoma de M\'exico, Cuernavaca, 
Morelos, 62210, M\'exico}

\author{Miguel A. Zapata\,\orcidlink{0009-0004-8575-4844}}
\email{miguel\_delacruz@icf.unam.mx}
\affiliation{Instituto de Ciencias F\'isicas, Universidad Nacional Aut\'onoma de M\'exico, 
Cuernavaca, Morelos, 62210, M\'exico}

\author{\"{O}zg\"{u}r Akarsu\,\orcidlink{0000-0001-6917-6176}}
\email{akarsuo@itu.edu.tr}
\affiliation{Department of Physics, Istanbul Technical University, Maslak 34469 Istanbul, Turkey}

\author{J. Alberto Vazquez\,\orcidlink{0000-0002-7401-0864}}
\email{javazquez@icf.unam.mx}
\affiliation{Instituto de Ciencias F\'isicas, Universidad Nacional Aut\'onoma de M\'exico, Cuernavaca, 
Morelos, 62210, M\'exico}



%----------------------------------------------------------
% DATES
%----------------------------------------------------------

%=== TITLE & AUTHORS ====================================================================
%\bstctlcite{IEEEexample:BSTcontrol}
    %  Tibault~Reveyrand,~\IEEEmembership{Member,~IEEE,}\\


% ====================================================================

\begin{abstract}
We present a unified \emph{background-level} framework that maps a prescribed late-time dark-energy density history $\rho_{\rm de}(z)$ onto an effective scalar-field description in a spatially flat Friedmann--Lema\^{i}tre--Robertson--Walker universe. Working directly with $\rho_{\rm de}(z)$ and $\dd\rho_{\rm de}/\dd z$, we reconstruct the associated pressure $p_{\rm de}(z)$, kinetic contribution $K(z)$, field trajectory $\phi(z)$, and field-space potential $V(\phi)$, without relying on the ratio $w_{\rm de}=p_{\rm de}/\rho_{\rm de}$ in regimes where it is ill-defined (e.g.\ at a density zero crossing). The physically relevant classifier is instead the null-energy-condition (NEC) combination $\rho_{\rm de}+p_{\rm de}=\tfrac13(1+z)\,\dd\rho_{\rm de}/\dd z$, whose sign controls the kinetic sector and provides a direct single-field consistency check. We apply the method to three benchmark histories: (i) the Chevallier--Polarski--Linder (CPL) form; (ii) a smooth mirror AdS$\rightarrow$dS sign-switching profile in which $\rho_{\rm de}$ crosses zero at $z_\dagger$, interpolating between a positive late-time plateau and a negative high-$z$ plateau ($\Lambda_{\rm s}$CDM-like at the background level); and (iii) a shifted-$\tanh$ emergent profile that remains positive definite and approaches $\rho_{\rm de}\to 0^{+}$ at high redshift. For CPL, $\rho_{\rm de}>0$ by construction but the evolution crosses the NEC boundary $\rho_{\rm de}+p_{\rm de}=0$ (equivalently $w_{\rm de}=-1$), so $K(z)$ changes sign and the reconstructed mapping becomes multivalued in field space; consequently, the full CPL history cannot be realized globally by a single minimally coupled real scalar field with fixed kinetic signature and is naturally interpreted as an effective one-dimensional description of an extended sector. By contrast, the $\tanh$-based transition histories satisfy the sign-consistency condition over the reconstruction range, selecting a single-field realization on the phantom branch ($\epsilon=-1$) at the background level. Finally, treating the reconstructed potential, $V_{\rm tar}(\phi)$, as a target, we perform Bayesian model comparison directly in \emph{potential space} and rank representative analytic potential families by their Bayesian evidence. For CPL (restricting to the single-valued phantom branch for the potential-space comparison), the exponential potential has the highest evidence in the baseline analysis, while the shifted-$\tanh$ and hilltop quartic forms remain close competitors; for the sign-switching $\tanh$ target, the shifted-$\tanh$ potential is strongly preferred, and the emergent profile yields the same qualitative ranking. These results provide a practical dictionary between phenomenological expansion histories and the scalar-field potential shapes required to reproduce them at the background level.
\end{abstract}

\maketitle

%%============================================================================================================%%
\section{Introduction}
\label{sec:intro}
%%============================================================================================================%%

Late-time cosmic acceleration is supported by a broad and mutually reinforcing set of observations, including cosmic microwave background (CMB) anisotropies, baryon acoustic oscillations (BAO), Type~Ia supernova distances, and increasingly precise probes of large-scale structure and weak lensing~\cite{Planck:2018vyg,eBOSS:2020yzd,Scolnic:2021amr,Brout:2022vxf,Rubin:2023jdq,DESI:2024mwx,AtacamaCosmologyTelescope:2025blo,DESI:2025zgx,SPT-3G:2025bzu,DES:2025sig}.
Within general relativity (GR), the minimal phenomenological description is the spatially flat $\Lambda$ cold dark matter model ($\Lambda$CDM), in which the current acceleration is driven by a strictly constant vacuum-energy density with equation of state $w=-1$.
However, despite its empirical success, the physical origin of this component remains unknown.
The cosmological-constant problem, and more broadly the question of whether late-time data truly require a rigid $\Lambda$ or instead allow nontrivial dynamics in an effective dark sector, continue to motivate precision tests of departures from the minimal picture~\cite{Weinberg:1988cp,Sahni:1999gb,Sahni:2002kh}.
From a model-building perspective, this naturally leads to a constructive inverse problem: given a phenomenological late-time history for the dark-energy density $\rho_{\rm de}(z)$, what scalar-field structure is required to reproduce it at the background level?

A second and complementary motivation comes from internal consistency in the era of precision cosmology.
As cosmological measurements have entered the percent regime, several parameters inferred from different probes exhibit mild-to-moderate discordances that may reflect residual systematics, underestimated covariances, or limitations of the minimal model~\cite{Verde:2019ivm,DiValentino:2020zio,DiValentino:2021izs,Perivolaropoulos:2021jda,Schoneberg:2021qvd,Abdalla:2022yfr,DiValentino:2022fjm,Vagnozzi:2023nrq,Khalife:2023qbu,Akarsu:2024qiq,CosmoVerseNetwork:2025alb}.
The best-known example is the Hubble-constant tension between local determinations---for example Cepheid-calibrated SN~Ia distance-ladder measurements~\cite{Riess:2021jrx,Breuval:2024lsv} and community syntheses such as H0DN~\cite{H0DN:2025lyy}---and values inferred from early-universe data analysed within $\Lambda$CDM, most notably CMB-based determinations~\cite{Planck:2018vyg,SPT-3G:2025bzu}.
Another, milder but persistent discrepancy is the so-called $S_8$ tension, associated with growth and clustering observables~\cite{DiValentino:2020vvd,Nunes:2021ipq,Adil:2023jtu,Akarsu:2024hsu,CosmoVerseNetwork:2025alb,DAmico:2019fhj,Troster:2019ean,Heymans:2020gsg,KiDS:2020suj,DES:2021bvc,DES:2021vln,DES:2021wwk,Dalal:2023olq,Kilo-DegreeSurvey:2023gfr,Chen:2024vvk,Wright:2025xka,DES:2026fyc}.
Although the ultimate origin of any individual anomaly remains debated, one robust empirical lesson is that the \emph{inferred} late-time behavior can depend sensitively on how one represents the dark sector and on how one propagates early-time calibrations---most notably the sound horizon at baryon drag---into late-time distances~\cite{CosmoVerseNetwork:2025alb}.
This makes it especially valuable to construct frameworks that translate \emph{phenomenological expansion histories} into \emph{concrete theory-space realizations} in a way that remains meaningful even when the effective dark-energy sector exhibits nonstandard behavior.

In this broader landscape, proposed resolutions of $H_0$ and related late-time discrepancies are often classified according to \emph{when} they modify the expansion history~\cite{CosmoVerseNetwork:2025alb}.
\emph{Early-time} solutions alter pre-recombination physics and typically shift the sound-horizon scale at baryon drag, $r_{\rm d}$, as in early dark energy (EDE) and related constructions~\cite{Garcia-Arroyo:2024tqq, Karwal:2016vyq,Poulin:2018cxd,Smith:2019ihp,Sakstein:2019fmf,Hill:2020osr,Ivanov:2020ril,Kamionkowski:2022pkx,Poulin:2023lkg,Niedermann:2019olb,Niedermann:2020dwg,Smith:2025grk,Poulin:2025nfb,SPT-3G:2025vyw}.
By contrast, \emph{late-time} solutions modify the post-recombination expansion history while leaving $r_{\rm d}$ essentially unchanged, assuming standard pre-recombination physics.
These include interacting dark-energy (IDE) scenarios~\cite{Kumar:2017dnp,DiValentino:2017iww,Yang:2018euj,Yang:2018uae,Pan:2019gop,Kumar:2019wfs,DiValentino:2019jae,DiValentino:2019ffd,Gomez-Valent:2020mqn,Lucca:2020zjb,Pan:2020zza,Gao:2021xnk,Kumar:2021eev,Yang:2021hxg,Nunes:2022bhn,Bernui:2023byc,Escamilla:2023shf,Giare:2024smz,Li:2024qso,Li:2025owk,Sabogal:2025mkp,Silva:2025hxw,Yang:2025uyv,vanderWesthuizen:2025rip,Li:2026xaz}
and dynamical dark-energy scenarios with late AdS-to-dS(-like) transitions, such as $\Lambda_{\rm s}$CDM~\cite{Akarsu:2019hmw,Akarsu:2021fol,Akarsu:2022typ,Akarsu:2023mfb,Paraskevas:2024ytz,Yadav:2024duq,Akarsu:2024qsi,Akarsu:2024eoo,Akarsu:2024nas,Souza:2024qwd,Akarsu:2025gwi,Akarsu:2025dmj,Akarsu:2025ijk,Escamilla:2025imi,Akarsu:2025nns,Kibris:2026cqq,Akarsu:2026lva} (see also Refs.~\cite{Anchordoqui:2023woo,Anchordoqui:2024gfa,Anchordoqui:2024dqc,Soriano:2025gxd}).
One phenomenological attraction of some models in this latter class---most notably sign-switching vacuum-energy constructions such as $\Lambda_{\rm s}$CDM---is that they can alleviate the $H_0$ and $S_8$ tensions simultaneously, rather than addressing $H_0$ primarily through a shift in $r_{\rm d}$.
More general effective-fluid constructions have also been proposed, including omnipotent dark energy, which allows richer phenomenology such as phantom crossing and sign-changing effective dark-energy density~\cite{DiValentino:2020naf,Adil:2023exv,Specogna:2025guo}.
For broader theoretical and observational discussions, including model-agnostic reconstructions and related late-time analyses, see Refs.~\cite{AlbertoVazquez:2012ofj, Sahni:2002dx,Vazquez:2012ag,BOSS:2014hwf,Sahni:2014ooa,BOSS:2014hhw,DiValentino:2017rcr,Mortsell:2018mfj,Poulin:2018zxs,Wang:2018fng,Banihashemi:2018oxo,Dutta:2018vmq,Banihashemi:2018has,Akarsu:2019ygx,Li:2019yem,Visinelli:2019qqu,Perez:2020cwa,Akarsu:2020yqa,Ruchika:2020avj,Yang:2020ope,Calderon:2020hoc,DeFelice:2020cpt,Paliathanasis:2020sfe,Bonilla:2020wbn,Acquaviva:2021jov,Bag:2021cqm,Bernardo:2021cxi,Escamilla:2021uoj,Sen:2021wld,Ozulker:2022slu,DiGennaro:2022ykp,Akarsu:2022lhx,Moshafi:2022mva,vandeVenn:2022gvl,Ong:2022wrs,Tiwari:2023jle,Malekjani:2023ple,Ben-Dayan:2023rgt,Vazquez:2023kyx,Alexandre:2023nmh,Adil:2023ara,Paraskevas:2023itu,Gomez-Valent:2023uof,Wen:2023wes,DeFelice:2023bwq,Menci:2024rbq,Gomez-Valent:2024tdb,DESI:2024aqx,Bousis:2024rnb,Wang:2024hwd,Colgain:2024ksa,Tyagi:2024cqp,Toda:2024ncp,Sabogal:2024qxs,Dwivedi:2024okk,Escamilla:2024ahl,Pai:2024ydi,Wen:2024orc,Gomez-Valent:2024ejh,Manoharan:2024thb,Mukherjee:2025myk,Efstratiou:2025xou,Tamayo:2025xci,Wang:2025dtk,Gonzalez-Fuentes:2025lei,Bouhmadi-Lopez:2025spo,Bouhmadi-Lopez:2025ggl,Hogas:2025ahb,Gomez-Valent:2025mfl,Tan:2025xas,Yadav:2025vpx,Pedrotti:2025ccw,Forconi:2025gwo,Nyergesy:2025lyi,Ghafari:2025eql,Akarsu:2026anp,Akarsu:2026pom,Gokcen:2026pkq}.
 Similarly, model-independent reconstructions of IDE kernels do not rule out negative effective dark-energy densities at $z\gtrsim 2$~\cite{Escamilla:2023shf}.

The phenomenological interest in such histories is reinforced by theoretical considerations.
In quantum field theory, naive estimates of vacuum energy exceed the observed value by many orders of magnitude, encapsulating the cosmological constant problem~\cite{Weinberg:1988cp,Sahni:1999gb,Sahni:2002kh}.
From the perspective of string theory, constructing a fully controlled metastable de~Sitter vacuum remains notoriously difficult, whereas anti-de~Sitter vacua arise more naturally in the landscape~\cite{Maldacena:1997re,Witten:1998qj,Bousso:2000xa,Vafa:2005ui,Obied:2018sgi,Garg:2018reu,Lehnert:2025izp}.
At the same time, recent work has argued that, despite the AdS swampland conjecture suggesting that a late-universe AdS-to-dS transition is unlikely due to the arbitrarily large separation between AdS and dS vacua in moduli space, such a transition can nonetheless be realized through the Casimir forces of fields inhabiting the bulk~\cite{Anchordoqui:2023woo,Anchordoqui:2024gfa,Anchordoqui:2024dqc,Soriano:2025gxd} (see also~\cite{Lehnert:2025izp}).
Taken together, these considerations motivate taking seriously effective late-time histories in which $\rho_{\rm de}(z)$ is dynamical, may approach zero or may even change sign.

Scalar-field frameworks provide a natural language for model building in this setting.
Quintessence, phantom, and more general scalar constructions offer flexible realizations of evolving dark energy, with the potential $V(\phi)$ controlling the cosmic dynamics~\cite{Copeland:2006wr,Tsujikawa:2013fta}.
One may also reverse the usual logic and begin from a \emph{phenomenological} dark-energy history, asking what field-space structure is required to reproduce it; this inverse problem has been explored in various forms in the literature~\cite{Kamenshchik:2001cp,Bilic:2001cg,Bento:2002ps}.
However, a central subtlety is that the usual equation-of-state ratio $w_{\rm de}(z)=p_{\rm de}(z)/\rho_{\rm de}(z)$ is not always a reliable global diagnostic~\cite{Akarsu:2025gwi,Akarsu:2026anp,Akarsu:2026pom,Gokcen:2026pkq}.
If $\rho_{\rm de}(z)$ approaches zero and changes sign---as can happen in sign-switching scenarios and in broader effective reconstructions---then $w_{\rm de}$ becomes ill-defined at the crossing even when the background expansion remains perfectly regular.
In a minimally coupled scalar-field realization, the physically meaningful quantity is instead the null-energy-condition (NEC) combination $\rho_{\rm de}+p_{\rm de}$, since the scalar identity $\rho_\phi+p_\phi=\epsilon\dot\phi^{\,2}$ implies that for a single real field with fixed kinetic signature the sign of $\rho_\phi+p_\phi$ cannot change.
Consequently, a NEC-boundary crossing provides an immediate \emph{single-field consistency check}, whereas a density zero crossing $\rho_{\rm de}=0$ does not need to signal any physical singularity at the level of the homogeneous background.
This distinction is central for any attempt to translate an effective late-time history into scalar-field theory space.
For related discussions, see, e.g., Refs.~\cite{Ozulker:2022slu,Adil:2023exv,Paraskevas:2024ytz,Akarsu:2025gwi,Akarsu:2026anp,Akarsu:2026pom,Gokcen:2026pkq}.

The issue has become especially timely in the DESI era.
Recent BAO measurements from DESI~\cite{DESI:2024mwx,DESI:2025zgx} provide high-precision late-time distance information over a wide redshift range and have sharpened the discussion of dynamical dark-energy (DDE) extensions~\cite{Cortes:2024lgw,Wang:2024dka,Dinda:2024kjf,Roy:2024kni,Gialamas:2024lyw,Najafi:2024qzm,Giare:2024gpk,Wolf:2024eph,RoyChoudhury:2024wri,Wolf:2024stt,Giare:2024oil,Wolf:2025jlc,Giare:2025pzu,Ormondroyd:2025iaf,DESI:2025fii,Pang:2025lvh,Kessler:2025kju,Wolf:2025jed,Li:2025cxn,Teixeira:2025czm,RoyChoudhury:2025dhe,Scherer:2025esj,Cheng:2025lod,Cheng:2025hug,Sabogal:2025jbo,Herold:2025hkb,Ozulker:2025ehg,Gialamas:2025pwv,Mishra:2025goj,Lee:2025pzo,Silva:2025twg,Fazzari:2025lzd,Wolf:2025acj,RoyChoudhury:2025iis,Li:2025vuh,Bouhmadi-Lopez:2025lzm,Xu:2026sbw,Akarsu:2026pom}.
In combinations including CMB and BAO, evolving dark energy can be preferred for some dataset choices, while other analyses emphasize the role of priors, degeneracies, and the precise selection of low-redshift data~\cite{DESI:2025zgx,Giare:2025pzu,Xu:2026sbw}.
A widely used phenomenological description in this context is the Chevallier--Polarski--Linder (CPL) parametrization~\cite{Chevallier:2000qy,Linder:2002et}, $w(a)=w_0+w_a(1-a)$; see, e.g., Refs.~\cite{DESI:2024mwx,DESI:2025zgx,DES:2025sig}.
Its compactness makes it useful for comparing analyses, but it also means that degeneracy directions in parameter space can correspond to qualitatively different density histories when extrapolated beyond the redshift range directly anchored by data.
This point has been emphasized in DESI-era discussions of phantom-crossing behavior and parameter degeneracies~\cite{Giare:2025pzu,Ozulker:2025ehg}.
Relatedly, several alternative two-parameter ans\"atze---for example JBP, BA, logarithmic, and exponential forms---can yield qualitatively similar ``evolving dark energy'' preferences for certain dataset combinations~\cite{Efstathiou:1999tm,Jassal:2005qc,Barboza:2008rh,Dimakis:2023oje,Giare:2024gpk,Akarsu:2026pom}.
For these reasons, CPL and similar effective-fluid parametrizations are best regarded as diagnostic benchmarks whose implications are most transparently analysed in terms of $\rho_{\rm de}(z)$ and $\dd\rho_{\rm de}/\dd z$, rather than through $w_{\rm de}$ alone.

A useful phenomenological lesson comes from recent work revisiting CPL-like dark energy in the presence of a sign-switching density phase~\cite{Gokcen:2026pkq}  (see also Ref.~\cite{Akarsu:2026pom}). Once the effective dark-energy density is allowed to become negative at higher redshift, the apparent significance of deviations from a pure cosmological constant can be reduced, while current late-time BAO data tend to push the negative-density phase beyond the redshift interval they directly probe. This suggests that a positive-density CPL history remains a useful effective benchmark in the low-redshift domain where it is commonly inferred, although more general completions may admit $\rho_{\rm de}<0$ at earlier times. This observation motivates the benchmark set adopted here: besides the standard CPL case with $\rho_{\rm de}>0$, we consider a smooth sign-switching $\tanh$ history as a prototype sigmoid realization of a mirror AdS-to-dS transition in the effective dark-energy density, in which the early negative plateau mirrors the late positive plateau in magnitude, and a shifted-$\tanh$ profile that instead implements a smooth uplift from an early near-zero plateau to the observed late positive one. Taken together, these three cases allow us to disentangle which features are specific to the standard CPL effective description and which are generic to smooth transition histories with or without a sign change in $\rho_{\rm de}$.

Against this backdrop, the aim of the present work is not to perform a new global cosmological parameter fit, but to develop a statistically grounded \emph{background-level dictionary} that maps an assumed late-time dark-energy density history $\rho_{\rm de}(z)$ into an effective scalar-field description.
Working directly with $\rho_{\rm de}(z)$ and $\dd\rho_{\rm de}/\dd z$, we reconstruct the associated pressure $p_{\rm de}(z)$, kinetic contribution $K(z)$, field trajectory $\phi(z)$, and field-space potential $V(\phi)$ at the background level, in a formulation that remains well defined even when the effective dark-energy density crosses zero.
We apply this reconstruction to three benchmark histories chosen to span both standard phenomenological practice and transition scenarios motivated by sign-switching vacuum-energy ideas:
(i) CPL form~\cite{Chevallier:2000qy,Linder:2002et,DESI:2024mwx,DESI:2025zgx,DES:2025sig};
(ii) a smooth sign-switching $\tanh$ transition that interpolates between a positive late-time plateau and a negative high-$z$ plateau, yielding a $\Lambda_{\rm s}$CDM-like background history~\cite{Akarsu:2019hmw,Akarsu:2021fol,Akarsu:2022typ,Akarsu:2023mfb,Paraskevas:2024ytz,Yadav:2024duq,Akarsu:2024qsi,Akarsu:2024eoo,Akarsu:2024nas,Souza:2024qwd,Akarsu:2025gwi,Akarsu:2025dmj,Akarsu:2025ijk,Escamilla:2025imi,Akarsu:2025nns,Kibris:2026cqq,Akarsu:2026lva};
and (iii) a shifted-$\tanh$ ``emergent'' profile~\cite{Li:2019yem,Yang:2020ope,DeFelice:2020cpt,Ben-Dayan:2023rgt} that remains positive definite and approaches $\rho_{\rm de}\to 0^{+}$ at high redshift.
For the benchmark choice adopted here, CPL provides a particularly instructive stress test:
it crosses the NEC boundary $\rho_{\rm de}+p_{\rm de}=0$ (equivalently the usual phantom-divide line when $\rho_{\rm de}>0$), implying a sign change in $K(z)$ and obstructing a global realization by a single minimally coupled scalar with fixed kinetic signature.
In that case, the reconstruction is naturally interpreted as an effective one-dimensional projection of an extended sector, such as a quintom-like or non-canonical completion~\cite{Vazquez:2023kyx}.
By contrast, the $\tanh$-based transition histories satisfy the single-field sign-consistency condition over the reconstruction range and admit a consistent phantom-branch realization at the background level.

In the second stage of the analysis, we treat the reconstructed target potential $V_{\rm tar}(\phi)$ as the object to be reproduced and perform Bayesian model comparison directly in \emph{potential space}.
We confront the reconstructed targets with a representative set of analytic potentials---exponential, PNGB, hilltop quartic, inverse power law, Gaussian bump, and shifted-$\tanh$---and rank them by their Bayesian evidence.
This step is intentionally distinct from a direct fit to cosmological observables:
the likelihood quantifies agreement with the reconstructed target in field space under an adopted noise model, thereby providing a controlled ``theory-space filter'' once a background history has been specified.

The paper is organized as follows.
In~\cref{sec:background} we summarize the background cosmological equations.
In~\cref{sec:scalarfield} we derive the effective scalar-field reconstruction and the single-field consistency criterion.
The reconstruction pipeline and the potential-space Bayesian framework are presented in~\cref{sec:method}.
 Results for the benchmark histories and the model comparison are given in~\cref{sec:results}.
We conclude in~\cref{sec:conclude} with a summary and outlook.

%%============================================================================================================%%
\section{Background}
\label{sec:background}
%%============================================================================================================%%

We work within a spatially flat Friedmann--Lema\^{i}tre--Robertson--Walker (FLRW) spacetime in general relativity (GR), whose constant-time hypersurfaces are spatially homogeneous and isotropic (maximally symmetric). The line element in comoving coordinates $(t,\vec{x})$ is
\begin{equation}
 \dd s^2=-\dd t^2+a^2(t)\,\dd\vec{x}^{\,2},
\end{equation}
where $a(t)$ is the scale factor and $t$ denotes the cosmic time. We set $a_0\equiv a(t_0)=1$, so that $1+z=a^{-1}$.

For spatially flat FLRW, Einstein's field equations, $G_{\mu\nu}=8\pi G\,T_{\mu\nu}$, reduce to the Friedmann equations
\begin{align}
3M_{\rm Pl}^2H^2 &= \sum_i \rho_i,
\label{eq:Friedmann1}
\\[3pt]
-2M_{\rm Pl}^2\dot{H} &= \sum_i(\rho_i+p_i),
\label{eq:Friedmann2}
\end{align}
where $H\equiv \dot{a}/a$ is the Hubble rate, $M_{\rm Pl}^2\equiv(8\pi G)^{-1}$ is the reduced Planck mass, and the sum runs over $i\in\{{\rm r},{\rm m},{\rm de}\}$ (radiation, pressureless matter and dark energy). We work in natural units $c=\hbar=1$ and keep $M_{\rm Pl}$ explicit; when discussing the scalar-field reconstruction and the potential-space analysis we often use the dimensionless field variable $\phi/M_{\rm Pl}$.

Assuming that each component is conserved separately at the background level, $\nabla_{\mu}T^{\mu\nu}_{(i)}=0$, one obtains the continuity equation
\begin{equation}
\dot{\rho}_i+3H(\rho_i+p_i)=0.
\label{eq:continuity_i}
\end{equation}
For non-relativistic matter ($p_{\rm m}=0$) and radiation ($p_{\rm r}=\rho_{\rm r}/3$), Eq.~\eqref{eq:continuity_i} gives
\begin{equation}
\rho_{\rm m}(z)=\rho_{\rm m0}(1+z)^3,
\qquad
\rho_{\rm r}(z)=\rho_{\rm r0}(1+z)^4.
\label{eq:rhom_rhor}
\end{equation}
For the dark-energy sector we keep $\rho_{\rm de}(z)$ generic. If, instead, one parameterizes it through an equation-of-state function $w(z)\equiv p_{\rm de}/\rho_{\rm de}$, then Eq.~\eqref{eq:continuity_i} implies
\begin{equation}
\rho_{\rm de}(a)=\rho_{{\rm de}0}\,
\exp\!\left[-3\!\int_{1}^{a}\bigl(1+w(\tilde a)\bigr)\,\frac{\dd\tilde a}{\tilde a}\right].
\label{eq:rho_de_w}
\end{equation}
However, in our reconstruction approach, $\rho_{\rm de}(z)$ is specified directly and the associated $p_{\rm de}(z)$ is obtained from the continuity equation (see~\cref{sec:scalarfield}).

Introducing the present-day critical density $\rho_{\rm c0}\equiv 3H_0^2M_{\rm Pl}^2$ and the present-day density parameters $\Omega_{i0}\equiv \rho_{i0}/\rho_{\rm c0}$, the first Friedmann equation can be written in dimensionless form as
\begin{equation}
E^2(z)\equiv \frac{H^2(z)}{H_0^2}
= \Omega_{\rm r0}(1+z)^4 + \Omega_{\rm m0}(1+z)^3 + \tilde\Omega_{\rm de}(z),
\label{eq:E2_from_rho_norm}
\end{equation}
where $\tilde\Omega_{\rm de}(z)\equiv \rho_{\rm de}(z)/\rho_{\rm c0}$ and $\Omega_{{\rm de}0}\equiv\tilde\Omega_{\rm de}(0)$. Spatial flatness implies
\begin{equation}
\Omega_{\rm m0}+\Omega_{\rm r0}+\Omega_{{\rm de}0}=1.
\end{equation}

For the present-day radiation (photons plus relativistic neutrinos) density parameter, we use
\begin{equation}
\Omega_{\rm r0}
  = \frac{2.469\times10^{-5}}{h^2}
    \left(\frac{T_{\mathrm{CMB}}}{2.7255~{\rm K}}\right)^{4}
    \left[1+\frac{7}{8}\left(\frac{4}{11}\right)^{4/3}N_{\mathrm{eff}}\right],
\label{eq:omega_r0}
\end{equation}
where $h\equiv H_0/(100~{\rm km\,s^{-1}\,Mpc^{-1}})$ is the dimensionless Hubble parameter, $T_{\rm CMB}$ is the present-day CMB monopole temperature, and we adopt the standard value $N_{\mathrm{eff}}=3.046$ unless stated otherwise. Equation~\eqref{eq:E2_from_rho_norm} therefore provides $H(z)$ once $\rho_{\rm de}(z)$ is specified; this is the only background input needed for the scalar-field reconstruction in the following.

For orientation,~\cref{fig:cpl_de_profile1,fig:cpl_de_profile2,fig:tanh_de_profile1,fig:tanh_de_profile2,fig:shift_de_profile1,fig:shift_de_profile2} display the reconstructed background quantities for the CPL benchmark and for the two switching profiles used in this work; the mapping to an effective scalar-field description is developed in~\cref{sec:scalarfield}.

%========================================================
\section{Effective scalar-field reconstruction}
\label{sec:scalarfield}
%%============================================================================================================%%

In this section, we map a phenomenological dark-energy (DE) history, specified at the background level through $\rho_{\rm de}(z)$, onto an \emph{effective} scalar-field description on a spatially flat FLRW background. Our reconstruction concerns only the homogeneous dynamics. Perturbative stability and UV completion issues are discussed only insofar as they affect the interpretation of the mapping.

\subsection*{Minimally coupled scalar field and Klein--Gordon equation}

We consider a spatially homogeneous, minimally coupled scalar field $\phi(t)$ with kinetic term and its corresponding Lagrangian as
\begin{equation}
X \equiv -\frac12\,g^{\mu\nu}\partial_\mu\phi\,\partial_\nu\phi,
\qquad
{\cal L}_\phi=\epsilon\,X - V(\phi),
\label{eq:Lagrangian}
\end{equation} 
where $\epsilon=+1$ corresponds to a canonical (quintessence-type) field and $\epsilon=-1$ to a phantom field~\cite{Vazquez:2020ani,Akarsu:2025gwi}. For a homogeneous configuration, $X=\dot\phi^2/2$, and the associated energy density and pressure are
\begin{equation}
\begin{aligned}
\rho_\phi &= \epsilon X+V(\phi)=\frac{\epsilon}{2}\dot{\phi}^2+V(\phi),\\
p_\phi   &= \epsilon X-V(\phi)=\frac{\epsilon}{2}\dot{\phi}^2-V(\phi).
\end{aligned}
\label{eq:rhophiphi}
\end{equation}
A key identity that will be used repeatedly is
\begin{equation}
\rho_\phi+p_\phi=\epsilon\,\dot\phi^{\,2},
\label{eq:rho_p_scalar}
\end{equation}
so for a real minimally coupled field with fixed $\epsilon$ the sign of $\rho_\phi+p_\phi$ is fixed. For $\epsilon=-1$, the kinetic term enters with the opposite sign and the field is ghost-like as a fundamental QFT degree of freedom. Accordingly, whenever a phantom realization is invoked below, it should be understood as an \emph{effective} background description (with an implied cutoff/UV completion), not as a fundamental low-energy theory.

The homogeneous field evolution obeys the Klein--Gordon equation, obtained by varying the action corresponding to Eq.~\eqref{eq:Lagrangian},
\begin{equation}
\ddot\phi+3H\dot\phi+\epsilon\,V_{,\phi}=0,
\label{eq:KG}
\end{equation}
where $H=\dot a/a$. Writing primes as derivatives with respect to redshift (${}'\equiv \dd/\dd z$) and using
$\dd/\dd t=-(1+z)H\,\dd/\dd z$, Eq.~\eqref{eq:KG} can be recast as
\begin{equation}
\phi''+\left(\frac{H'}{H}-\frac{2}{1+z}\right)\phi'
+\frac{\epsilon}{(1+z)^2H^2}\,\frac{\dd V}{\dd\phi}=0.
\label{eq:KG_z}
\end{equation}
For cases that admit a single-field realization with fixed $\epsilon$, we verify Eq.~\eqref{eq:KG_z} numerically via a Klein--Gordon residual (\cref{app:KGcheck}).

%%============================================================================================================%%
\subsection*{\texorpdfstring{DE fluid $\leftrightarrow$ scalar field mapping}{DE fluid to scalar field mapping}}
%%============================================================================================================%%

At the background level, the DE component can be treated as an effective fluid with density $\rho_{\rm de}(z)$ and pressure $p_{\rm de}(z)$ that obeys energy conservation,
\begin{equation}
\frac{\dd\rho_{\rm de}}{\dd\ln a}=-3\bigl(\rho_{\rm de}+p_{\rm de}\bigr).
\end{equation}
In terms of redshift $z=a^{-1}-1$, this implies the \emph{signed} relation
\begin{equation}
\rho_{\rm de}+p_{\rm de}=\frac{1+z}{3}\,\frac{\dd\rho_{\rm de}}{\dd z}.
\label{eq:rhop_de_signed}
\end{equation}
The combination $\rho_{\rm de}+p_{\rm de}$ is the null-energy-condition (NEC) combination (often called the enthalpy or inertial mass density). It is the quantity that controls NEC satisfaction/violation and it will be the key source for the scalar-field kinetic term below.

A central point for the present work is that we do \emph{not} require $\rho_{\rm de}$ to be positive definite: in sign-switching scenarios, $\rho_{\rm de}(z)$ may become negative and can cross zero. Consequently, the line $w_{\rm de}=-1$ is not a global separator of physical regimes in the $(w,z)$ plane once sign-changing densities are admitted~\cite{Akarsu:2025gwi,Akarsu:2026anp,Gokcen:2026pkq}. For convenience we sometimes display the derived equation-of-state parameter $w_{\rm de}\equiv p_{\rm de}/\rho_{\rm de}$; this ratio is ill-defined at a density zero-crossing and may diverge there purely kinematically. The physically relevant discriminator is instead the sign of $\rho_{\rm de}+p_{\rm de}$ (equivalently the sign of $(1+w_{\rm de})\rho_{\rm de}$ where $w_{\rm de}$ is defined).

Whenever $w_{\rm de}$ is well defined one has $\rho_{\rm de}+p_{\rm de}=\rho_{\rm de}(1+w_{\rm de})$, and ``phantom-like'' behavior corresponds to $\rho_{\rm de}+p_{\rm de}<0$, equivalently $(1+w_{\rm de})\rho_{\rm de}<0$. This implies
\begin{equation}
\rho_{\rm de}+p_{\rm de}<0
\quad\Longleftrightarrow\quad
\begin{cases}
w_{\rm de}<-1, & \rho_{\rm de}>0,\\[2pt]
w_{\rm de}>-1, & \rho_{\rm de}<0.
\end{cases}
\label{eq:phantom_inequalities}
\end{equation}
For later reference, \cref{tab:pq-branches} summarizes the relation between the sign of $\rho_{\rm de}$, the location relative to $w_{\rm de}=-1$, and the sign of the NEC combination $\rho_{\rm de}+p_{\rm de}$ (the true physical classifier)~\cite{Akarsu:2025gwi,Akarsu:2026anp,Gokcen:2026pkq}.

\begin{table}[t]
    \centering
    \setlength{\tabcolsep}{10pt}
    \renewcommand{\arraystretch}{1.15}
    \begin{tabular}{cccc}
        \toprule
         $\rho_{\rm de}$ & $w_{\rm de}$ & $\rho_{\rm de}+p_{\rm de}$ & Classification \\
        \midrule
        $>0$ & $>-1$ & $>0$ & $p$-quintessence \\
        $<0$ & $<-1$ & $>0$ & $n$-quintessence \\
        \addlinespace
        $>0$ & $<-1$ & $<0$ & $p$-phantom \\
        $<0$ & $>-1$ & $<0$ & $n$-phantom \\
        \bottomrule
    \end{tabular}
    \caption{Classification of quintessence-like and phantom-like branches when $\rho_{\rm de}$ is allowed to take either sign. The prefixes $p$ and $n$ denote $\rho_{\rm de}>0$ and $\rho_{\rm de}<0$, respectively, while ``phantom''/``quintessence'' refer to the sign of the NEC combination $\rho_{\rm de}+p_{\rm de}$. See, e.g., Refs.~\cite{Akarsu:2025gwi,Akarsu:2026anp,Gokcen:2026pkq}.}
    \label{tab:pq-branches}
\end{table}

Defining the present critical density $\rho_{\rm c0}\equiv 3H_0^2M_{\rm Pl}^2$, the dimensionless Hubble function $E(z)\equiv H(z)/H_0$, and $\tilde\Omega_{\rm de}(z)\equiv\rho_{\rm de}(z)/\rho_{\rm c0}$, Eq.~\eqref{eq:rhop_de_signed} can be written as
\begin{equation}
\frac{\rho_{\rm de}+p_{\rm de}}{H^2M_{\rm Pl}^2}
=\frac{(1+z)\,\dd\tilde\Omega_{\rm de}/\dd z}{E^2(z)}.
\label{eq:rho_de_diff}
\end{equation}

%%============================================================================================================%%
\subsection*{\texorpdfstring{Reconstructing $\phi(z)$ and the single-field consistency condition}{Reconstructing phi(z) and the single-field consistency condition}}
%%============================================================================================================%%

Define the kinetic contribution
\begin{equation}
K \equiv \frac{\epsilon}{2}\dot{\phi}^2,
\end{equation}
so that Eq.~\eqref{eq:rhophiphi} implies
\begin{equation}
\rho_\phi+p_\phi=2K
\quad\Rightarrow\quad
\dot{\phi}^2=\epsilon(\rho_\phi+p_\phi).
\label{eq:phidot2}
\end{equation}
It follows that
\begin{equation}
\left(\frac{\dd}{\dd\ln a}\frac{\phi}{M_{\rm Pl}}\right)^2
=\frac{\dot{\phi}^2}{H^2M_{\rm Pl}^2}
=\frac{\epsilon(\rho_\phi+p_\phi)}{H^2M_{\rm Pl}^2}.
\label{eq:phi_diff}
\end{equation}
Identifying $\rho_\phi\leftrightarrow\rho_{\rm de}$ and $p_\phi\leftrightarrow p_{\rm de}$ and combining
Eqs.~\eqref{eq:rho_de_diff} and \eqref{eq:phi_diff}, we obtain the following
\begin{equation}
\epsilon\left(\frac{\dd}{\dd\ln a}\frac{\phi}{M_{\rm Pl}}\right)^2
=\frac{(1+z)\,\dd\tilde\Omega_{\rm de}/\dd z}{E^2(z)}.
\label{eq:phi_de}
\end{equation}
Equivalently, using $\dd\ln a=-\dd z/(1+z)$,
\begin{equation}
\left(\frac{\dd}{\dd z}\frac{\phi}{M_{\rm Pl}}\right)^2
=\frac{1}{\epsilon}\,\frac{1}{1+z}\,
\frac{\dd\tilde\Omega_{\rm de}/\dd z}{E^2(z)}.
\label{eq:phi_zform}
\end{equation}
Equations~\eqref{eq:phi_de}--\eqref{eq:phi_zform} determine $\phi(z)$ up to the degeneracies $\phi\to-\phi$ and $\phi\to\phi+\mathrm{const}$.

\paragraph{Single-field consistency.}
For a minimally coupled \emph{single} field with fixed $\epsilon$, the right-hand side of Eq.~\eqref{eq:phi_de} must maintain a definite sign over the reconstruction range:
\begin{equation}
\epsilon\,\frac{(1+z)\,\dd\tilde\Omega_{\rm de}/\dd z}{E^2(z)}\ge 0
\quad \text{for all } z \text{ in the range considered.}
\label{eq:sign_consistency}
\end{equation}
Thus, $\dd\tilde\Omega_{\rm de}/\dd z>0$ selects a canonical realization ($\epsilon=+1$), while $\dd\tilde\Omega_{\rm de}/\dd z<0$ selects a phantom realization ($\epsilon=-1$). If $(1+z)\,\dd\tilde\Omega_{\rm de}/\dd z$ changes sign---equivalently, if the NEC boundary $\rho_{\rm de}+p_{\rm de}=0$ is crossed---then a single minimally coupled real scalar field with fixed $\epsilon$ cannot reproduce the full history, since $\rho_\phi+p_\phi=\epsilon\dot\phi^2$ has a fixed sign [Eq.~\eqref{eq:rho_p_scalar}], see Ref.~\cite{Akarsu:2025gwi,Akarsu:2026anp}. In that case the reconstruction must be interpreted as \emph{effective} (e.g.\ arising from a quintom sector, non-canonical $P(X,\phi)$ dynamics, or another complete theory).

For an \emph{effective} one-dimensional field-space representation of histories that change sign in $\dd\tilde\Omega_{\rm de}/\dd z$, we may define a formal reconstruction variable by writing
\begin{equation}
\left(\frac{\dd}{\dd z}\frac{\phi}{M_{\rm Pl}}\right)^2
=
\frac{1}{1+z}\,
\frac{\left|\dd\tilde\Omega_{\rm de}/\dd z\right|}{E^2(z)} \, ,
\label{eq:phi_zform_multi}
\end{equation}
which simply enforces $\bigl(\dd\phi/\dd z\bigr)^2\ge 0$ at the background
level, with the right-hand side understood as the continuous extension through
points where $\dd\tilde\Omega_{\rm de}/\dd z=0$.

We stress that Eq.~\eqref{eq:phi_zform_multi} does \emph{not} define a single fundamental minimally coupled scalar theory; rather, it serves as a convenient effective mapping and naturally points toward multi-field (quintom-like) realizations when sign changes occur.

\subsection*{\texorpdfstring{Deriving $p_{\rm de}(z)$, $K(z)$, and $V(z)$}{Deriving dark-energy pressure and scalar-field functions}}

Given $\rho_{\rm de}(z)$, the associated pressure follows directly from the continuity equation:
\begin{equation}
p_{\rm de}(z)
=-\rho_{\rm de}(z)+\frac{1+z}{3}\,\frac{\dd \rho_{\rm de}(z)}{\dd z}.
\label{eq:pressure}
\end{equation}
The kinetic contribution and potential as functions of redshift are then
\begin{align}
K(z)
&=\frac12\big[\rho_{\rm de}(z)+p_{\rm de}(z)\big]
=\frac{1}{6}(1+z)\,\frac{\dd\rho_{\rm de}(z)}{\dd z},
\label{eq:K_z}
\\[3pt]
V(z)
&=\frac12\big[\rho_{\rm de}(z)-p_{\rm de}(z)\big]
=\rho_{\rm de}(z)-\frac{1+z}{6}\,\frac{\dd\rho_{\rm de}(z)}{\dd z}.
\label{eq:potential_z}
\end{align}
These relations remain regular even if $\rho_{\rm de}$ crosses zero (where $w_{\rm de}=p_{\rm de}/\rho_{\rm de}$ can diverge kinematically), and provide a robust route for reconstructing $V(\phi)$ by eliminating $z$ between $\phi(z)$ of Eq.~\eqref{eq:phi_de} and $V(z)$ of Eq.~\eqref{eq:potential_z}.

Different choices of $\rho_{\rm de}(z)$ therefore correspond to distinct field trajectories $\phi(z)$ and reconstructed potentials $V(\phi)$. When the sign-consistency condition~\eqref{eq:sign_consistency} is met, the reconstruction admits a single-field interpretation with fixed $\epsilon$; otherwise, it should be viewed as an effective one-dimensional representation of the background DE history, as in Eq.~\eqref{eq:phi_zform_multi}.

\subsection*{Perturbative interpretation and limitations}

The reconstruction developed here is a background-level mapping. At linear order, and neglecting perturbative backreaction, a fixed homogeneous input history $\rho_{\rm de}(z)$ determines the relations used to obtain $p_{\rm de}(z)$, $K(z)$, $\phi(z)$ and $V(\phi)$; these background relations are not modified by adding linear perturbations. Perturbations instead provide additional viability and stability conditions on any microscopic realization of the reconstructed background.

For a specified minimally coupled single-field completion, the perturbative sector may be introduced in the usual way by writing
\begin{equation}
\phi(t,\mathbf{x})=\bar{\phi}(t)+\delta\phi(t,\mathbf{x}) .
\end{equation}
In Newtonian gauge~\cite{Ma:1995ey},
\begin{equation}
\dd s^2=-(1+2\Phi)\dd t^2+a^2(t)(1-2\Psi)\dd\mathbf{x}^2 ,
\end{equation}
the scalar-field density and pressure perturbations associated with ${\cal L}_\phi=\epsilon X-V(\phi)$ are
\begin{align}
\delta\rho_\phi
&=
\epsilon\left(
\dot{\bar\phi}\,\delta\dot\phi
-\dot{\bar\phi}^{\,2}\Phi
\right)
+V_{,\phi}\delta\phi ,
\\
\delta p_\phi
&=
\epsilon\left(
\dot{\bar\phi}\,\delta\dot\phi
-\dot{\bar\phi}^{\,2}\Phi
\right)
-V_{,\phi}\delta\phi .
\end{align}
Equivalently, in Fourier space the field fluctuation obeys the corresponding perturbed Klein--Gordon equation,
\begin{equation}
\delta\ddot\phi_k+3H\delta\dot\phi_k+
\left(\frac{k^2}{a^2}+\epsilon V_{,\phi\phi}\right)\delta\phi_k
=
\dot{\bar\phi}\left(\dot\Phi_k+3\dot\Psi_k\right)
-2\epsilon V_{,\phi}\Phi_k ,
\end{equation}
with $\Phi=\Psi$ when anisotropic stress is negligible. Thus, once a specific single-field completion is chosen, the reconstructed background trajectory and potential provide the quantities entering the linear perturbation equations, including the effective mass term $\epsilon V_{,\phi\phi}$.

It is useful to distinguish this scalar-field perturbative description from a purely adiabatic effective-fluid closure. Although a minimally coupled scalar field can be written in fluid language when the field gradient is timelike, its pressure perturbation is not in general determined by the adiabatic relation $\delta p_{\rm de}=c_a^2\delta\rho_{\rm de}$, with $c_a^2\equiv\dot p_{\rm de}/\dot\rho_{\rm de}$. For ${\cal L}_\phi=\epsilon X-V(\phi)$ the scalar-field rest-frame sound speed is
\begin{equation}
c_{s,\phi}^2=1
\end{equation}
for either sign of $\epsilon$. For the phantom branch, $\epsilon=-1$, this statement concerns the propagation speed only and does not remove the usual ghost/effective-theory caveat associated with the wrong-sign kinetic term; see Refs.~\cite{Akarsu:2025gwi,Akarsu:2026lva} and references therein. The adiabatic sound speed is instead
\begin{equation}
c_a^2
=
\frac{\dot p_\phi}{\dot\rho_\phi}
=
1+\frac{2V_{,\phi}}{3\epsilon H\dot{\bar\phi}} ,
\end{equation}
where the last expression is understood on intervals where $\dot{\bar\phi}\neq0$, equivalently away from NEC-boundary points for a single-field branch. Therefore $c_{s,\phi}^2$ and $c_a^2$ generally differ, and an equivalent fluid description of the scalar field contains the corresponding non-adiabatic pressure contribution. Imposing a purely adiabatic barotropic fluid closure would therefore define a perturbative model that is background-degenerate with, but physically distinct from, the scalar-field completion considered here.

There are, however, two important caveats. First, for dark-energy histories with a density zero crossing, fractional variables such as $\delta_{\rm de}\equiv\delta\rho_{\rm de}/\rho_{\rm de}$ become ill-defined at $\rho_{\rm de}=0$, even when $\delta\rho_{\rm de}$, $\delta p_{\rm de}$, $\delta\phi$ and the metric perturbations remain finite. A perturbative analysis of such models should therefore be formulated in terms of regular field, density, pressure or gauge-invariant variables rather than ratios that divide by $\rho_{\rm de}$~\cite{Bouhmadi-Lopez:2025spo}. Second, for histories that cross the NEC boundary and require an effective extended-sector interpretation, such as the full \texttt{CPL} benchmark considered here, the perturbation dynamics are not uniquely fixed by the one-dimensional background reconstruction alone. They depend on the chosen quintom, non-canonical, interacting, or otherwise extended completion, including its sound speed, entropy perturbations and stability properties. Consequently, linear perturbations do not change the reconstructed potential for a fixed background history, but they can provide additional observational and theoretical filters that distinguish between background-degenerate realizations. A full perturbative treatment, including growth and stability constraints, is left for future work.

\section{Methodology}
\label{sec:method}

Our analysis proceeds in two stages, separating the background-level reconstruction of an effective scalar description from the subsequent model-selection step performed directly in field space. 
In Stage~1, for a prescribed dark-energy density history $\rho_{\rm de}(z)$ (equivalently
$\tilde\Omega_{\rm de}(z)\equiv\rho_{\rm de}(z)/\rho_{\rm c0}$, normalised so that $\tilde\Omega_{\rm de}(0)=\Omega_{\rm de0}$),
we reconstruct an effective homogeneous scalar-field representation by determining $p_{\rm de}(z)$, $K(z)$, $\phi(z)$, and the associated potential $V(\phi)$ using the relations derived in~\cref{sec:scalarfield}. 
The benchmark histories considered here include the Chevallier--Polarski--Linder (CPL) parametrisation~\cite{Chevallier:2000qy,Linder:2002et}, widely used in recent analyses~\cite{DESI:2024mwx,DESI:2025zgx,DES:2025sig}, as well as smooth $\tanh$-based switching/emergent profiles motivated by recent studies of cosmological tensions and late-time transitions~\cite{Akarsu:2019hmw,Akarsu:2021fol,Akarsu:2022typ,Akarsu:2023mfb,Paraskevas:2024ytz,Yadav:2024duq,Akarsu:2024qsi,Akarsu:2024eoo,Akarsu:2024nas,Souza:2024qwd,Akarsu:2025gwi,Akarsu:2025dmj,Akarsu:2025ijk,Escamilla:2025imi,Akarsu:2025nns,Kibris:2026cqq,Li:2019yem,Yang:2020ope,DeFelice:2020cpt,Ben-Dayan:2023rgt}. In Stage~2, treating the reconstructed $V(\phi)$ as a \emph{target function}, we perform Bayesian model comparison in \emph{potential space} across a set of analytic potentials, quantifying which potential shapes best reproduce the reconstructed target.

\paragraph*{Stage 1: Background reconstruction.}
Given $\tilde\Omega_{\rm de}(z)$ and the fixed background parameters $(H_0,\Omega_{\rm m0},\Omega_{\rm r0})$, we compute $E(z)$ from Eq.~\eqref{eq:E2_from_rho_norm} and obtain $p_{\rm de}(z)$, $K(z)$, and $V(z)$ from Eqs.~\eqref{eq:pressure}--\eqref{eq:potential_z}. 
We then reconstruct the field trajectory by integrating Eq.~\eqref{eq:phi_de}, fixing the integration constant by the convention $\phi(0)=0$ (the overall sign is chosen for plotting convenience).
Whenever the single-field sign-consistency condition, Eq.~\eqref{eq:sign_consistency}, holds over the redshift interval considered, the reconstruction admits a consistent single-field realization with fixed kinetic signature $\epsilon$. 
If the sign condition fails (as in the CPL benchmark, which crosses the NEC boundary $\rho_{\rm de}+p_{\rm de}=0$ while $\rho_{\rm de}>0$), the mapping cannot arise from a single minimally coupled real scalar field with fixed $\epsilon$; in that case we interpret $V(\phi)$ as an \emph{effective} one-dimensional representation of an extended scalar sector (e.g.\ a minimal quintom-like or non-canonical completion), and (where needed below) we restrict to a monotonic branch on which $V(\phi)$ is single-valued.

\paragraph*{Stage 2: Bayesian comparison in potential space.}
From the reconstructed target $V_{\rm tar}(\phi)$ we construct mock potential-space data points $\{(\phi_i,V_i^{\rm(mock)})\}$ by sampling $\phi_i$ throughout the reconstructed field range and drawing $V_i^{\rm(mock)} = V_{\rm tar}(\phi_i) + \delta V_i$ with Gaussian scatter $\delta V_i\sim \mathcal{N}(0,\sigma_i^2)$, where $\sigma_i^2=(\sigma_{\rm rel}|V_{\rm tar}(\phi_i)|)^2+\sigma_{\rm abs}^2$. We then fit each analytic candidate $V(\phi;\boldsymbol{\theta},M)$ using nested sampling and rank models by the Bayesian evidence $\log\mathcal{Z}_M$ (\cref{subsec:potentials,subsec:bayes,subsec:prior}). We emphasize that this evidence ranking quantifies agreement \emph{with the reconstructed target in field space} under the adopted noise model, rather than a direct fit to cosmological observations.

%%============================================================================================================%%
\subsection{Dark-energy density profiles}
\label{subsec:de_profiles}
%%============================================================================================================%%

We consider phenomenological prescriptions for the normalised dark-energy density $\tilde\Omega_{\rm de}(z)\equiv\rho_{\rm de}(z)/\rho_{\rm c0}$, normalised such that $\tilde\Omega_{\rm de}(0)=\Omega_{\rm de0}$.
These benchmark histories are used as inputs to the reconstruction pipeline described above and are chosen to span both standard smooth dynamical behavior and late-time transition scenarios (including sign-switching and emergent profiles).

\paragraph{Chevallier--Polarski--Linder (\texttt{CPL}) profile.}
The \texttt{CPL} parametrisation assumes an evolving dark-energy equation-of-state (EoS) parameter~\cite{Chevallier:2000qy,Linder:2002et}
\begin{equation}
w(a)=w_0+w_a(1-a), \qquad a=\frac{1}{1+z},
\end{equation}
which implies the density evolution
\begin{equation}
\tilde\Omega_{\mathrm{de}}^{\rm CPL}(z)
  = \Omega_{\mathrm{de}0}\,
    (1+z)^{3(1+w_0+w_a)}
    \exp\!\left[-3w_a\,\frac{z}{1+z}\right].
\label{eq:rho_cpl}
\end{equation}
Here, $w_0$ and $w_a$ denote the present-day value and first-order redshift evolution of the EoS parameter, respectively.
For the benchmark parameter choice adopted in this work, the \texttt{CPL} history crosses the NEC boundary $\rho_{\rm de}+p_{\rm de}=0$ while $\rho_{\rm de}>0$ by construction; equivalently, $(1+z)\,\dd\tilde\Omega_{\rm de}/\dd z$ changes sign over the redshift interval.
In this case the NEC-boundary crossing coincides with an EoS crossing of $w_{\rm de}=-1$ (the usual ``phantom-divide'' line), i.e.\ the evolution transitions from $w_{\rm de}<-1$ at higher redshift to $w_{\rm de}>-1$ toward the late universe.
Consequently, a single minimally coupled real scalar field with fixed kinetic signature cannot realise the full \texttt{CPL} evolution.
We therefore treat the \texttt{CPL} reconstruction as an \emph{effective} scalar-sector mapping and interpret the reconstructed $K(z)$ and $V(z)$ as total/effective quantities of an extended scalar sector (e.g.\ a minimal quintom completion), leaving an explicit multifield construction for future work.

\paragraph{Sign-changing $\tanh$ profile.}
We consider a smooth sign-switching history
\begin{equation}
\tilde\Omega_{\mathrm{de}}(z)
  = \Omega_{\mathrm{de}0}\,
    \frac{\tanh[\eta(z_{\dagger}-z)]}{\tanh(\eta z_{\dagger})},
\label{eq:rho_sign}
\end{equation}
where $z_{\dagger}$ denotes the transition redshift and $\eta$ controls the sharpness~\cite{Akarsu:2022typ,Akarsu:2024qsi,Akarsu:2024eoo}.
This profile allows $\rho_{\mathrm{de}}(z)$ to cross zero near $z_{\dagger}$, i.e.\ it realises a \emph{sign transition} in the dark-energy density.
The reconstruction is formulated in terms of $\rho_{\rm de}(z)$ and $\dd\rho_{\rm de}/\dd z$ (and hence $p_{\rm de}(z)$ via Eq.~\eqref{eq:pressure}), so it remains well defined through the zero-crossing; any divergence of the derived EoS parameter $w_{\rm de}(z)=p_{\rm de}/\rho_{\rm de}$ at $\rho_{\rm de}=0$ is purely kinematic.
Moreover, for $\eta>0$ one has $\dd\tilde\Omega_{\rm de}/\dd z<0$ throughout the redshift range considered, so the background evolution is consistent with a single-field realization on the phantom branch ($\epsilon=-1$).

\paragraph{Shifted $\tanh$ profile.}
We also consider a shifted profile that preserves the smooth transition behavior while remaining strictly positive,
\begin{equation}
\tilde\Omega_{\mathrm{de}}(z)
  = \Omega_{\mathrm{de}0}\,
    \frac{1+\tanh[\eta(z_{\dagger}-z)]}{1+\tanh(\eta z_{\dagger})}.
\label{eq:rho_shift}
\end{equation}
By construction, $\rho_{\mathrm{de}}(z)>0$ for all finite redshifts and $\tilde\Omega_{\rm de}\to 0$ at high redshift, realising an emergent late-time dark-energy component, see, e.g., Ref.~\cite{DeFelice:2020cpt}. As in the unshifted case, $\dd\tilde\Omega_{\rm de}/\dd z<0$ over the range considered (for $\eta>0$), so the dynamics can again be
modeled consistently by a single-field phantom-branch realization at the background level.

\medskip
\noindent
Together, these benchmark histories provide complementary test cases for the reconstruction and subsequent potential-space comparison.
In particular, they allow us to contrast (i) a standard phenomenological history that crosses the NEC boundary (CPL)---requiring an effective extended-sector interpretation---with (ii) smooth $\tanh$-based transition histories that satisfy the single-field sign-consistency condition.
By comparing the reconstructed $V(\phi)$ targets with analytic potential families in~\cref{subsec:potentials,subsec:bayes,subsec:prior}, we can quantify which potential shapes best reproduce each benchmark evolution.\footnote{In our work, we fix different dark-energy profiles to representative values from the literature: $h=0.7$, $T_{\rm CMB}=2.7255\,\rm K$, $N_{\rm eff}=3.046$, $\Omega_{\rm m}=0.31$, $w_{0}=-0.838$, $w_{a}=-0.62$, $z_{\dagger}=1.8$, and $\eta=5$, respectively~\cite{Planck:2018vyg,DESI:2025zgx,Akarsu:2023mfb}.}

%%============================================================================================================%%
\subsection{Scalar-field potentials}
\label{subsec:potentials}
%%============================================================================================================%%

To carry out the potential-space comparison described above, we consider a representative set of scalar-field potentials that are widely used in model building for late-time acceleration and span a broad range of theoretical motivations and phenomenological behaviors~\cite{Vazquez:2020ani}. The set is deliberately heterogeneous: it includes monotonic potentials that admit scaling or tracking behavior, periodic (axion-like) potentials, hilltop (thawing-type) potentials, and templates with localised or smooth transition features. This provides a compact basis for testing which classes of field-space shapes can reproduce the reconstructed target $V(\phi)$ associated with each benchmark history. In the template potentials below, $\phi$ denotes the reduced-Planck-unit field, i.e.\ $\phi\equiv \phi_{\rm phys}/M_{\rm Pl}$, so the arguments of $\exp$, $\cos$, and $\tanh$ are dimensionless and the parameters $(\lambda,\nu,f,w,\phi_{\rm c},\ldots)$ are dimensionless.

Among the simplest and most studied is the \textit{exponential potential},
\begin{equation}
V(\phi)=V_{0}e^{-\lambda\phi}+V_{1},
\end{equation}
originally introduced in the context of scaling solutions and attractor behavior~\cite{Wetterich:1987fm,Ratra:1987rm}. Here, $V_{0}$ sets the overall scale, $\lambda$ controls the steepness (and hence the rate of field evolution), and $V_{1}$ provides an effective constant offset. In canonical quintessence, exponential forms can yield scaling/attractor trajectories in which the field evolution becomes only weakly sensitive to initial conditions, making them a useful baseline template for monotonic, smooth dynamics.

We also consider the \textit{pseudo-Nambu--Goldstone boson} (PNGB) or \textit{cosine potential},
\begin{equation}
V(\phi)=A\left[1+\cos\!\left(\tfrac{\phi-\phi_{\rm c}}{f}\right)\right],
\end{equation}
motivated by axion-like fields arising from spontaneous symmetry breaking~\cite{Frieman1995,Caldwell:1997ii,Choi2000}. The amplitude $A$ fixes the dark-energy scale, $f$ is the (decay) constant that controls the periodicity and curvature, and $\phi_{\rm c}$ sets the location of the minimum in field space. Such potentials naturally realise ``thawing'' behavior when the field starts near a flat region and subsequently rolls as Hubble friction decreases.

As a flexible template for thawing dynamics we include the \textit{hilltop quartic potential},
\begin{equation}
V(\phi)=V_{0}-\tfrac{1}{2}m^{2}(\phi-\phi_{\rm c})^{2}
        +\tfrac{\lambda_{4}}{4}(\phi-\phi_{\rm c})^{4},
\end{equation}
which corresponds to an expansion around a local maximum and is commonly used in small-field inflation and thawing quintessence studies~\cite{Dutta:2008qn,Chiba2009,Matos2009}. Here $V_{0}$ is the potential height at the hilltop, $m^{2}$ controls the curvature near the maximum, and $\lambda_{4}$ stabilises the potential at large field values (with $\lambda_{4}>0$ ensuring boundedness). In the late-time context, this form captures the possibility that the field remains near the hilltop for an extended period before evolving more rapidly.

We further adopt a regularised \textit{inverse power-law potential},
\begin{equation}
V(\phi)=V_{0}\bigl(|\phi-\phi_{\rm c}|+\varepsilon\bigr)^{-\alpha},
\end{equation}
originally proposed in early quintessence constructions and later developed as a tracker potential~\cite{Peebles:1987ek,Steinhardt:1999nw}. This class admits tracker-like solutions in which a wide range of initial conditions converge to a common late-time trajectory, thus alleviating fine tuning of initial conditions~\cite{Ratra:1987rm,Caldwell:1997ii,Scherrer:2007pu}. The parameter $\alpha$ controls the steepness and tracking behavior, while the small regulator $\varepsilon$ ensures regularity at $\phi=\phi_{\rm c}$.

To capture localised features in field space we include the \textit{Gaussian bump} potential,
\begin{equation}
V(\phi)=A\exp\!\left[-\frac{(\phi-\phi_{\rm c})^{2}}{2w^{2}}\right],
\end{equation}
characterised by amplitude $A$, width $w$, and centre $\phi_{\rm c}$. This template provides a simple phenomenological description of a transient feature (localised deformation of the potential) that can induce a brief departure from otherwise smooth evolution.

Finally, we consider a smooth \textit{shifted-$\tanh$ potential},
\begin{equation}
V(\phi)=\frac{\Lambda(\xi_{1}+1)}{2}
         -\frac{\Lambda(\xi_{1}-1)}{2}\tanh[\nu(\phi-\phi_{\rm c})],
\end{equation}
which interpolates between two asymptotic plateau values and has been used to model smooth late-time transitions in the dark-energy sector~\cite{Akarsu:2025dmj,Akarsu:2026lva}. Here $\Lambda$ sets the overall scale, $\xi_{1}$ controls the relative heights of the plateaus, $\nu$ controls the sharpness of the transition, and $\phi_{\rm c}$ sets the transition location in field space. This form is particularly well suited to emulate reconstructed histories with a rapid but continuous change in the effective vacuum-energy level.

Together, these potentials provide a compact yet diverse basis for our Bayesian model comparison, allowing us to quantify which classes of field-space shapes best reproduce the reconstructed target $V(\phi)$ associated with each benchmark dark-energy history.

%%============================================================================================================%%
\subsection{Bayesian comparison framework}
\label{subsec:bayes}
%%============================================================================================================%%

To quantify which analytic scalar-field potentials can best reproduce a reconstructed target potential $V_{\rm tar}(\phi)$,
we perform Bayesian inference directly in \emph{potential space}. Each candidate model $M$ is specified by a parametric form
$V(\phi;\boldsymbol{\theta},M)$ with parameters $\boldsymbol{\theta}$.

Given $V_{\rm tar}(\phi)$ over the field range probed by the reconstruction, we construct a set of \emph{mock}
potential-space data points by sampling $\{\phi_i\}_{i=1}^{N}$ across this field range and assuming a Gaussian distribution
around the true potential values,
\begin{equation}
V_i^{\rm (mock)} = V_{\rm tar}(\phi_i) + \delta V_i,
\qquad
\delta V_i \sim \mathcal{N}(0,\sigma_i^2),
\label{eq:data_mock}
\end{equation}
where the variance includes both a relative and an absolute contribution,
\begin{equation}
\sigma_i^2 = \bigl(\sigma_{\rm rel}|V_{\rm tar}(\phi_i)|\bigr)^2 + \sigma_{\rm abs}^2.
\end{equation}
We denote the resulting dataset by
\begin{equation}
D \equiv \{(\phi_i,\,V_i^{\rm (mock)},\,\sigma_i)\}_{i=1}^{N}.
\end{equation}
We stress that this likelihood quantifies agreement with the reconstructed target \emph{in field space} under the adopted noise model, rather than a direct fit to cosmological observables. Accordingly, the resulting evidence ranking is conditional on the adopted potential-space noise prescription and on the prior ranges assigned to the analytic potentials; in the present work it should therefore be interpreted as a controlled baseline comparison within this setup.

For each model, we adopt weakly informative priors $\pi(\boldsymbol{\theta}|M)$ implemented via mappings from a unit hypercube to physical parameter space: parameters that can take either sign (e.g. offsets or field centres) follow symmetric uniform priors, while positive-definite quantities (e.g. amplitudes, widths, and couplings) follow log-uniform priors. This ensures a wide, yet physically consistent coverage of the parameter space. The explicit prior ranges used in this work are given in~\cref{subsec:prior}.

Assuming independent Gaussian errors, the log-likelihood function is
\begin{equation}
\begin{aligned}
\log\mathcal{L}(D|\boldsymbol{\theta},M)
= -\frac{1}{2}\sum_{i=1}^{N}
\Bigg[ &
\frac{\bigl(V_i^{\rm (mock)}-V(\phi_i;\boldsymbol{\theta},M)\bigr)^2}{\sigma_i^2}\\
&+ \log\!\bigl(2\pi\sigma_i^2\bigr)
\Bigg].
\end{aligned}
\end{equation}
The corresponding posterior distribution,
\begin{equation}
p(\boldsymbol{\theta}|D,M) \propto \mathcal{L}(D|\boldsymbol{\theta},M)\,\pi(\boldsymbol{\theta}|M),
\end{equation}
and the Bayesian evidence,
\begin{equation}
\mathcal{Z}(D|M) =
\int \mathcal{L}(D|\boldsymbol{\theta},M)\,\pi(\boldsymbol{\theta}|M)\,\dd\boldsymbol{\theta},
\end{equation}
are evaluated using nested sampling, which efficiently explores multi-modal posteriors and provides accurate estimates of $\log\mathcal{Z}(D|M)$ for model comparison~\cite{Padilla:2019mgi}.

From posterior-weighted samples, we compute predictive realizations of the potential $V^{(s)}(\phi)$ and derive the median reconstruction $V_{\rm med}(\phi)$ together with its $68\%$ and $95\%$ credible intervals, representing the posterior uncertainty on the inferred potential. Models are ranked according to their Bayesian evidences to identify the potential families that best reproduce $V_{\rm tar}(\phi)$ while accounting for model complexity. For likelihood and evidence evaluation we use the \texttt{UltraNest} nested sampler algorithm~\cite{Buchner1,Buchner2,Buchner3,Buchner4,nested1}. For posterior contours, we use \texttt{ChainConsumer}~\cite{2016JOSS....1...45H}.

\subsection{Priors}
\label{subsec:prior}

\begin{table*}[!t]
\centering
\renewcommand{\arraystretch}{1.25}
\setlength{\tabcolsep}{4pt}
\begin{tabular}{l l}
\hline\hline
\textbf{Model} & \textbf{Prior Ranges} \\
\hline
\textbf{Exponential} &
$V_0,\,V_1 \in [-4V_{\rm sc},\,4V_{\rm sc}],\quad 
 \lambda \in [10^{-3},\,50]_{\log}$ \\
\textbf{Gaussian Bump} &
$A \in [-4V_{\rm sc},\,4V_{\rm sc}],\;
 \phi_{\rm c} \in [-2\phi_{\max},\,2\phi_{\max}],\;
 w \in [10^{-4},\,\max(5\times10^{-3},\,3\phi_{\max})]_{\log}$ \\
\textbf{Shifted Tanh} &
$\Lambda \in [-4V_{\rm sc},\,4V_{\rm sc}],\;
 \xi_1 \in [-4,4],\;
 \nu \in [10^{-2},\,10^{2}]_{\log},\;
 \phi_{\rm c} \in [-2\phi_{\max},\,2\phi_{\max}]$ \\
\textbf{PNGB} &
$A \in [0,\,4V_{\rm sc}],\;
 f \in [10^{-3},\,\max(3\times10^{-3},\,3\phi_{\max})]_{\log},\;
 \phi_{\rm c} \in [-2\phi_{\max},\,2\phi_{\max}]$ \\
\textbf{Inverse Power Law} &
$V_0\in[0,\,4V_{\rm sc}],\;
 \alpha \in [0.1,5],\;
 \phi_{\rm c} \in [-2\phi_{\max},\,2\phi_{\max}],\;
 \varepsilon \in [10^{-6},\,\max(10^{-3},\phi_{\max})]_{\log}$ \\
\textbf{Hilltop Quartic} &
$V_0 \in [-4V_{\rm sc},\,4V_{\rm sc}],\;
 m^{2} \in [10^{-6},\,14]_{\log},\;
 \lambda_4 \in [10^{-6},\,10^{-1}]_{\log},\;
 \phi_{\rm c} \in [-2\phi_{\max},\,2\phi_{\max}]$ \\
\hline\hline
\end{tabular}
\caption{Prior ranges adopted for each scalar-field potential. $V_{\rm sc}=\max(|V_{\min}|,|V_{\max}|)$ denotes the characteristic potential scale inferred
from the reconstructed $V(\phi)$, and $\phi_{\max}=\max|\phi|$ denotes the field-domain width. Log-uniform priors are indicated by the subscript ``$\log$''.}
\label{tab:prior_ranges}
\end{table*}

Since the reconstructed \emph{target} potentials $V_{\rm tar}(\phi)$ for the $\tanh$ and shifted-$\tanh$ histories share the same qualitative structure---smooth, single-peaked, and single-valued in field space (cf.\ \cref{fig:tanh_de_profile2,fig:shift_de_profile2})---we focus on one representative $\tanh$ case for the Bayesian analysis; repeating the same procedure for the shifted-$\tanh$ target leads to the same qualitative evidence ranking. For the \texttt{CPL} and $\tanh$ benchmarks we generate mock potential-space data points $V_i^{\rm (mock)}$ according to
Eq.~\eqref{eq:data_mock}, adopting $\sigma_{\rm rel}=0.1$ and $\sigma_{\rm abs}=0.05$. The resulting $\{V_i^{\rm (mock)}\}$ are shown in \cref{fig:band_cpl,fig:fit_band_tanh} (black points with error bars).
We fit these mock data using the analytic potential families described in~\cref{subsec:potentials}.
The prior ranges adopted for each potential are listed in~\cref{tab:prior_ranges}, where $V_{\rm sc}=\max(|V_{\min}|,|V_{\max}|)$ and $\phi_{\max}=\max|\phi|$ are computed from the corresponding reconstructed target $V_{\rm tar}(\phi)$ over the fitted field range.

The priors adopted for the scalar-field potentials follow a unified, weakly informative prescription, designed to be broad enough to cover the field range and potential scale implied by the reconstructed target while remaining otherwise agnostic:

\begin{enumerate}
\item \textbf{Amplitude and offset parameters.}
Parameters that set an overall scale or additive offset (e.g.\ $V_{0}$, $V_{1}$, $A$, $\Lambda$) are assigned broad symmetric uniform ranges whenever the model does not impose a sign restriction,
\[
    V_{0},\,V_{1},\,A,\,\Lambda \in [-4V_{\rm sc},\,4V_{\rm sc}],
\]
where $V_{\rm sc}\equiv \max\!\bigl(|V_{\min}|,|V_{\max}|\bigr)$ denotes a characteristic scale of the reconstructed potential over the fitted field range. 
For models requiring positivity (e.g.\ PNGB amplitude $A$ and the inverse-power-law normalisation $V_0$), we restrict to the corresponding positive subset, as in~\cref{tab:prior_ranges}.

\item \textbf{Field-location parameters.}
Parameters controlling the location of features in the field space (e.g.\ $\phi_{\rm c}$) are assigned uniform priors,
\[
    \phi_{\rm c} \in [-2\phi_{\max},\,2\phi_{\max}],
\]
with $\phi_{\max}\equiv \max|\phi|$ determined from the reconstructed field-domain width. This ensures that minima, maxima, transition points, or localised features can occur anywhere within (and modestly beyond) the field range probed by the reconstruction.

\item \textbf{Log-uniform priors for positive-definite shape parameters.}
Parameters that are strictly positive and span multiple orders of magnitude (e.g.\ $\lambda$, $w$, $\nu$, $\varepsilon$, $m^{2}$, $\lambda_{4}$) are assigned log-uniform (scale-invariant) priors,
\[
    x \in [x_{\min},\,x_{\max}]_{\log},
\]
which avoids privileging any particular scale \emph{a priori} and enables an unbiased exploration of both sharp and smooth potential features.

\item \textbf{Model-specific regularity constraints.}
Where required, we impose minimal physical constraints for internal consistency: for example, $A>0$ in PNGB and inverse-power-law models; $m^{2}>0$ and $\lambda_{4}>0$ in the hilltop quartic potential to ensure local stability/boundedness; and $\xi_{1}\in[-4,4]$ for the shifted-$\tanh$ potential to restrict the relative plateau asymmetry. These conditions encode only basic regularity and do not otherwise impose structure on the fitted $V(\phi)$.
\end{enumerate}

Overall, the priors are intentionally broad and weakly informative, so that the posterior constraints and evidence ranking are driven primarily by the reconstructed target in potential space (and the adopted noise model), rather than by prior boundaries.

%%============================================================================================================%%
\section{Results}
\label{sec:results}
%%============================================================================================================%%

\begin{figure*}[!t]
    \centering
    % First row
    \includegraphics[width=0.40\linewidth]{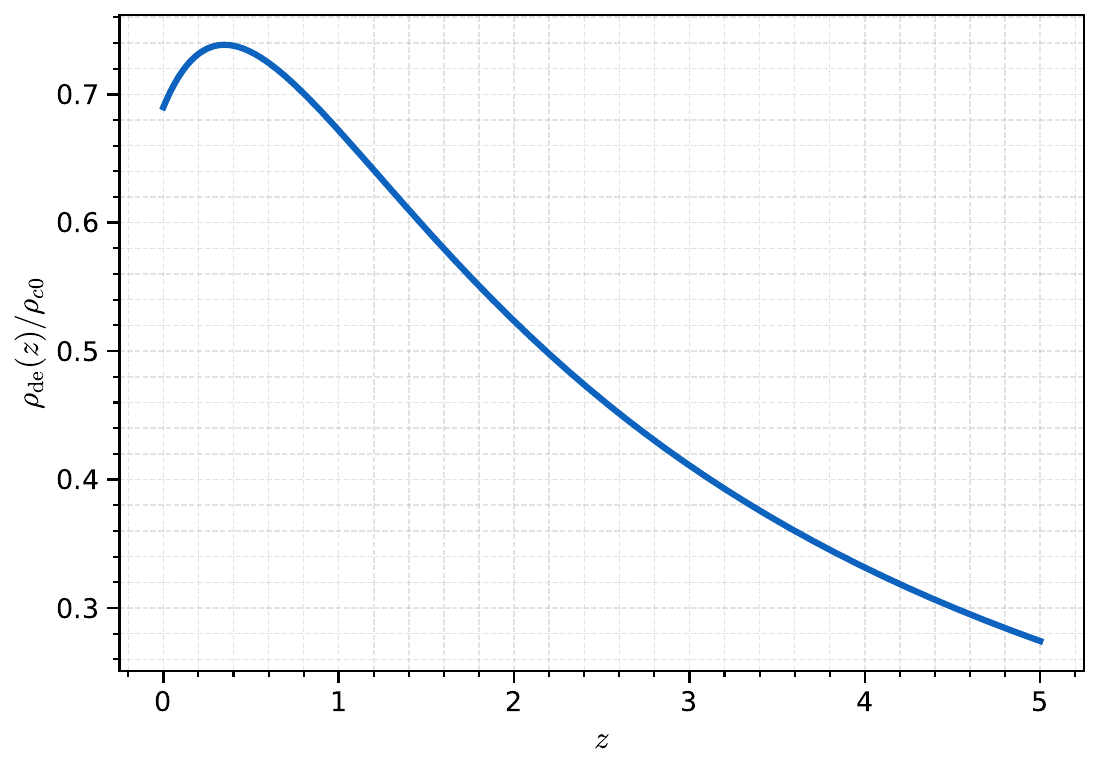}
     \includegraphics[width=0.40\linewidth]{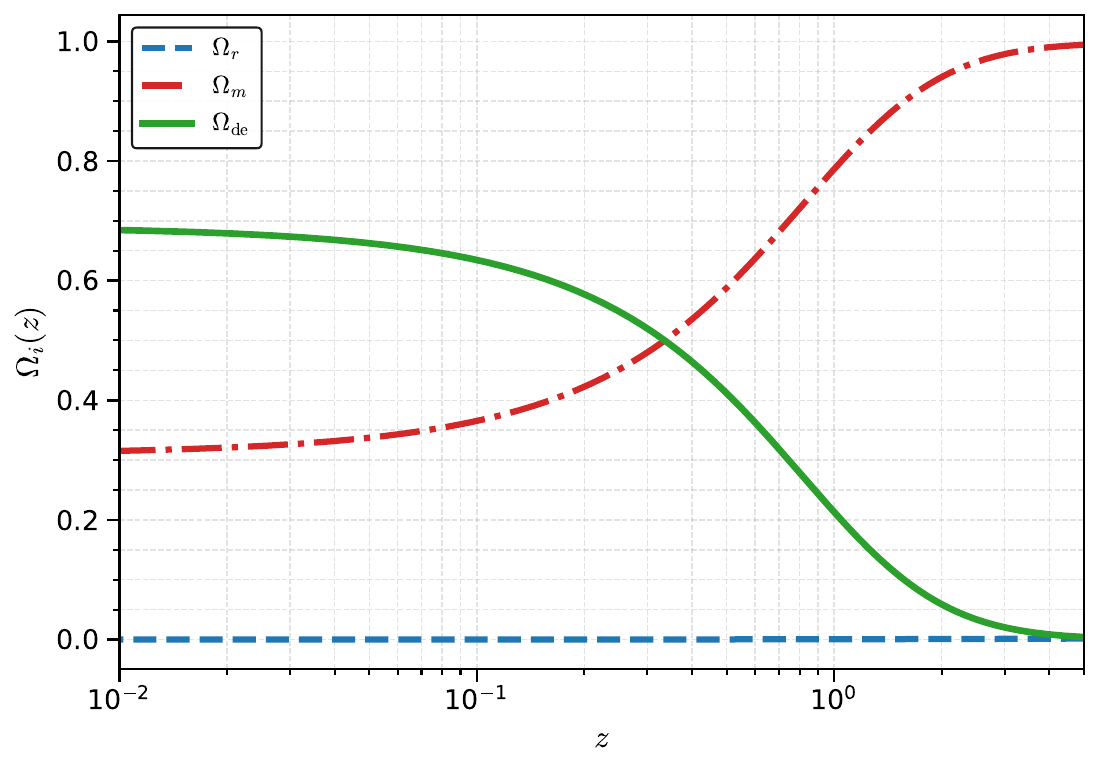}
      \includegraphics[width=0.40\linewidth]{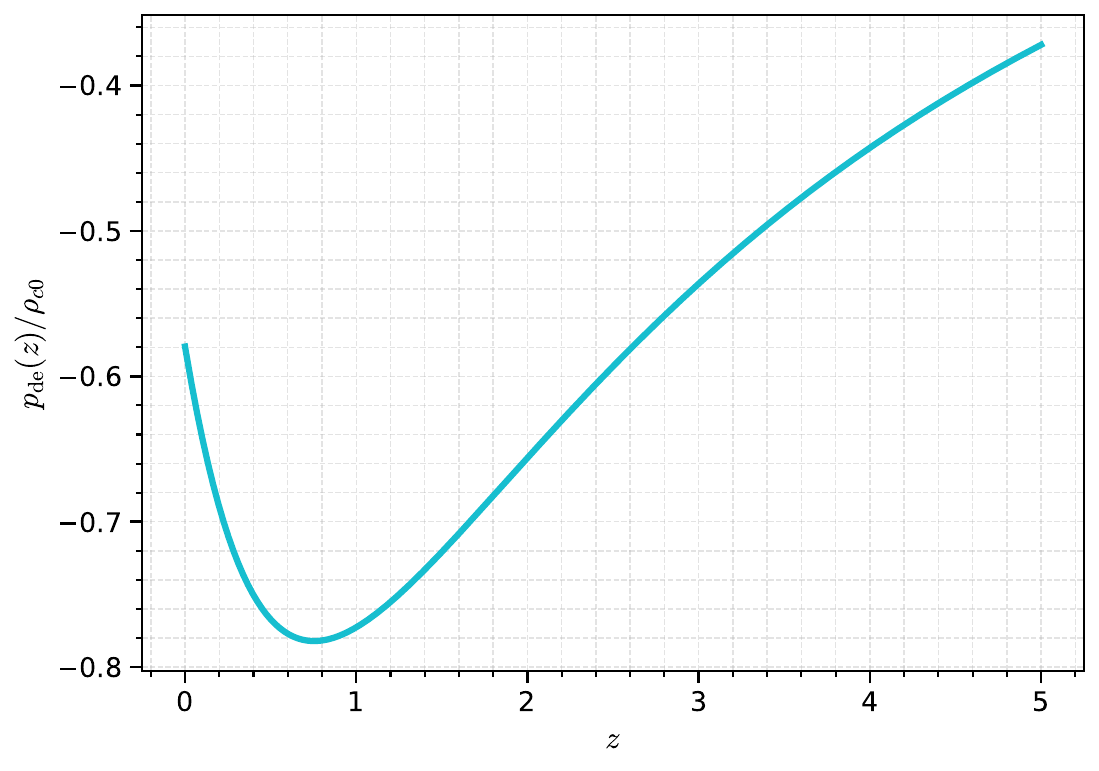}
       \includegraphics[width=0.40\linewidth]{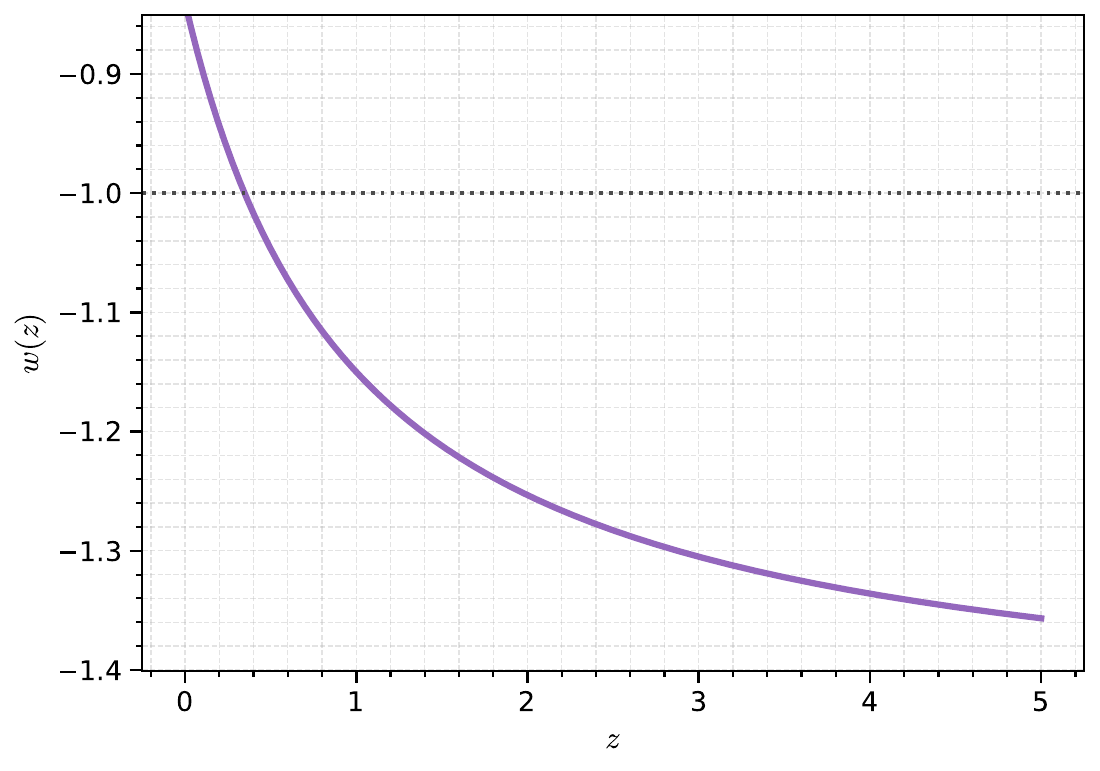}
 \caption{
Background evolution for the CPL dark-energy profile (effective-fluid benchmark).
\textbf{Top left:} normalized dark-energy density $\tilde\Omega_{\rm de}(z)=\rho_{\rm de}(z)/\rho_{\rm c0}$.
\textbf{Top right:} density parameters $\Omega_{\rm r}(z)$, $\Omega_{\rm m}(z)$, and $\Omega_{\rm de}(z)$.
\textbf{Bottom left:} dark-energy pressure $p_{\rm de}(z)/\rho_{\rm c0}$ obtained from the continuity equation.
\textbf{Bottom right:} effective equation-of-state parameter $w_{\rm de}(z)=p_{\rm de}(z)/\rho_{\rm de}(z)$ (dotted line marks $w_{\rm de}=-1$).
For CPL, $\rho_{\rm de}>0$ by construction, so the $w_{\rm de}=-1$ line coincides with the NEC boundary $\rho_{\rm de}+p_{\rm de}=0$.
In this benchmark, $\rho_{\rm de}$ remains positive while the DE history crosses the NEC boundary (i.e.\ the phantom-divide line) as $z$ decreases toward the present, transitioning from the $p$-phantom regime ($\rho_{\rm de}+p_{\rm de}<0$, $w_{\rm de}<-1$) to the $p$-quintessence regime ($\rho_{\rm de}+p_{\rm de}>0$, $w_{\rm de}>-1$).
}

    \label{fig:cpl_de_profile1}
\end{figure*}

\begin{figure*}[!t]
    \centering
       \includegraphics[width=0.40\linewidth]{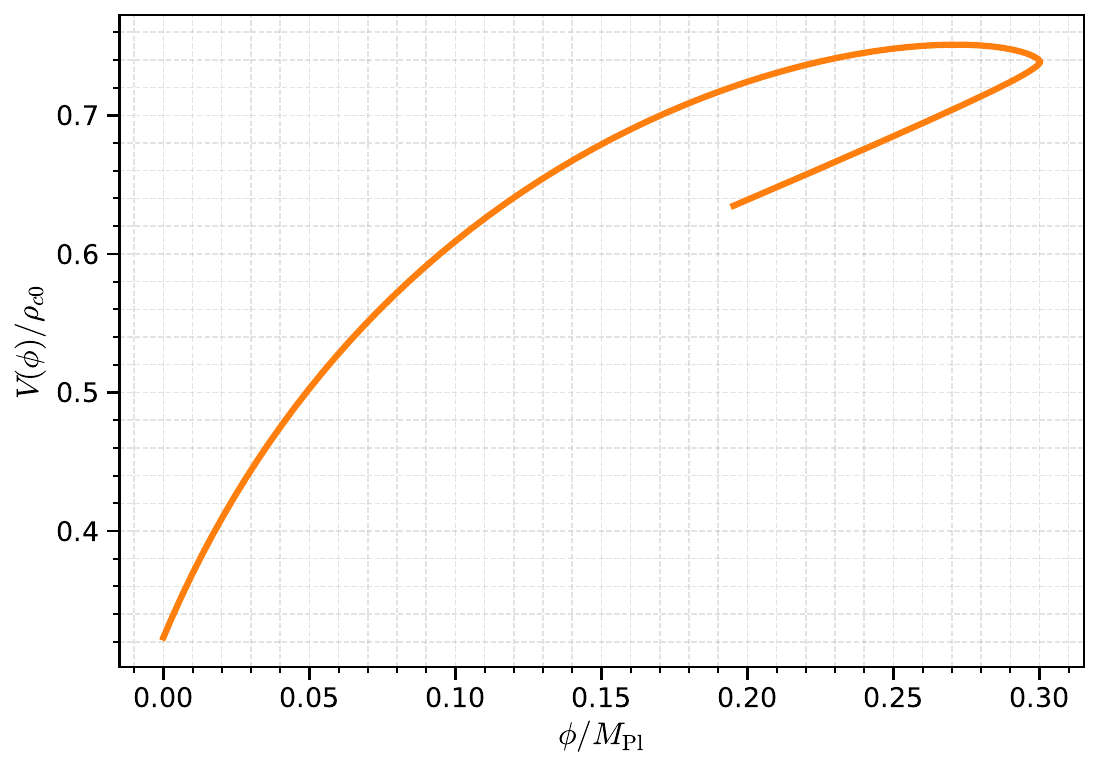}
    \includegraphics[width=0.40\linewidth]{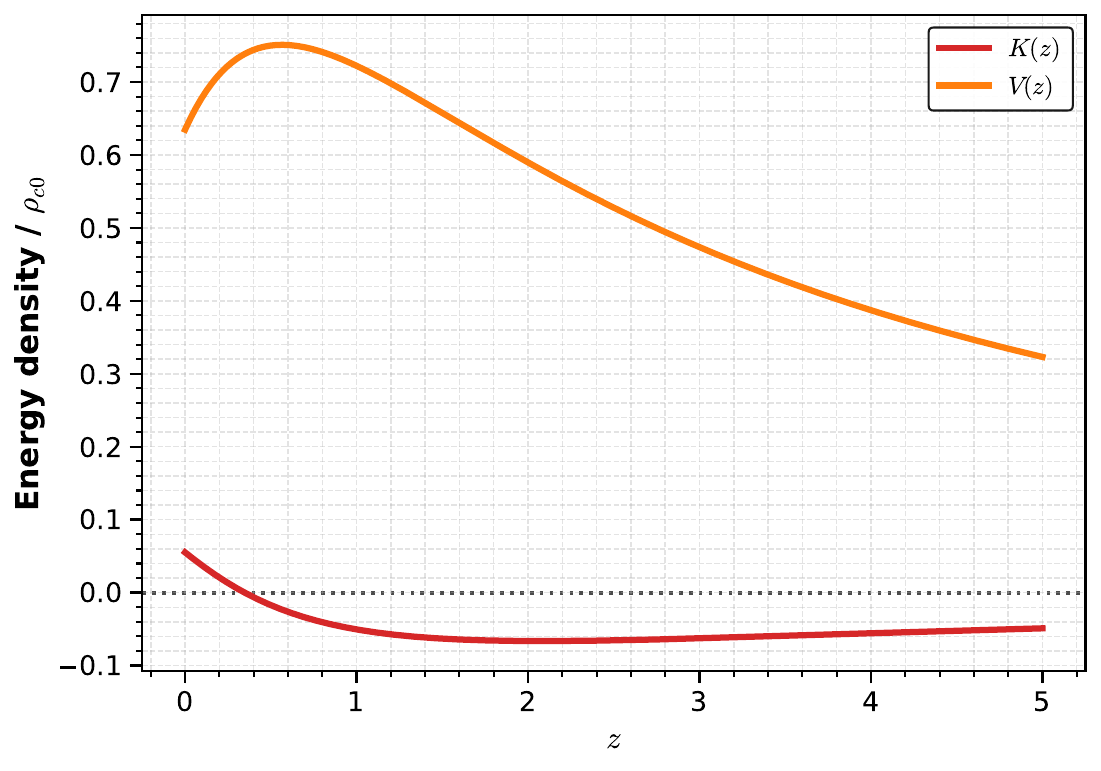}
        \includegraphics[width=0.40\linewidth]{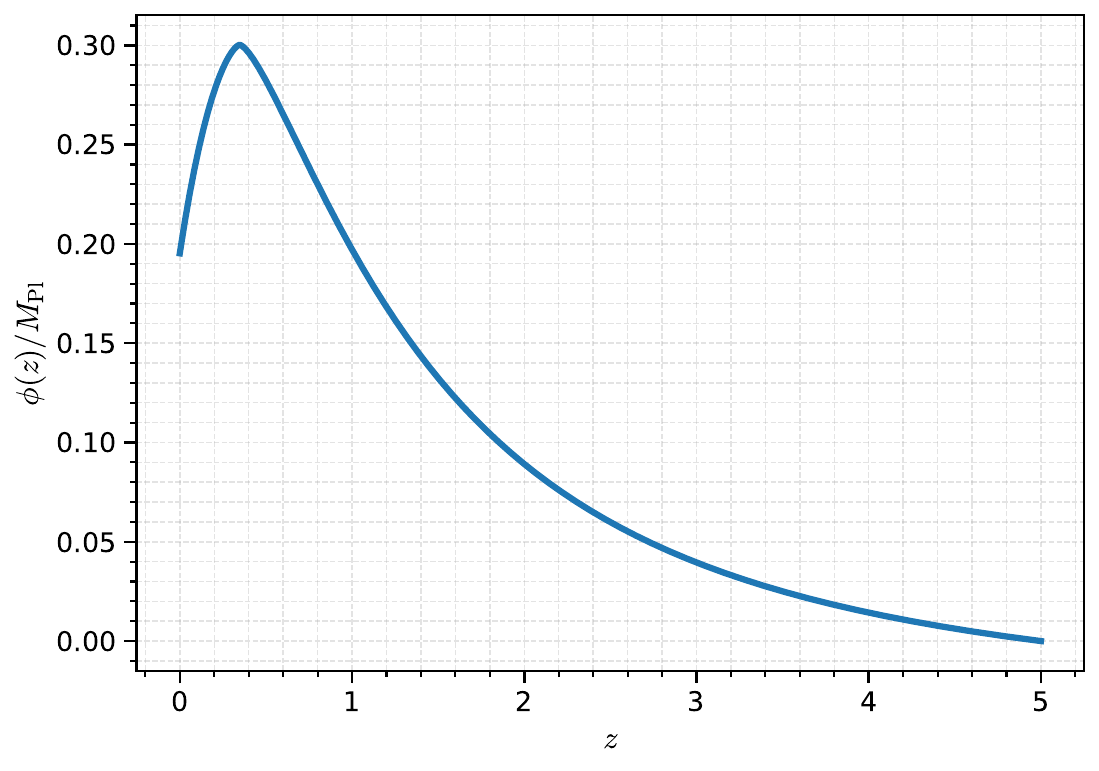}
    \includegraphics[width=0.40\linewidth]{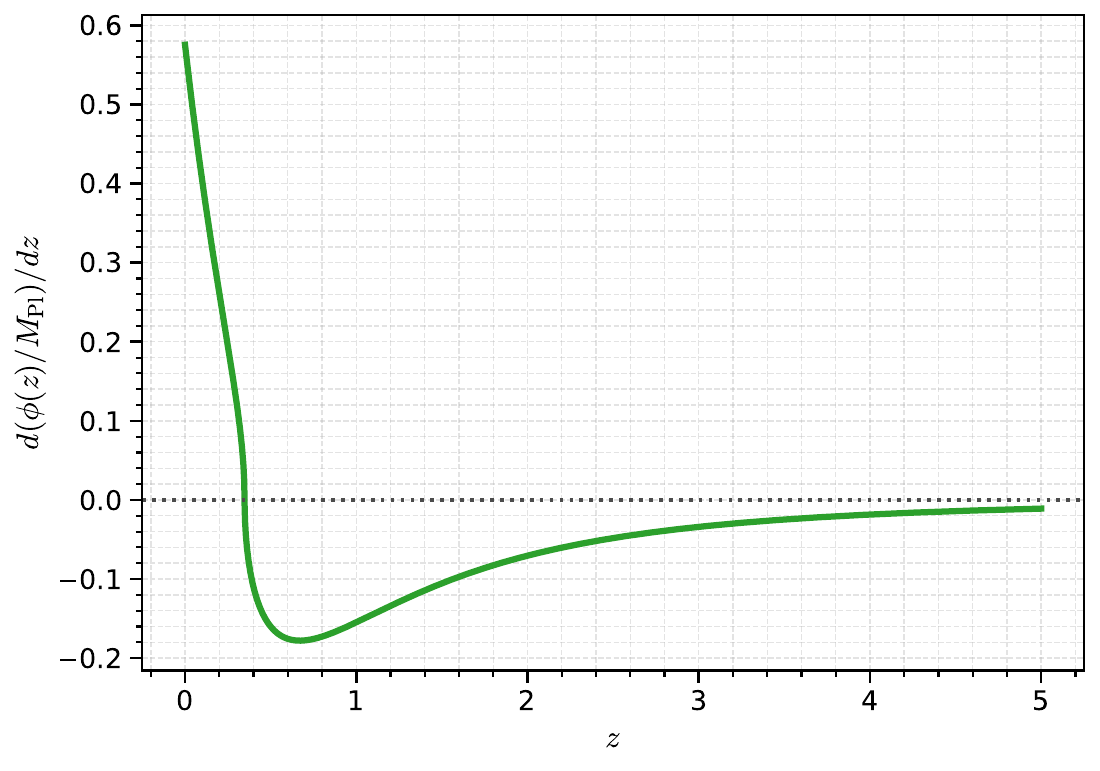}
\caption{
Effective scalar-sector reconstruction associated with the \texttt{CPL} benchmark (treated as an effective-fluid history).
\textbf{Top left:} reconstructed (effective) potential $V(\phi)/\rho_{\rm c0}$ obtained by eliminating $z$ between $V(z)$ and $\phi(z)$; the loop/multivalued relation reflects the non-monotonic $\phi(z)$ induced by the NEC-boundary crossing.
\textbf{Top right:} effective kinetic and potential contributions, $K(z)/\rho_{\rm c0}$ and $V(z)/\rho_{\rm c0}$, with $K(z)=\tfrac12(\rho_{\rm de}+p_{\rm de})$; the sign change of $K(z)$ marks the NEC boundary $\rho_{\rm de}+p_{\rm de}=0$, which for \texttt{CPL} (where $\rho_{\rm de}>0$ by construction) is equivalent to $w_{\rm de}=-1$ (the usual phantom-divide line).
\textbf{Bottom left:} reconstructed field trajectory $\phi(z)/M_{\rm Pl}$.
\textbf{Bottom right:} $\dd(\phi/M_{\rm Pl})/\dd z$, showing a turning point ($\dd\phi/\dd z=0$) coincident with the sign change in $K(z)$.
Because $K(z)$ changes sign from negative to positive as $z$ decreases toward the present (i.e.\ the evolution moves from the $p$-phantom to the $p$-quintessence regime in~\cref{tab:pq-branches}), a single minimally coupled real scalar field with fixed $\epsilon=\pm1$ cannot realise the full \texttt{CPL} history; the displayed $\phi(z)$ and $V(\phi)$ should therefore be interpreted as an effective one-dimensional mapping of an extended scalar sector (e.g.\ a quintom-like or non-canonical completion).
}

    \label{fig:cpl_de_profile2}
\end{figure*}

\begin{figure*}[!t]
    \centering
    \includegraphics[width=0.40\linewidth]{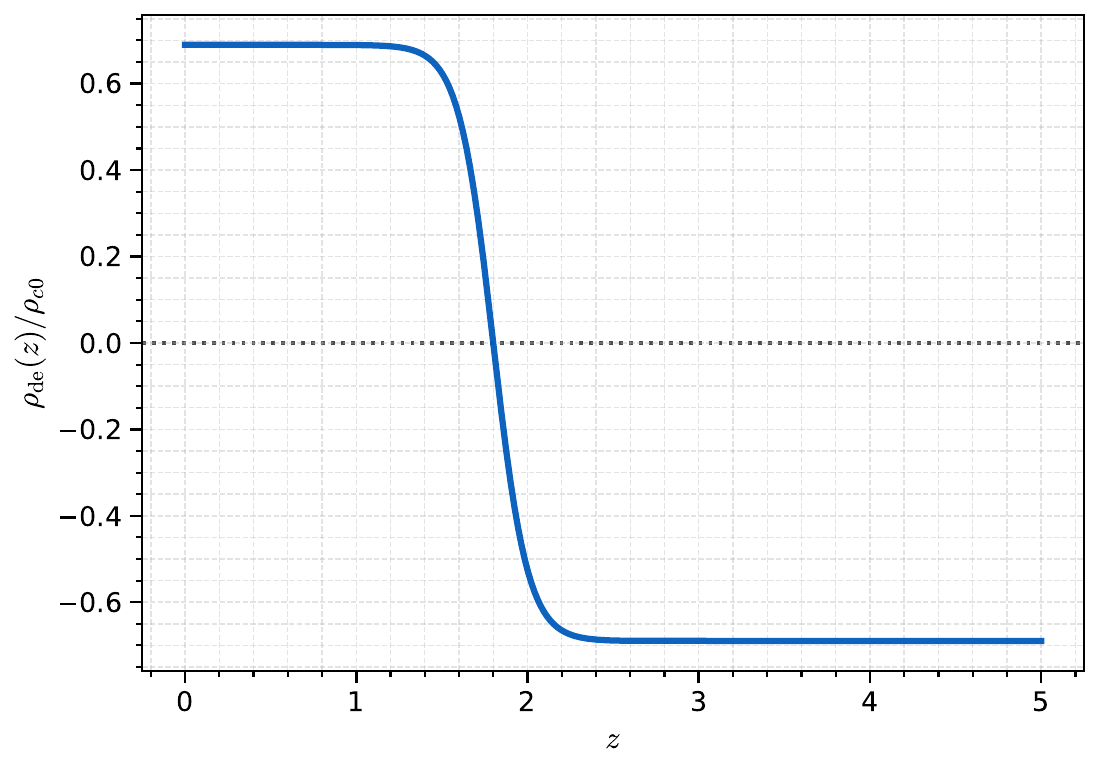}
     \includegraphics[width=0.40\linewidth]{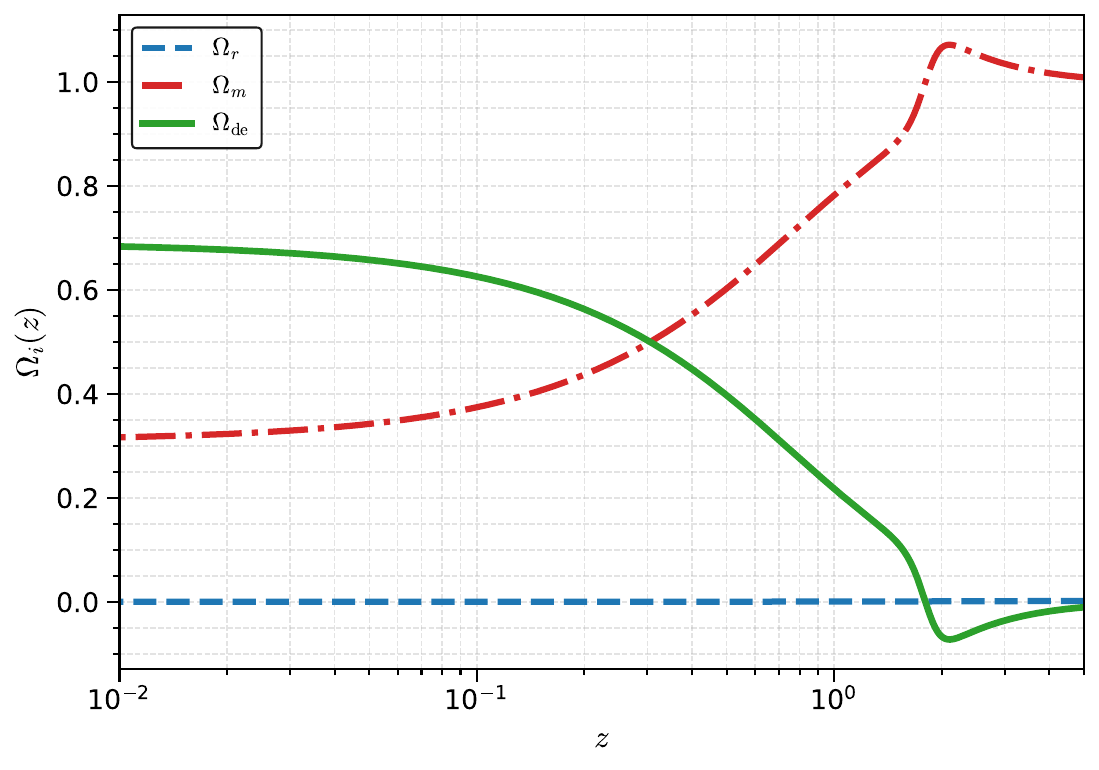}
    \includegraphics[width=0.40\linewidth]{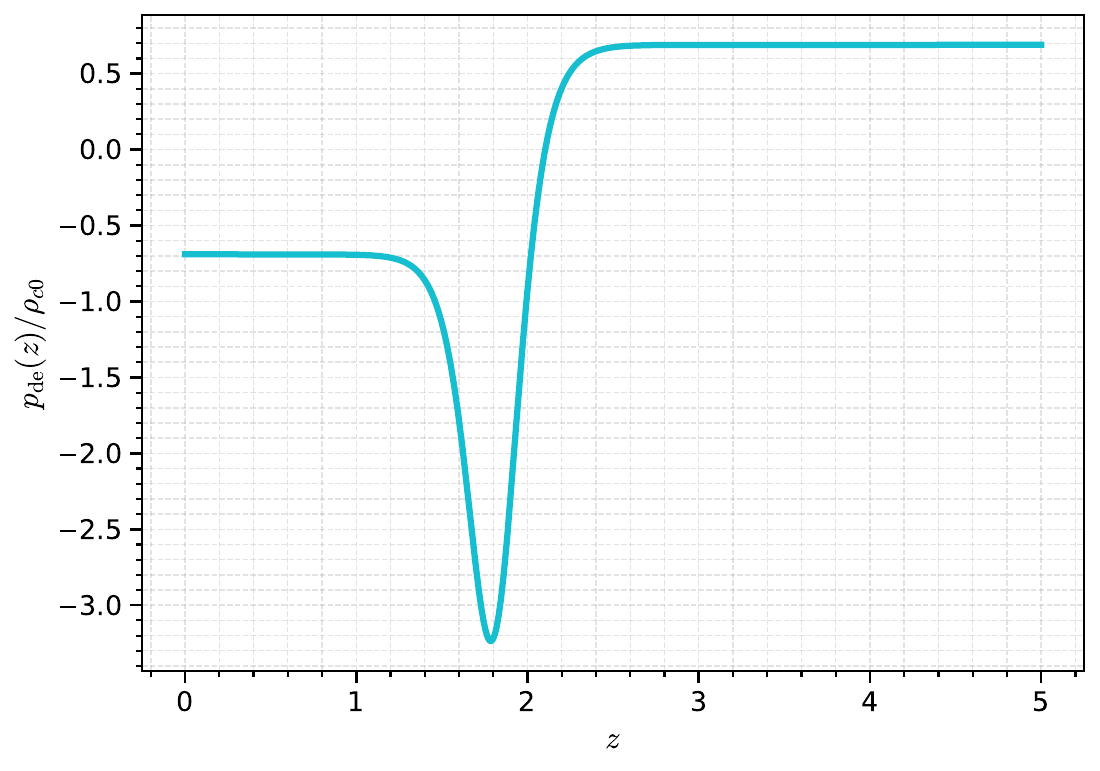}
    \includegraphics[width=0.40\linewidth]{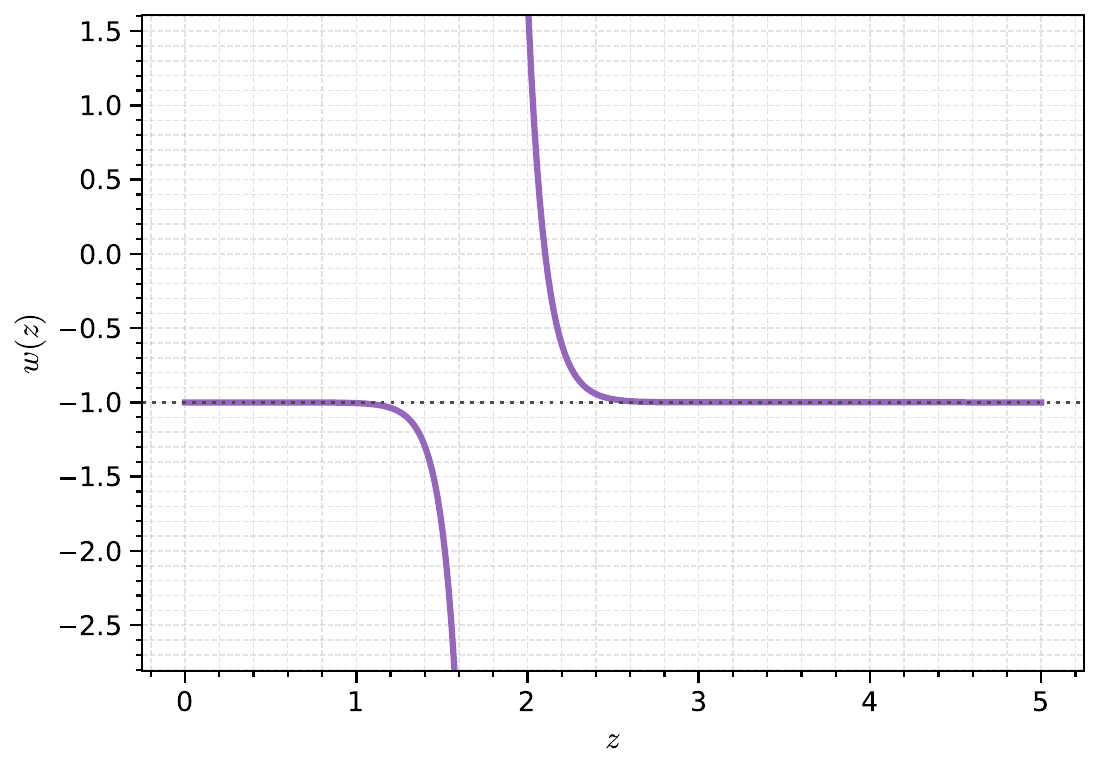}

\caption{
Background evolution for the sign-changing $\tanh$ dark-energy density (a smooth mirror AdS$\rightarrow$dS sign-switching transition).
\textbf{Top left:} normalized density $\tilde\Omega_{\rm de}(z)=\rho_{\rm de}(z)/\rho_{\rm c0}$, which decreases monotonically and smoothly interpolates from a positive late-time plateau to a negative high-$z$ plateau, crossing $\rho_{\rm de}=0$ at the transition redshift $z_\dagger$.
\textbf{Top right:} fractional densities $\Omega_{\rm r}(z)$, $\Omega_{\rm m}(z)$, and $\Omega_{\rm de}(z)$, showing standard radiation/matter domination at early times and the late-time emergence of the DE component; $\Omega_{\rm de}$ crosses zero and becomes negative beyond $z_\dagger$.
\textbf{Bottom left:} pressure $p_{\rm de}(z)/\rho_{\rm c0}$ inferred from the continuity equation, which remains finite and exhibits a rapid excursion across the transition.
\textbf{Bottom right:} derived equation-of-state parameter $w_{\rm de}(z)=p_{\rm de}/\rho_{\rm de}$; the divergence at $\rho_{\rm de}=0$ is purely kinematic (a ratio singularity). The physically meaningful classifier is instead the NEC combination $\rho_{\rm de}+p_{\rm de}=\tfrac13(1+z)\,\dd\rho_{\rm de}/\dd z$:
for $\eta>0$ one has $\dd\tilde\Omega_{\rm de}/\dd z<0$ over the range shown, implying $\rho_{\rm de}+p_{\rm de}<0$ throughout (phantom/NEC-violating).
In the notation of~\cref{tab:pq-branches}, the evolution lies on the $p$-phantom branch for $z<z_{\dagger}$ and on the $n$-phantom branch for $z>z_{\dagger}$.
}

    \label{fig:tanh_de_profile1}
\end{figure*}

\begin{figure*}[!t]
    \centering
    % 2nd row
    \includegraphics[width=0.40\linewidth]{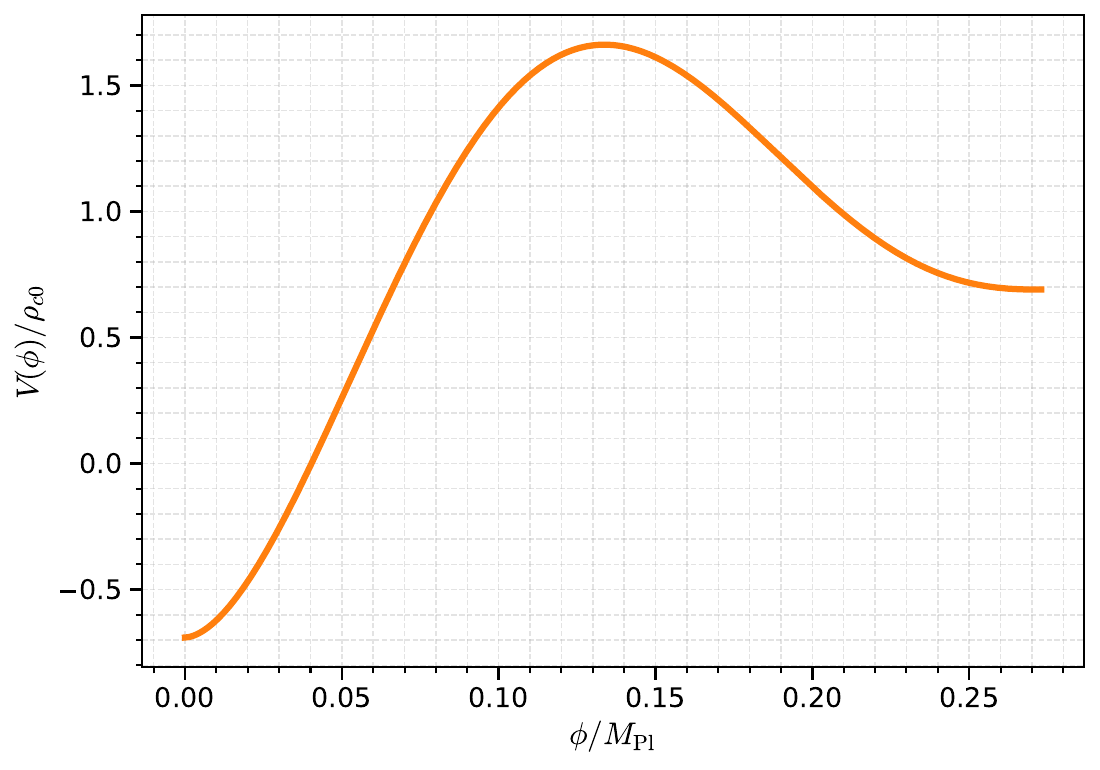}
    % 3rd row
    \includegraphics[width=0.40\linewidth]{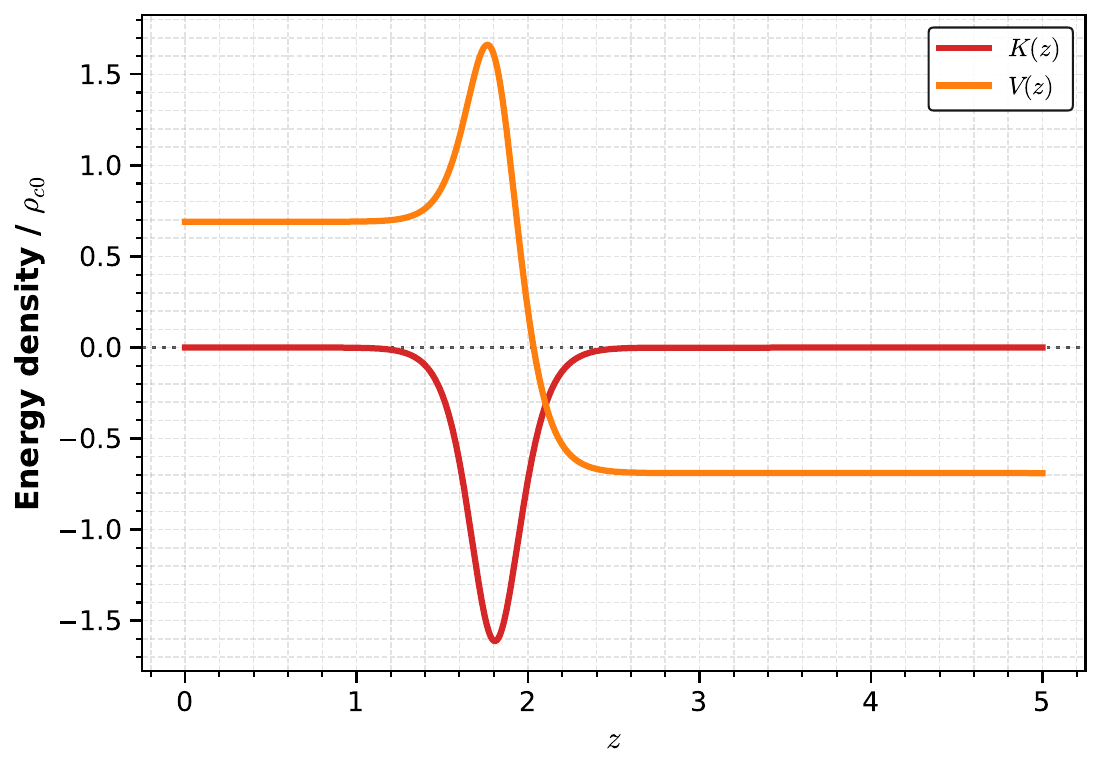}
     \includegraphics[width=0.40\linewidth]{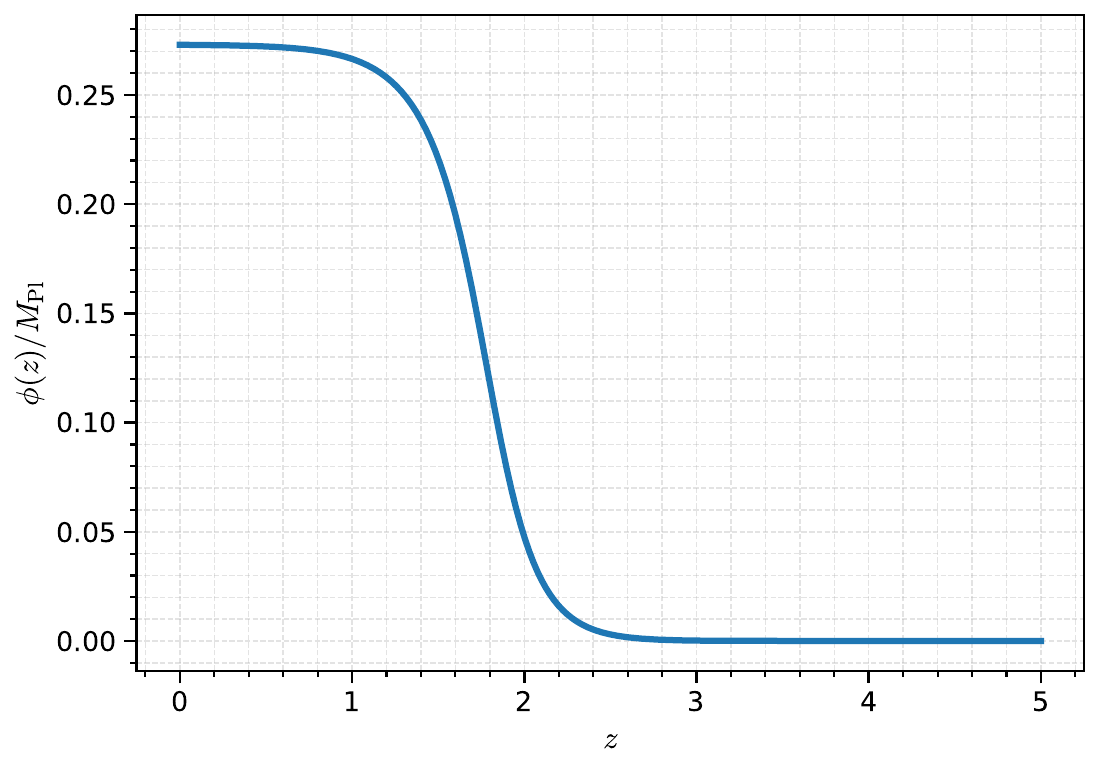}
    \includegraphics[width=0.40\linewidth]{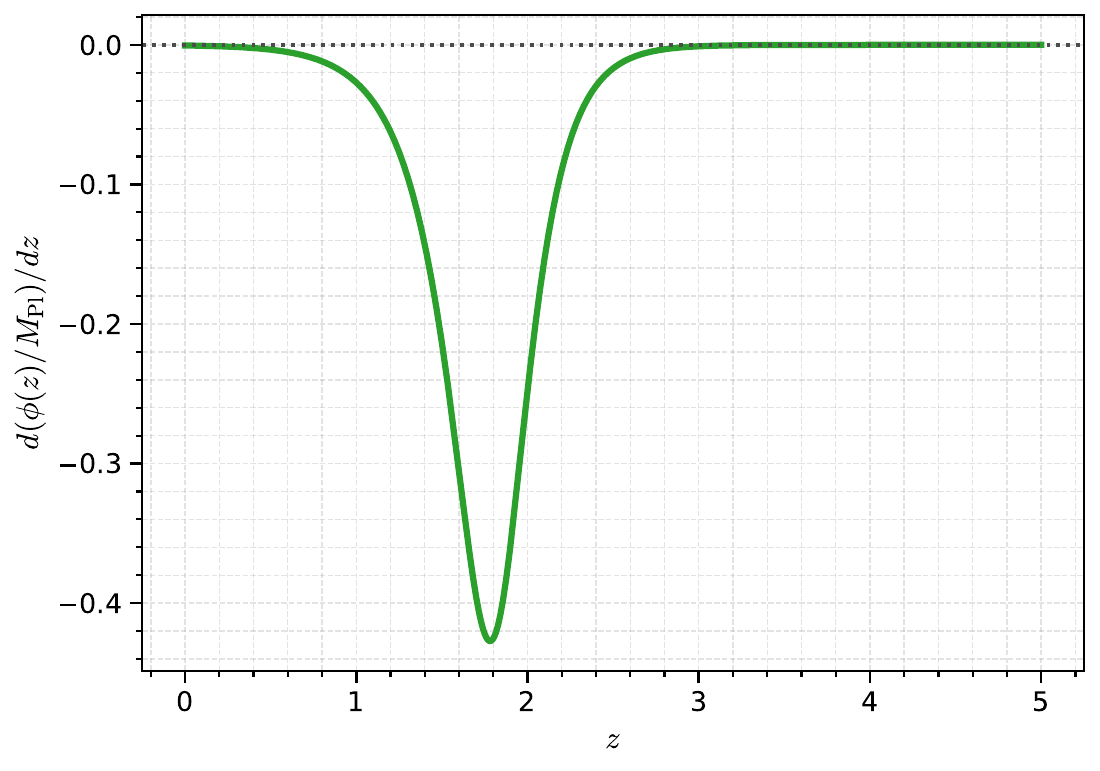} 

\caption{
Scalar-field reconstruction for the sign-changing $\tanh$ (mirror AdS$\rightarrow$dS) dark-energy profile.
\textbf{Top left:} reconstructed effective potential $V(\phi)/\rho_{\rm c0}$ obtained by eliminating $z$ between $V(z)$ and $\phi(z)$; the potential is smooth and single-valued and exhibits a broad maximum around the transition, interpolating between an AdS-like (negative) plateau at high redshift and a dS-like (positive) value at late times.
\textbf{Top right:} effective kinetic and potential contributions $K(z)/\rho_{\rm c0}$ and $V(z)/\rho_{\rm c0}$, with $K(z)=\tfrac12(\rho_{\rm de}+p_{\rm de})$; although $\rho_{\rm de}$ crosses zero, $K(z)$ remains negative throughout, indicating that the evolution stays on the phantom/NEC-violating branch and does not cross the NEC boundary.
\textbf{Bottom left:} reconstructed field trajectory $\phi(z)/M_{\rm Pl}$, showing that the field is nearly frozen at both early and late times and evolves mainly around the transition redshift.
\textbf{Bottom right:} $\dd(\phi/M_{\rm Pl})/\dd z$, sharply localized near the transition.
Over the range shown $\dd\tilde\Omega_{\rm de}/\dd z<0$, so the single-field sign-consistency condition selects a phantom realization with $\epsilon=-1$ at the effective background level.
}

    \label{fig:tanh_de_profile2}
\end{figure*}

\begin{figure*}[ht]
    \centering
    % First row
     \includegraphics[width=0.40\linewidth]{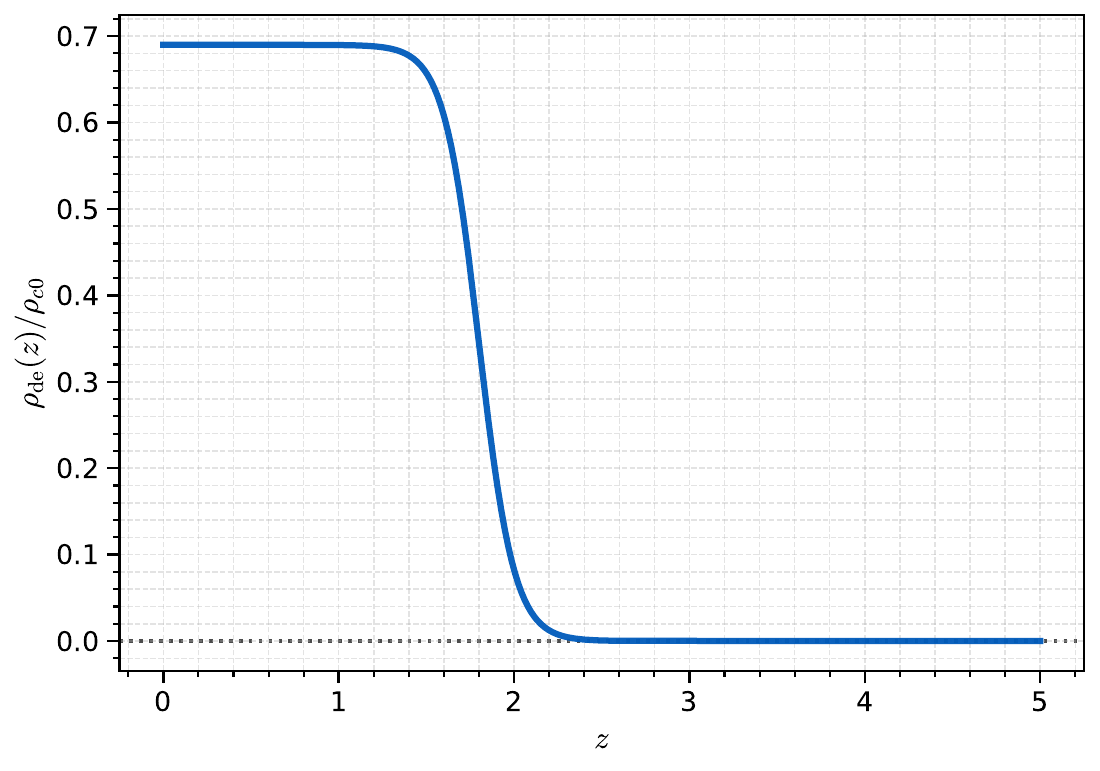}
      \includegraphics[width=0.40\linewidth]{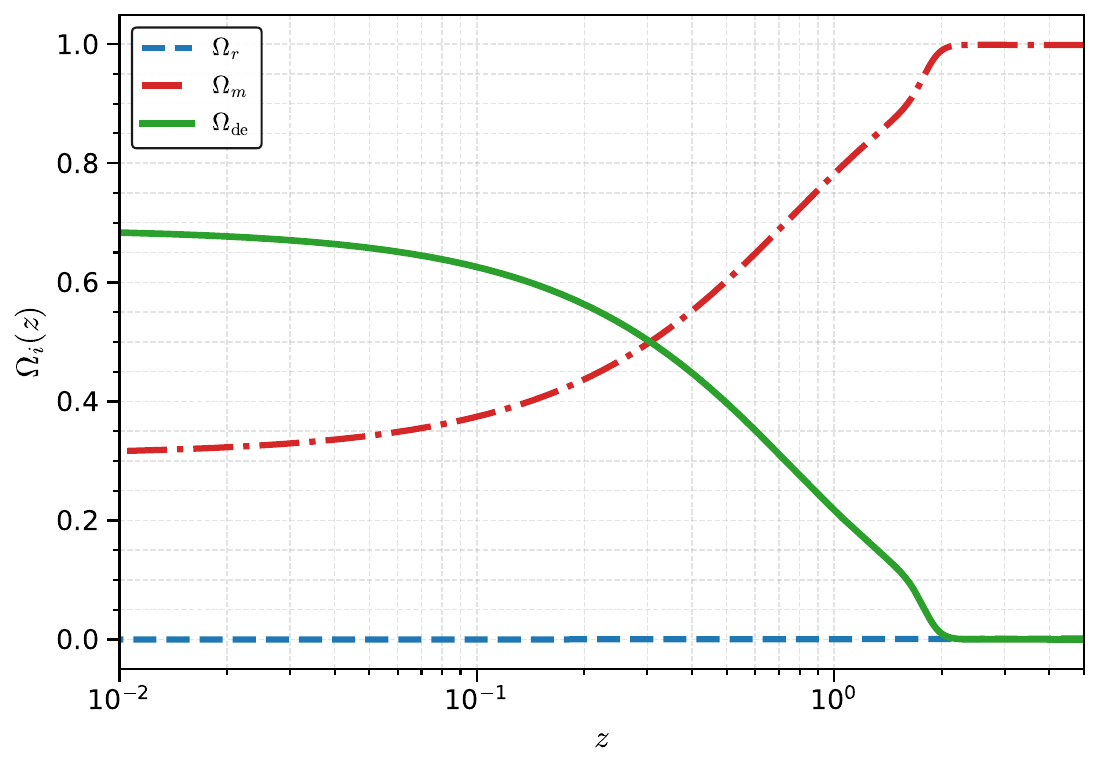}
    \includegraphics[width=0.40\linewidth]{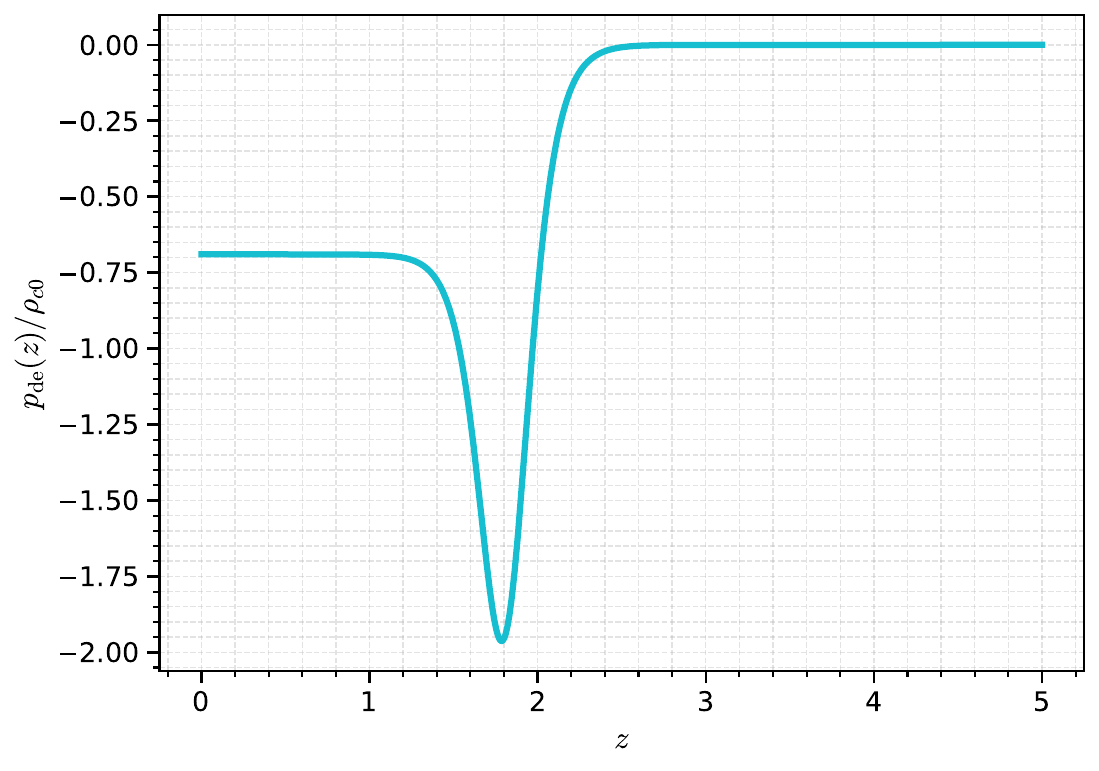}
    \includegraphics[width=0.40\linewidth]{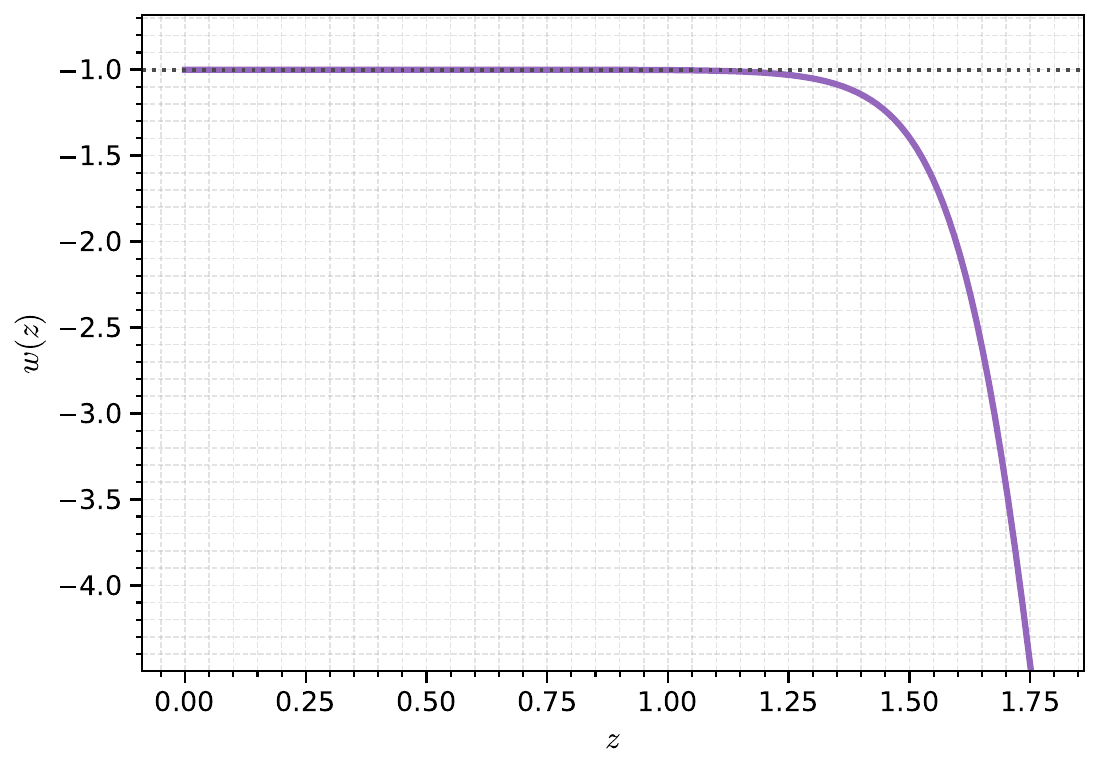}
\caption{
Background evolution for the shifted-$\tanh$ (emergent, positive-definite) dark-energy profile.
\textbf{Top left:} normalized density $\tilde\Omega_{\rm de}(z)=\rho_{\rm de}(z)/\rho_{\rm c0}$, which remains strictly positive and smoothly approaches $\tilde\Omega_{\rm de}\to 0$ at high redshift, realizing an emergent late-time DE component.
\textbf{Top right:} density parameters $\Omega_{\rm r}(z)$, $\Omega_{\rm m}(z)$, and $\Omega_{\rm de}(z)$, showing standard radiation/matter domination at early times and late-time DE domination.
\textbf{Bottom left:} dark-energy pressure $p_{\rm de}(z)/\rho_{\rm c0}$ obtained from the continuity equation.
\textbf{Bottom right:} equation-of-state parameter $w_{\rm de}(z)=p_{\rm de}/\rho_{\rm de}$, which is finite for all $z$ (no kinematic pole since $\rho_{\rm de}>0$), and becomes strongly phantom at high redshift as $\rho_{\rm de}\to 0^{+}$. For $\eta>0$ one has $\dd\tilde\Omega_{\rm de}/\dd z<0$ over the plotted range, hence $\rho_{\rm de}+p_{\rm de}<0$ throughout; in~\cref{tab:pq-branches} this corresponds to the $p$-phantom branch.
}

    \label{fig:shift_de_profile1}
\end{figure*}

\begin{figure*}[ht]
    \centering
     \includegraphics[width=0.40\linewidth]{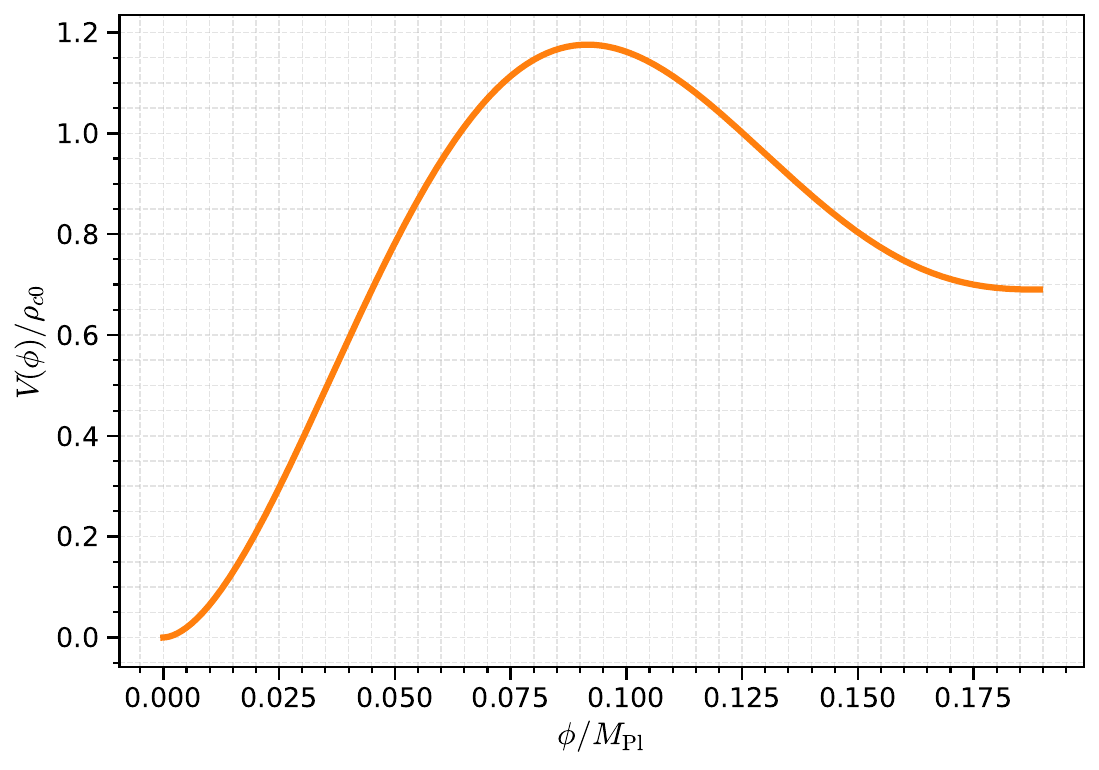}
    \includegraphics[width=0.40\linewidth]{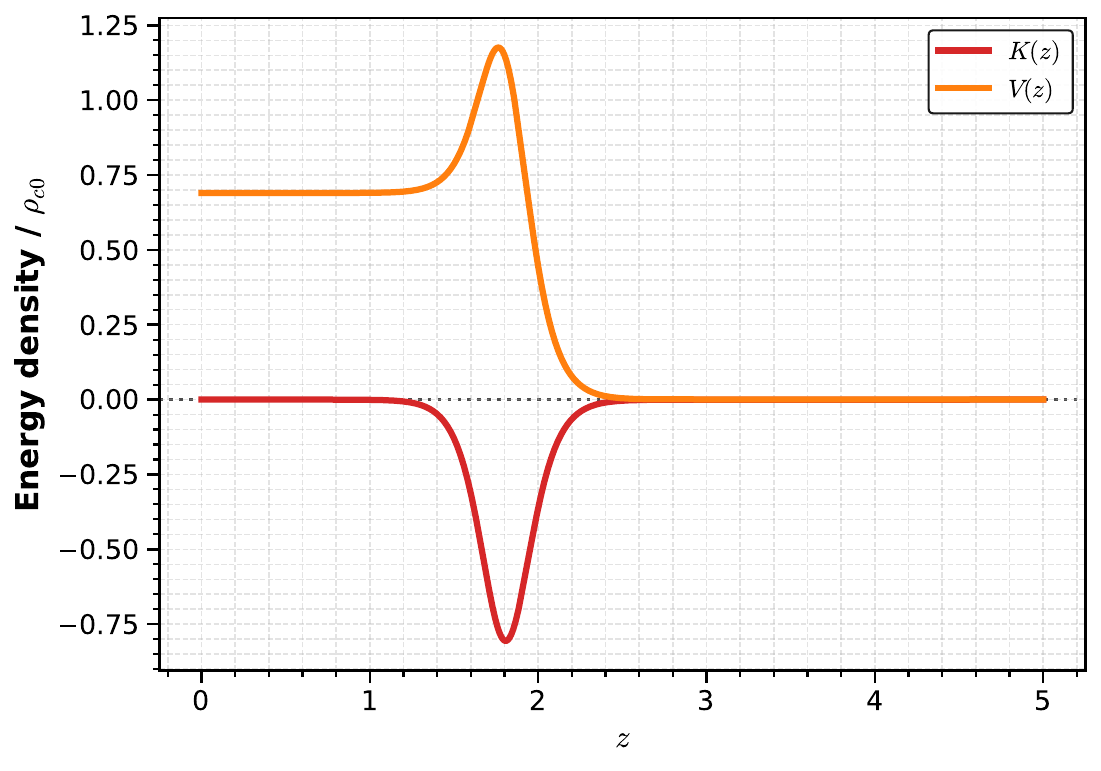}
     \includegraphics[width=0.40\linewidth]{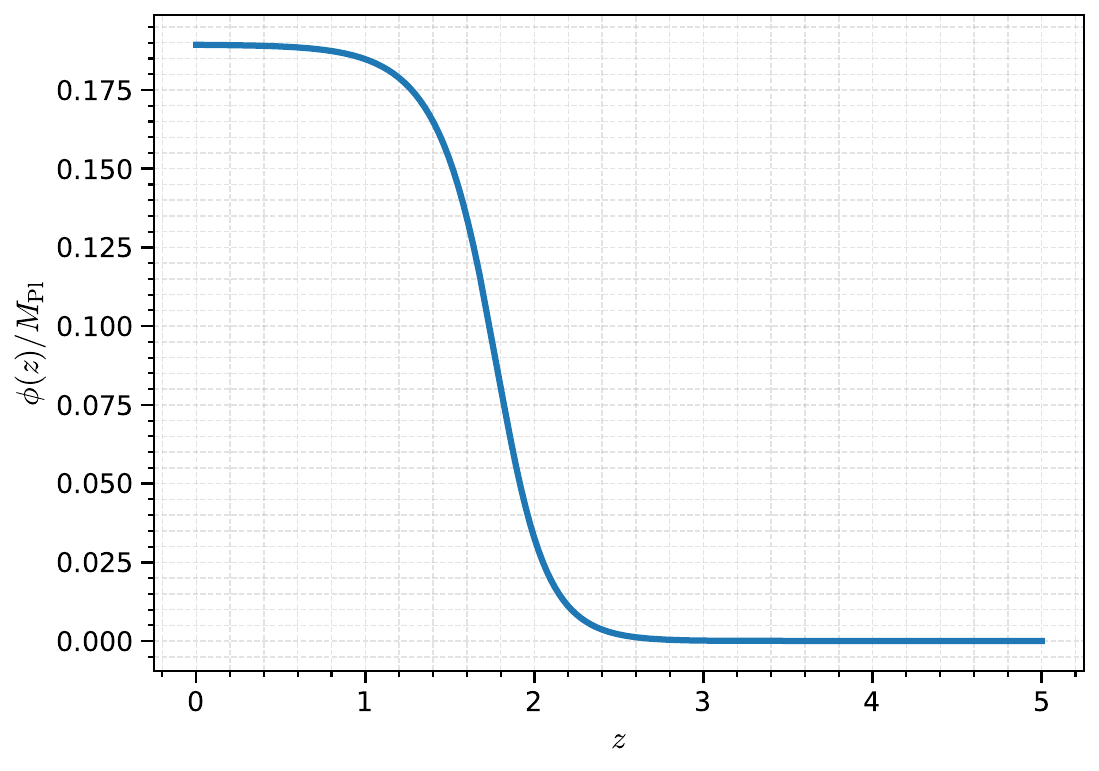}
    \includegraphics[width=0.40\linewidth]{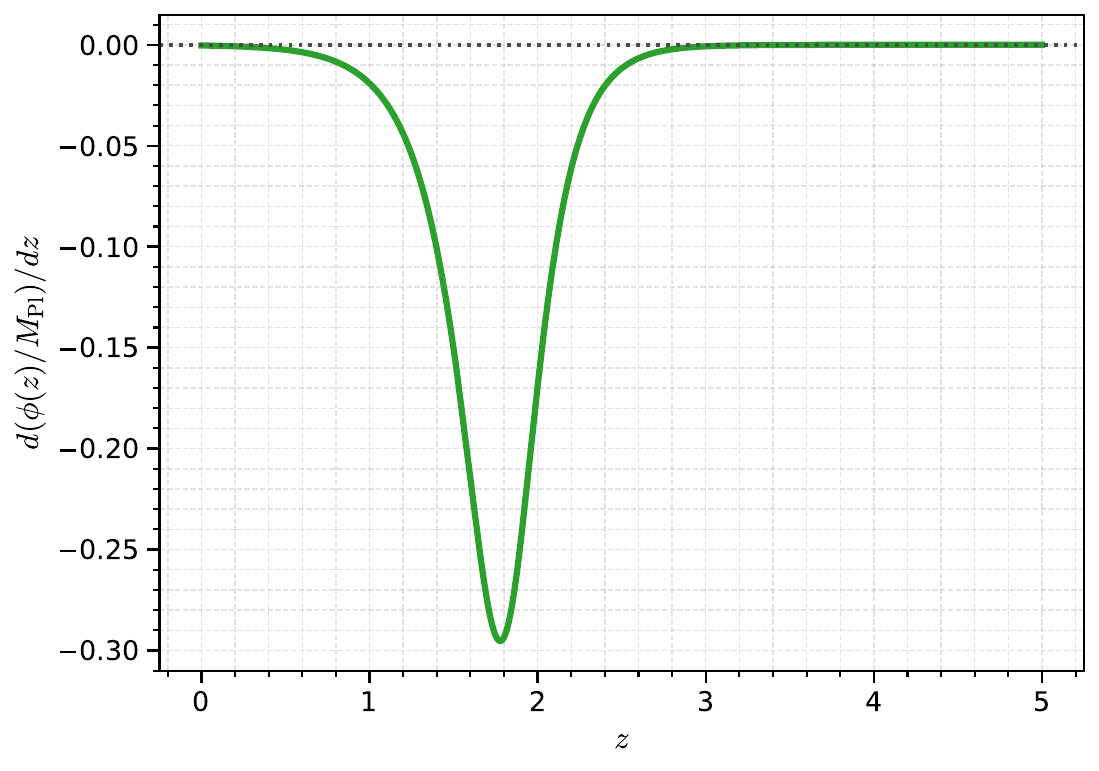}
\caption{
Scalar-field reconstruction for the shifted-$\tanh$ (emergent) profile.
\textbf{Top left:} reconstructed potential $V(\phi)/\rho_{\rm c0}$; the relation is smooth and single-valued (reflecting monotonic $\phi(z)$).
\textbf{Top right:} kinetic and potential contributions $K(z)/\rho_{\rm c0}$ and $V(z)/\rho_{\rm c0}$, with $K(z)=\tfrac12(\rho_{\rm de}+p_{\rm de})$ remaining negative throughout (no NEC-boundary crossing).
\textbf{Bottom left:} field trajectory $\phi(z)/M_{\rm Pl}$, showing that the field is nearly frozen away from the transition and evolves mainly in the intermediate-$z$ regime.
\textbf{Bottom right:} $\dd(\phi/M_{\rm Pl})/\dd z$, sharply localized around the transition.
Since $\dd\tilde\Omega_{\rm de}/\dd z<0$ over the range shown, the single-field sign-consistency condition selects a phantom-branch realization with $\epsilon=-1$; with $\rho_{\rm de}>0$ this corresponds to the $p$-phantom branch in~\cref{tab:pq-branches}.
}

    \label{fig:shift_de_profile2}
\end{figure*}

In this section, we present the results of the two-stage analysis described in~\cref{sec:method}.
We first summarise the illustrative benchmark reconstructions---the CPL history and the two $\tanh$-based transition histories---together with their associated effective scalar-field mappings (\cref{subsec:illustrative};~\cref{fig:cpl_de_profile1,fig:cpl_de_profile2,fig:tanh_de_profile1,fig:tanh_de_profile2,fig:shift_de_profile1,fig:shift_de_profile2}).
These reconstructions define the target potentials $V_{\rm tar}(\phi)$ that are subsequently confronted with analytic potential families.

We then report the results of the Bayesian model comparison performed directly in \emph{potential space}:
posterior constraints on the parameters of each candidate potential, posterior-predictive reconstructions of $V(\phi)$ (with residuals), and the Bayesian
evidences used to rank models.
Throughout this section, the black points with error bars shown in the fit-band figures correspond to the \emph{mock potential-space dataset} $D$ constructed
from $V_{\rm tar}(\phi)$ as defined in~\cref{subsec:bayes}; they are not direct cosmological measurements.

\subsection{Reconstructions of the benchmark DE profiles}
\label{subsec:illustrative}

We now highlight the characteristic features of the three benchmark histories and their associated scalar-sector reconstructions shown in
\cref{fig:cpl_de_profile1,fig:cpl_de_profile2,fig:tanh_de_profile1,fig:tanh_de_profile2,fig:shift_de_profile1,fig:shift_de_profile2}.

\cref{fig:cpl_de_profile1} (top-left panel) shows the normalised dark-energy density $\tilde\Omega_{\rm de}(z)$ for the \texttt{CPL} benchmark.
By construction $\rho_{\rm de}>0$ throughout and $\tilde\Omega_{\rm de}(z)$ exhibits a mild non-monotonic evolution: it increases from its present value,
reaches a maximum, and then decreases toward higher redshift. This behavior follows directly from Eq.~\eqref{eq:rho_cpl} for the adopted \texttt{CPL} parameters.
At the background level the physically relevant classifier of the regime is the NEC combination $\rho_{\rm de}+p_{\rm de}=\tfrac13(1+z)\,\dd\rho_{\rm de}/\dd z$ (cf.\ \cref{sec:scalarfield}), which changes sign whenever $(1+z)\,\dd\tilde\Omega_{\rm de}/\dd z$ changes sign.
Since $\rho_{\rm de}>0$ for \texttt{CPL}, the NEC boundary $\rho_{\rm de}+p_{\rm de}=0$ coincides with the usual phantom-divide line in the EoS plane, i.e.\ $w_{\rm de}=-1$.
This is reflected in \cref{fig:cpl_de_profile1} (bottom-right), where the derived EoS parameter $w_{\rm de}(z)=p_{\rm de}/\rho_{\rm de}$ crosses $-1$ (as $z$ increases): the evolution transitions from $p$-quintessence ($\rho_{\rm de}>0$, $w_{\rm de}>-1$) at very low redshift to $p$-phantom ($\rho_{\rm de}>0$, $w_{\rm de}<-1$) at higher redshift, reaching values as low as $w_{\rm de}(z)\sim -1.3$ by $z\simeq 5$ for the benchmark choice (see~\cref{tab:pq-branches} for the $p/n$-prefix convention and NEC-based classification).
The remaining panels provide complementary background information:
\cref{fig:cpl_de_profile1} (top-right) shows the standard radiation/matter scaling at high $z$ and the late-time emergence of dark energy, with dark energy overtaking matter around $z\lesssim 0.5$ in this benchmark, while the pressure $p_{\rm de}(z)$ (bottom-left) follows directly from the continuity equation and remains smooth across the crossing.

In \cref{fig:cpl_de_profile2}, we display the corresponding effective scalar-sector mapping.
The top-right panel shows $K(z)/\rho_{\rm c0}$ and $V(z)/\rho_{\rm c0}$, with $K(z)=\tfrac12(\rho_{\rm de}+p_{\rm de})$; hence the sign change of $K(z)$ marks the NEC-boundary crossing.
The bottom-left and bottom-right panels show that the reconstructed field trajectory $\phi(z)$ is non-monotonic and possesses a turning point ($\dd\phi/\dd z=0$) coincident with the sign change in $K(z)$.
As a consequence, eliminating $z$ between $V(z)$ and $\phi(z)$ produces a loop/multivalued relation in the top-left panel: the same field value is attained in different parts of the evolution with different $V$, reflecting the reversal of the field motion.
This behavior is precisely what obstructs a global realization by a single minimally coupled real scalar with fixed kinetic signature, since $\rho_\phi+p_\phi=\epsilon\dot\phi^{\,2}$ has a fixed sign for fixed $\epsilon$ (\cref{sec:scalarfield}). Such a crossing is nontrivial from a scalar-field perspective and typically requires either an extended (quintom-like) sector or non-canonical dynamics.
We therefore interpret the \texttt{CPL} reconstruction as an \emph{effective} one-dimensional representation of an extended sector (e.g.\ a quintom-like or non-canonical completion).
Where a single-valued target $V(\phi)$ is required for the potential-space comparison, we restrict attention to an appropriate monotonic branch, as described in~\cref{sec:method}.

We now repeat the reconstruction for the sign-switching $\tanh$ history in Eq.~\eqref{eq:rho_sign}.
The resulting background evolution is summarised in \cref{fig:tanh_de_profile1}.
The top-left panel shows that $\tilde\Omega_{\rm de}(z)$ decreases monotonically with redshift and crosses zero at the transition redshift $z_\dagger$, interpolating from a positive late-time plateau to a negative high-$z$ plateau (a smooth mirror AdS$\rightarrow$dS sign switch in $\rho_{\rm de}$).
The derived EoS parameter $w_{\rm de}=p_{\rm de}/\rho_{\rm de}$ (bottom-right) diverges at the zero-crossing, but this divergence is purely kinematic (a ratio singularity): the pressure $p_{\rm de}(z)$ (bottom-left) remains finite and the background evolution is smooth.
In this sign-changing case, $w_{\rm de}=-1$ is \emph{not} a global separator of physical regimes (\cref{sec:scalarfield}); the meaningful diagnostic is instead the sign of $\rho_{\rm de}+p_{\rm de}$.
For the profile in Eq.~\eqref{eq:rho_sign} with $\eta>0$, one has $\dd\tilde\Omega_{\rm de}/\dd z<0$ over the plotted range, implying $\rho_{\rm de}+p_{\rm de}<0$ throughout: the evolution therefore stays on the phantom/NEC-violating branch even though $w_{\rm de}$ may lie on either side of $-1$ depending on the sign of $\rho_{\rm de}$. In the terminology of~\cref{tab:pq-branches}, the evolution is \emph{phantom} throughout: it lies on the $p$-phantom branch for $z<z_\dagger$ (where $\rho_{\rm de}>0$) and on the $n$-phantom branch for $z>z_\dagger$ (where $\rho_{\rm de}<0$).
Accordingly, whenever $w_{\rm de}=p_{\rm de}/\rho_{\rm de}$ is well defined, the inequality relative to $-1$ reverses across the sign switch, with $w_{\rm de}<-1$ for $z<z_\dagger$ and $w_{\rm de}>-1$ for $z>z_\dagger$, consistent with the NEC criterion. The top-right panel confirms that the standard early-time expansion is recovered: although $\tilde\Omega_{\rm de}$ approaches an $\mathcal{O}(1)$ plateau in units of $\rho_{\rm c0}$, its \emph{fractional} contribution $\Omega_{\rm de}(z)=\rho_{\rm de}/(3M_{\rm Pl}^2H^2)$ is strongly suppressed at high redshift, while $\Omega_{\rm de}(z)$ crosses through zero and becomes negative beyond $z_\dagger$. Equivalently, the dark-energy contribution to the total energy budget is negligible at early times, despite $\tilde\Omega_{\rm de}$ approaching a constant in units of $\rho_{\rm c0}$. As the Universe evolves, $\Omega_{\rm de}(z)$ undergoes a smooth but relatively sharp change around $z_\dagger$ and becomes dominant toward the present epoch, consistent with late-time acceleration.

The associated scalar-field reconstruction is shown in \cref{fig:tanh_de_profile2}.
In contrast to the \texttt{CPL} case, $K(z)$ remains negative throughout (top-right), so there is no NEC-boundary crossing and the single-field sign-consistency condition selects the phantom branch $\epsilon=-1$ over the plotted range.
The field trajectory $\phi(z)$ (bottom-left) is monotonic and the evolution is sharply localised around the transition redshift, as seen in $\dd(\phi/M_{\rm Pl})/\dd z$ (bottom-right).
Eliminating $z$ produces a smooth single-valued $V(\phi)$ (top-left), which interpolates between an AdS-like (negative) value at high redshift and a dS-like (positive) value at late times, with a broad maximum around the transition. 
The absence of sharp features or multiple extrema in the reconstructed $V(\phi)$ supports the interpretation that this history admits a consistent single-field effective realization at the background level.

Finally, we consider the shifted-$\tanh$ (emergent) history in Eq.~\eqref{eq:rho_shift}, for which $\rho_{\rm de}(z)>0$ at all finite redshifts and $\tilde\Omega_{\rm de}\to 0$ at high redshift.
The background evolution is shown in \cref{fig:shift_de_profile1}.
Since $\rho_{\rm de}$ does not cross zero, the derived EoS parameter $w_{\rm de}(z)$ (bottom-right) remains finite for all $z$ (no kinematic pole).
At late times $w_{\rm de}(z)$ approaches a nearly constant value close to $-1$, mimicking a cosmological constant, while toward high redshift it evolves to strongly phantom values as $\rho_{\rm de}\to 0^{+}$.
As in the unshifted $\tanh$ case, $\dd\tilde\Omega_{\rm de}/\dd z<0$ over the plotted range (for $\eta>0$), implying $\rho_{\rm de}+p_{\rm de}<0$ throughout and hence a phantom/NEC-violating evolution. Since $\rho_{\rm de}>0$ over the full redshift range, this corresponds to the $p$-phantom branch in~\cref{tab:pq-branches}.
The remaining panels show that the early-time expansion is again radiation/matter dominated, with dark energy emerging only at late times.

The corresponding scalar-field mapping is displayed in \cref{fig:shift_de_profile2}.
The kinetic term $K(z)$ remains negative throughout (top-right), so there is no NEC-boundary crossing and the sign-consistency condition again selects $\epsilon=-1$.
The field trajectory is monotonic (bottom-left) and the evolution is concentrated around the transition region (bottom-right), yielding a smooth, single-valued reconstructed potential $V(\phi)$ (top-left).
Unlike the sign-switching case, $V(\phi)$ remains non-negative over the reconstructed field domain, consistent with the fact that
$\rho_{\rm de}(z)$ approaches zero from above at high redshift.

\subsection{Bayesian model comparisons in potential space}
\label{subsec:bayes_results}

We now confront the reconstructed target potentials $V_{\rm tar}(\phi)$ with the analytic potential families introduced in~\cref{subsec:potentials}, using the Bayesian framework of~\cref{subsec:bayes}.
For each target, we construct the mock potential-space dataset $D=\{(\phi_i,\,V_i^{\rm (mock)},\,\sigma_i)\}_{i=1}^{N}$ from $V_{\rm tar}(\phi)$ through Eq.~\eqref{eq:data_mock}, and infer the posterior $p(\boldsymbol{\theta}\,|\,D,M)$ and evidence $\mathcal{Z}(D\,|\,M)$ for each candidate model $M$ (with parameters $\boldsymbol{\theta}$) using nested sampling.
For the \texttt{CPL} benchmark, where the scalar-sector mapping is multivalued because $\phi(z)$ is non-monotonic, we restrict ourselves to the monotonic branch used to define a single-valued target in potential space (\cref{sec:method}).
In the fit-band figures, the black points with error bars are the mock dataset $D$ (i.e.\ not direct cosmological measurements), while the solid curve and shaded regions show the posterior-predictive median $V_{\rm med}(\phi)$ together with its $68\%$ and $95\%$ credible bands; the lower panels display the corresponding residuals with respect to $D$.
For compactness, the full parameter posteriors for each target are collected in~\cref{fig:corner_cpl_all,fig:corner_tanh_all}, and the evidence ranking is summarised in~\cref{tab:model_comparison_combined}.

%%============================================================================================================%%
\subsubsection{\texttt{CPL} reconstruction}
\label{subsubsec:cpl_potspace}
%%============================================================================================================%%

For the \texttt{CPL} benchmark, the reconstructed scalar-sector mapping is multivalued in field space because the trajectory $\phi(z)$ is non-monotonic(cf.\ \cref{fig:cpl_de_profile2}). 
To define a single-valued target potential for the potential-space comparison, we therefore restrict the analysis to a monotonic branch of the reconstruction (in practice the $p$-phantom branch with $K<0$; see~\cref{sec:method} and~\cref{tab:pq-branches}).

\begin{figure*}[!t]
    \centering

    % Row 1
    \includegraphics[width=0.46\linewidth]{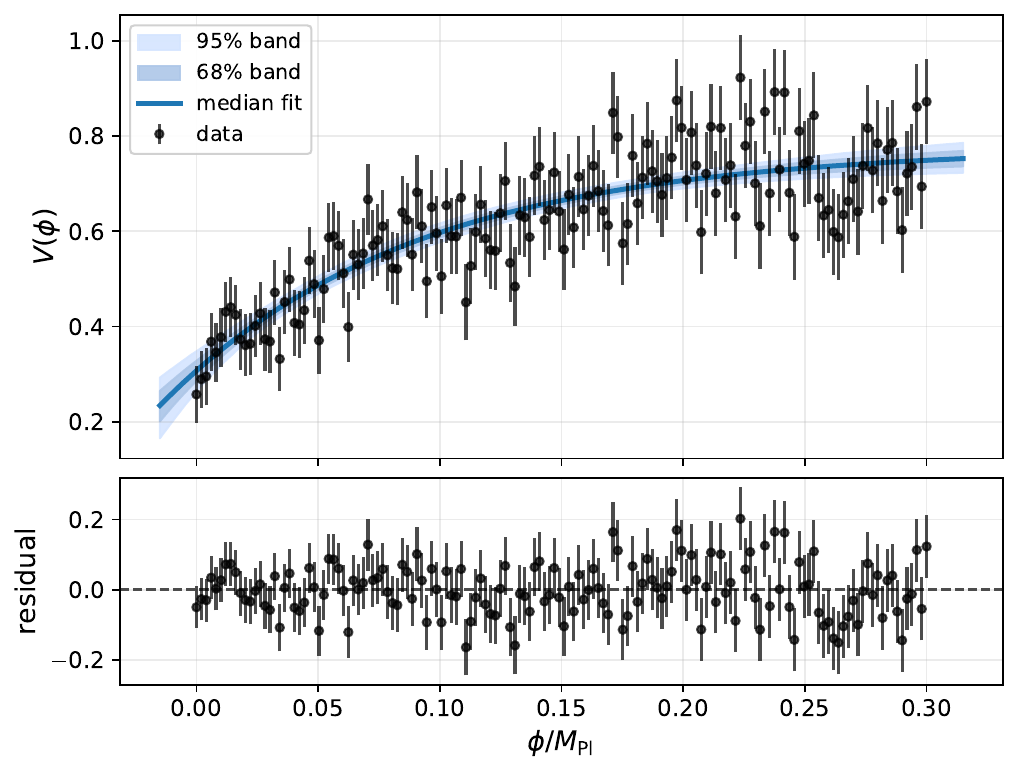}\hfill
    \includegraphics[width=0.46\linewidth]{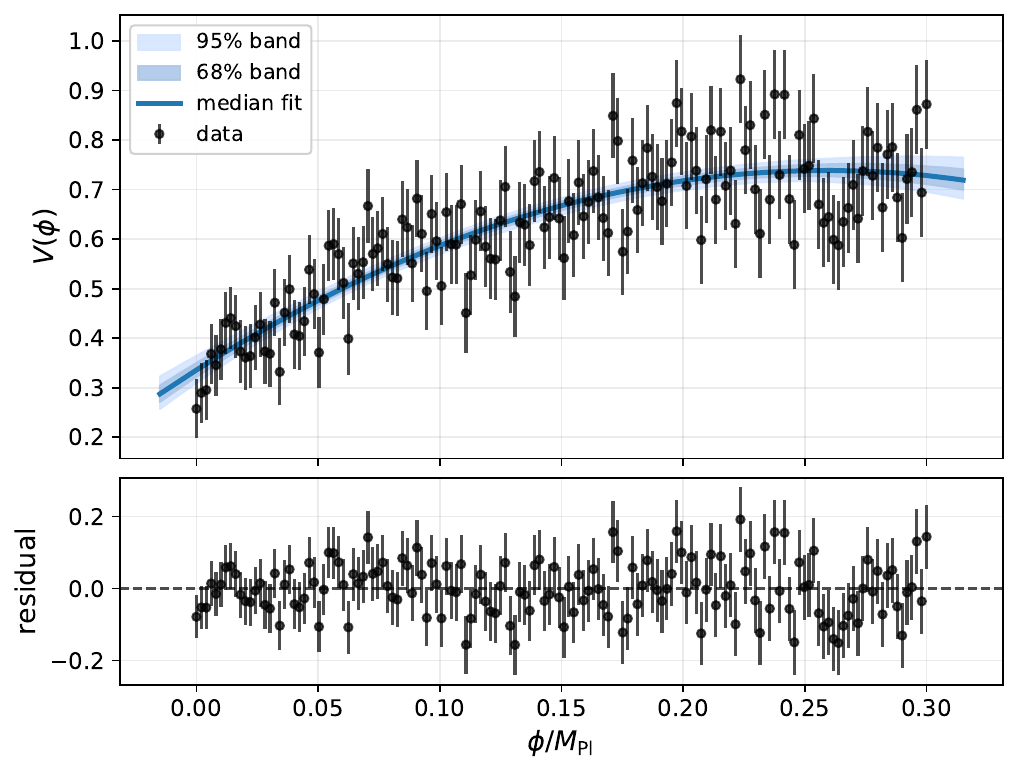}\\[-2pt]
    \parbox[t]{0.46\linewidth}{\centering (a) Exponential}\hfill
    \parbox[t]{0.46\linewidth}{\centering (b) Hilltop quartic}\\[6pt]

    % Row 2
    \includegraphics[width=0.46\linewidth]{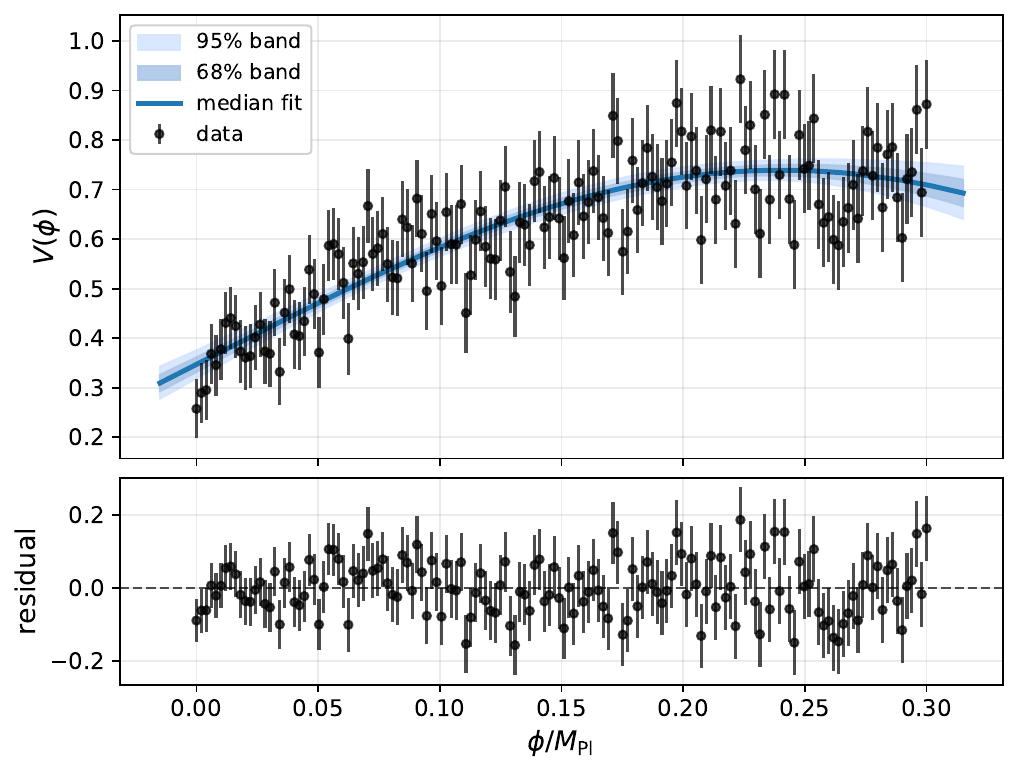}\hfill
    \includegraphics[width=0.46\linewidth]{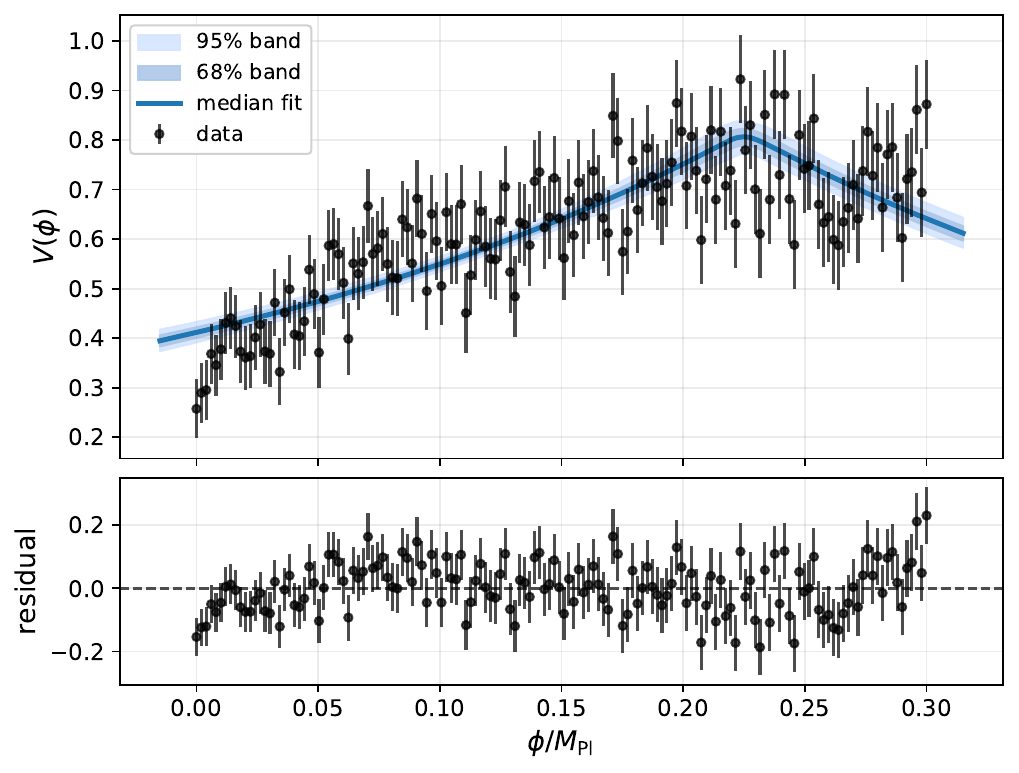}\\[-2pt]
    \parbox[t]{0.46\linewidth}{\centering (c) PNGB}\hfill
    \parbox[t]{0.46\linewidth}{\centering (d) Inverse power law}\\[6pt]

    % Row 3
    \includegraphics[width=0.46\linewidth]{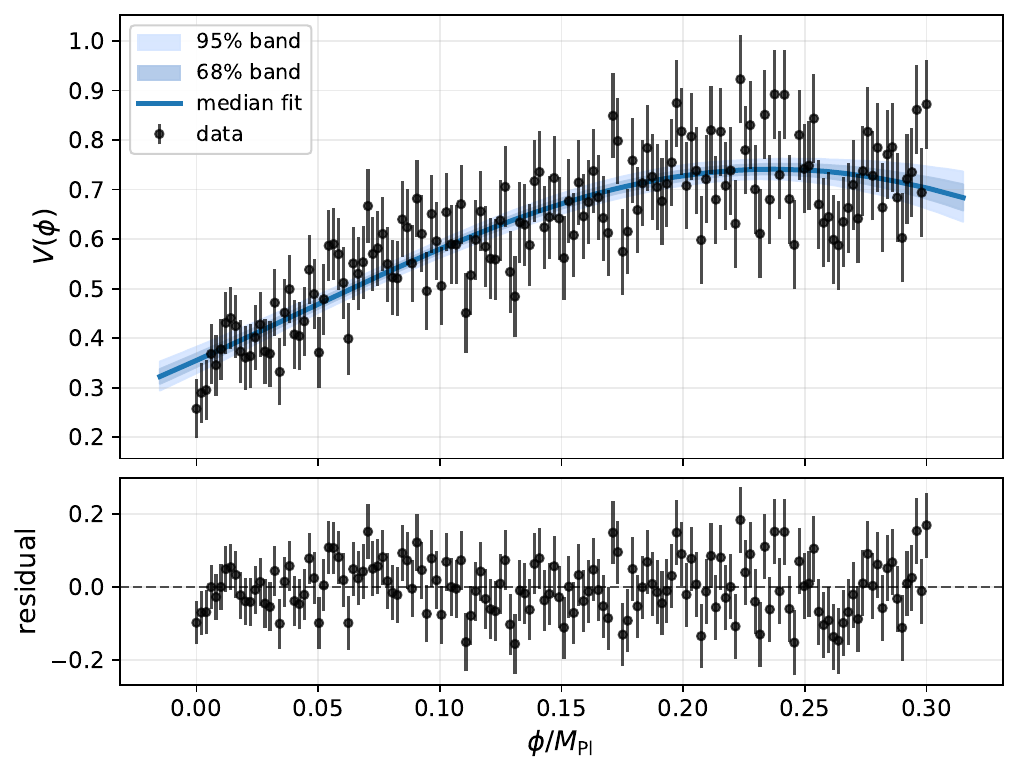}\hfill
    \includegraphics[width=0.46\linewidth]{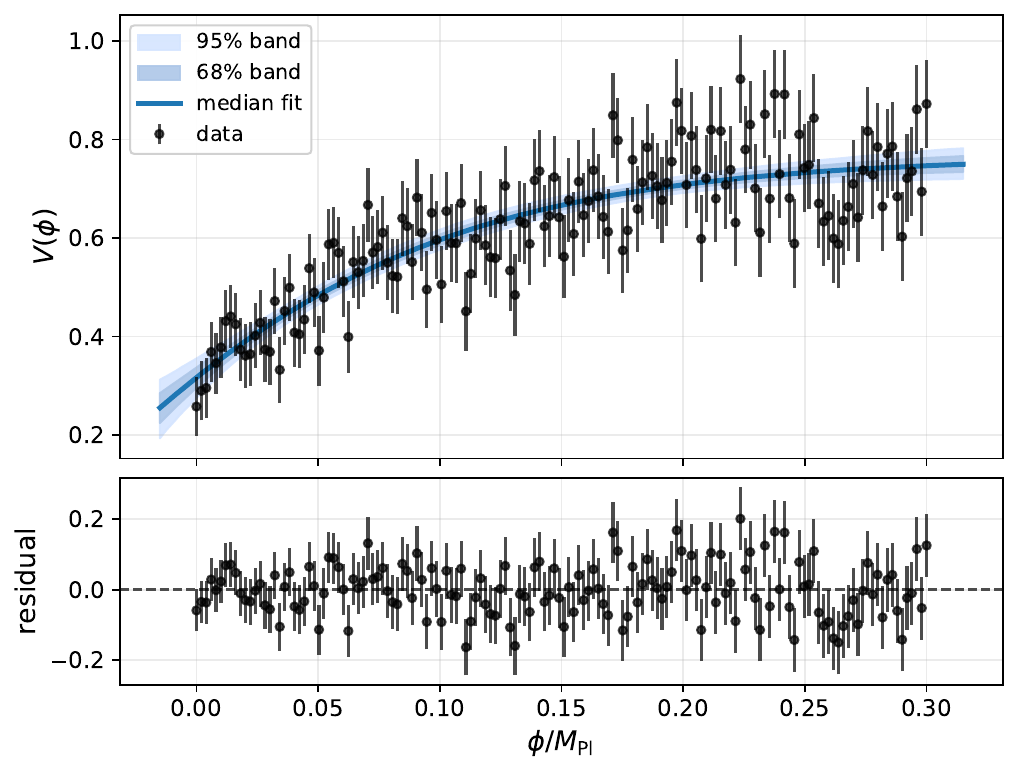}\\[-2pt]
    \parbox[t]{0.46\linewidth}{\centering (e) Gaussian bump}\hfill
    \parbox[t]{0.46\linewidth}{\centering (f) Shifted-$\tanh$}\\

\caption{
Posterior-predictive fit bands for the CPL target in \emph{potential space}, shown for six representative analytic potentials:
(a) Exponential, (b) Hilltop quartic, (c) PNGB, (d) Inverse power law, (e) Gaussian bump, and (f) Shifted-$\tanh$.
In each panel, the solid curve denotes the median reconstructed $V(\phi)$ and the shaded regions show the corresponding $68\%$ and $95\%$ credible bands.
Black points with error bars are the mock potential-space dataset $D$ constructed from $V_{\rm tar}(\phi)$ (\cref{subsec:bayes}), and the lower strips show the corresponding residuals.
}

    \label{fig:band_cpl}
\end{figure*}

\cref{fig:band_cpl} shows the posterior-predictive median potential and the corresponding $68\%$ and $95\%$ credible bands for each analytic model fitted to the \texttt{CPL} mock dataset. 
The black points with error bars are the mock potential-space data constructed from the target $V_{\rm tar}(\phi)$ via Eq.~\eqref{eq:data_mock}; they are not direct cosmological measurements. 
The lower panels show the residuals $\Delta V_i \equiv V_{\rm med}(\phi_i)-V_i^{\rm(mock)}$, which highlight localised mismatches.
Overall, the tested potentials reproduce the \texttt{CPL} mock dataset at the $\sim 2\sigma$ level over the fitted field interval.
A mild but systematic deviation is visible around $\phi/M_{\rm Pl}\simeq 0.25$ for most models, while at lower field values ($\phi/M_{\rm Pl}\sim 0.05$) the exponential (\cref{fig:band_cpl}a) and shifted-$\tanh$ (\cref{fig:band_cpl}f) potentials exhibit comparatively better agreement.

\begin{figure*}[!t]
    \centering

    % Row 1
    \includegraphics[width=0.32\linewidth]{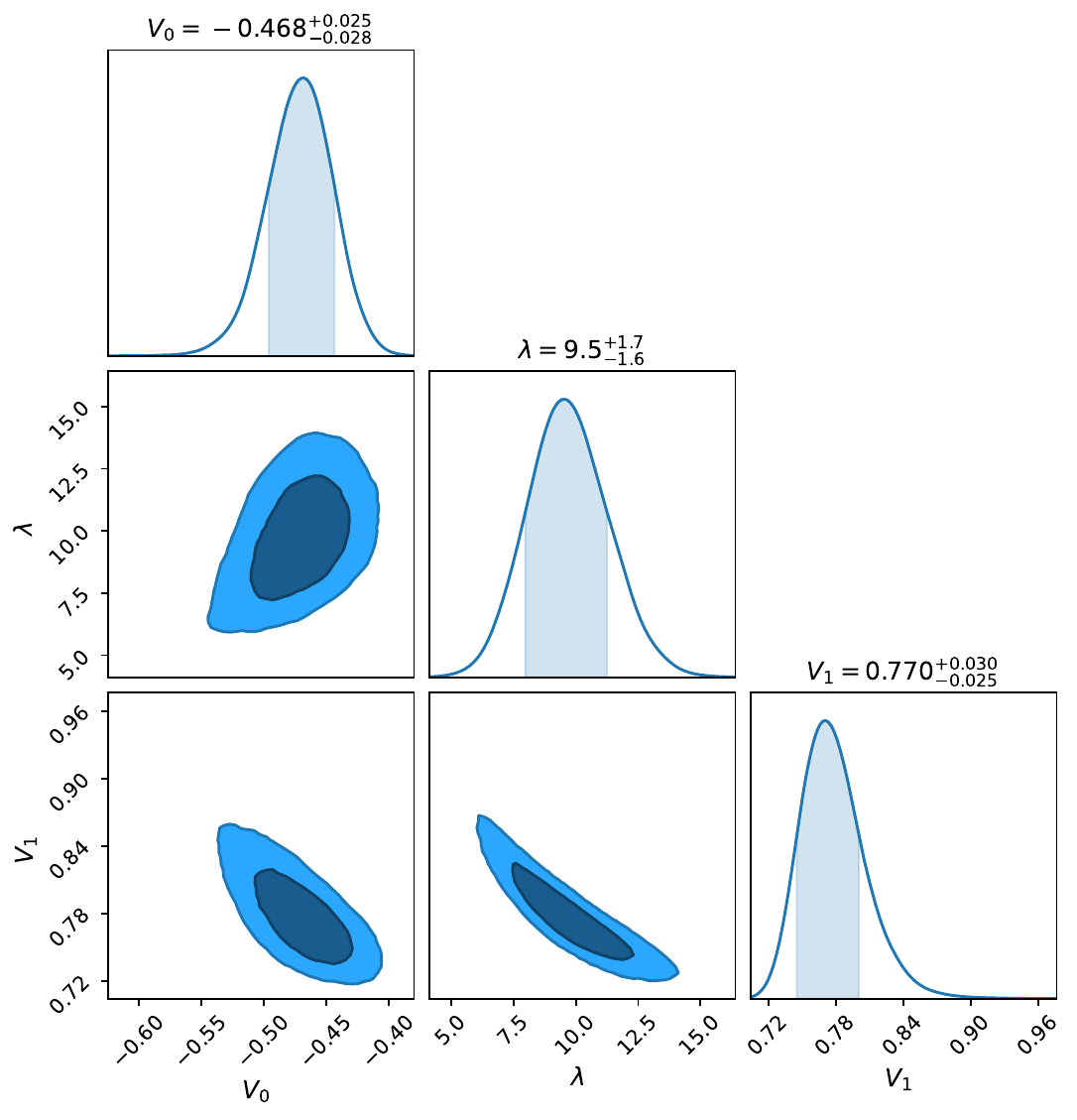}\hfill
    \includegraphics[width=0.32\linewidth]{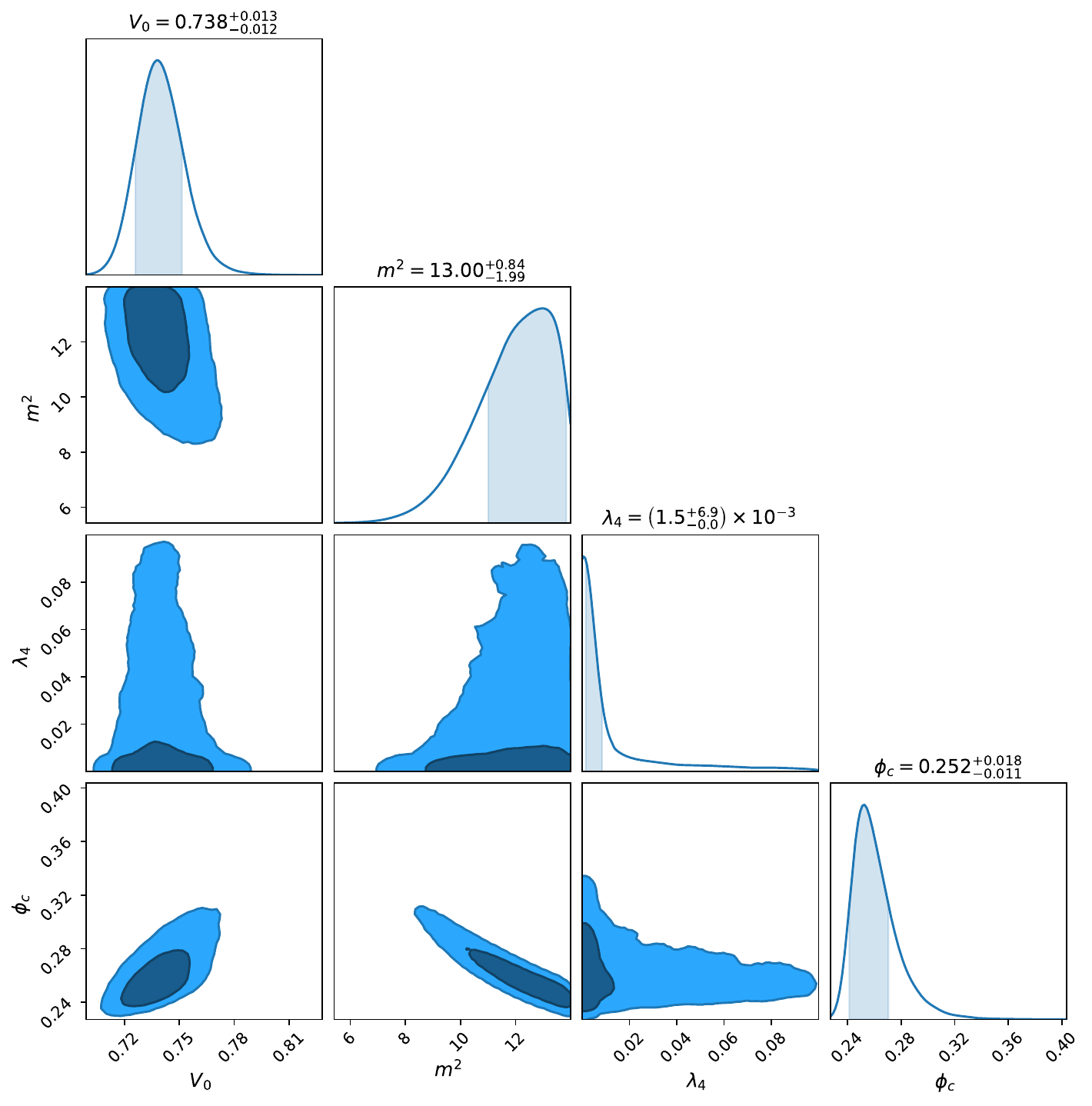}\hfill
    \includegraphics[width=0.32\linewidth]{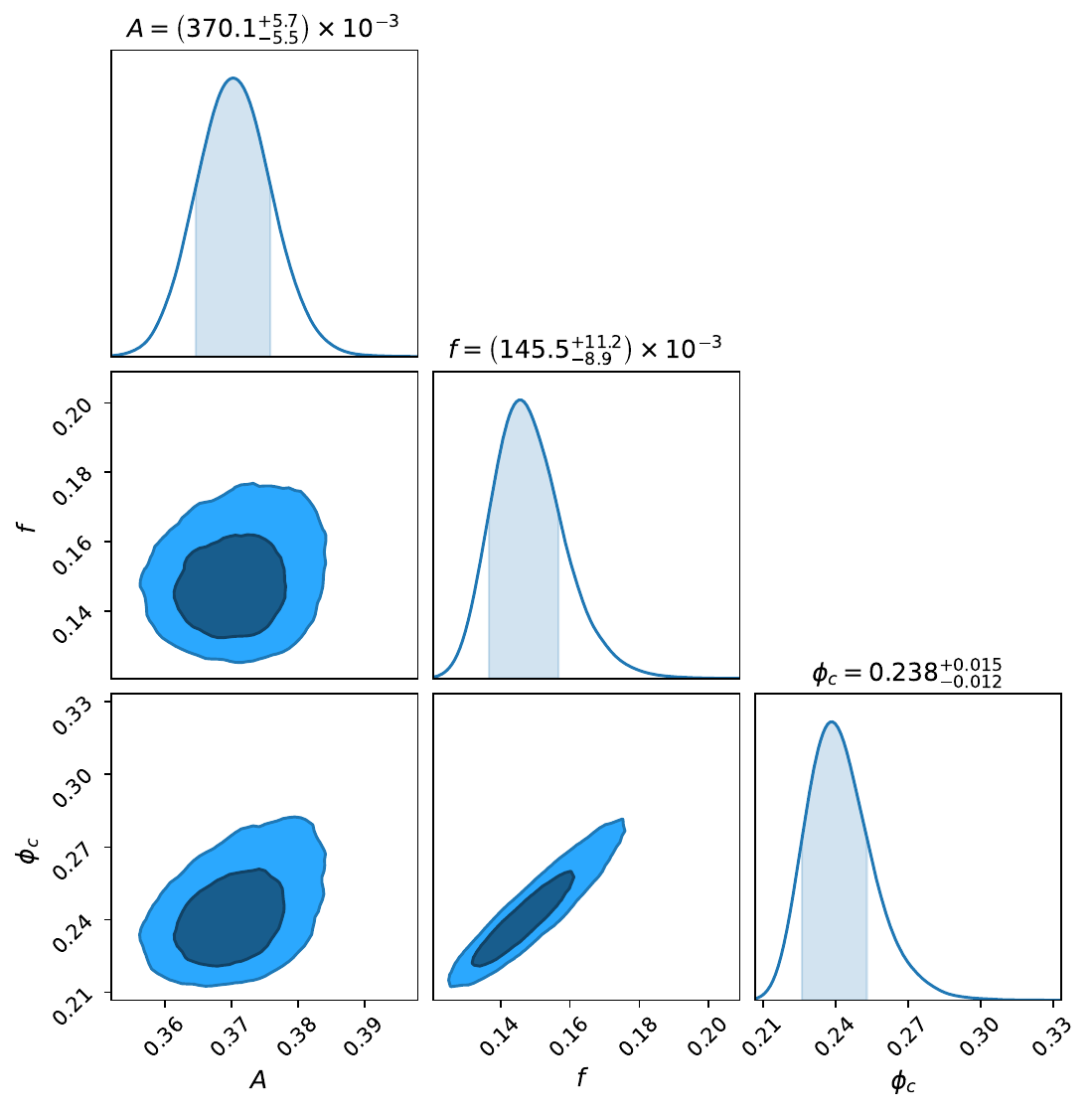}\\[-2pt]
    \parbox[t]{0.32\linewidth}{\centering (a) Exponential}\hfill
    \parbox[t]{0.32\linewidth}{\centering (b) Hilltop quartic}\hfill
    \parbox[t]{0.32\linewidth}{\centering (c) PNGB}\\[8pt]

    % Row 2
    \includegraphics[width=0.32\linewidth]{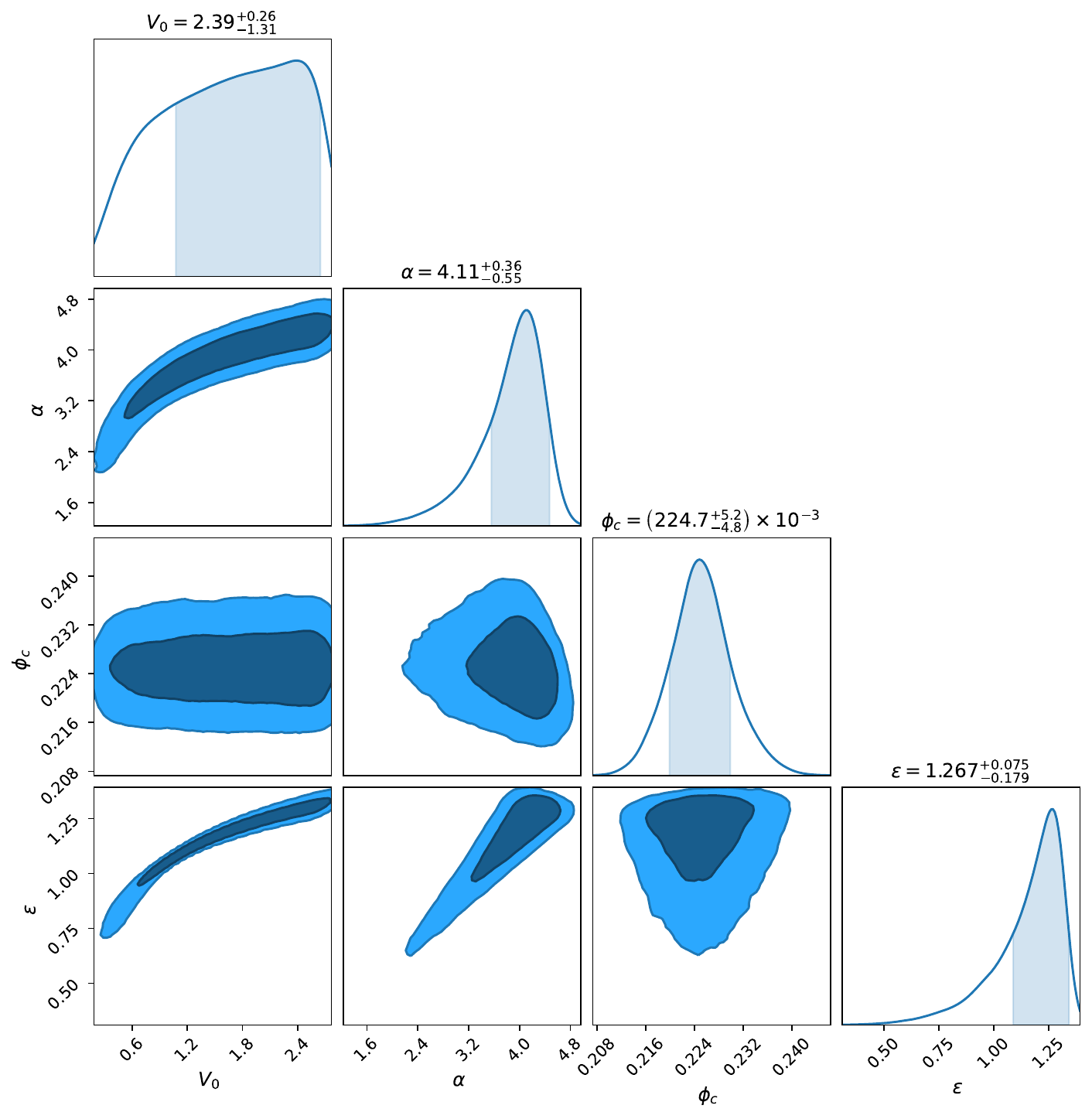}\hfill
    \includegraphics[width=0.32\linewidth]{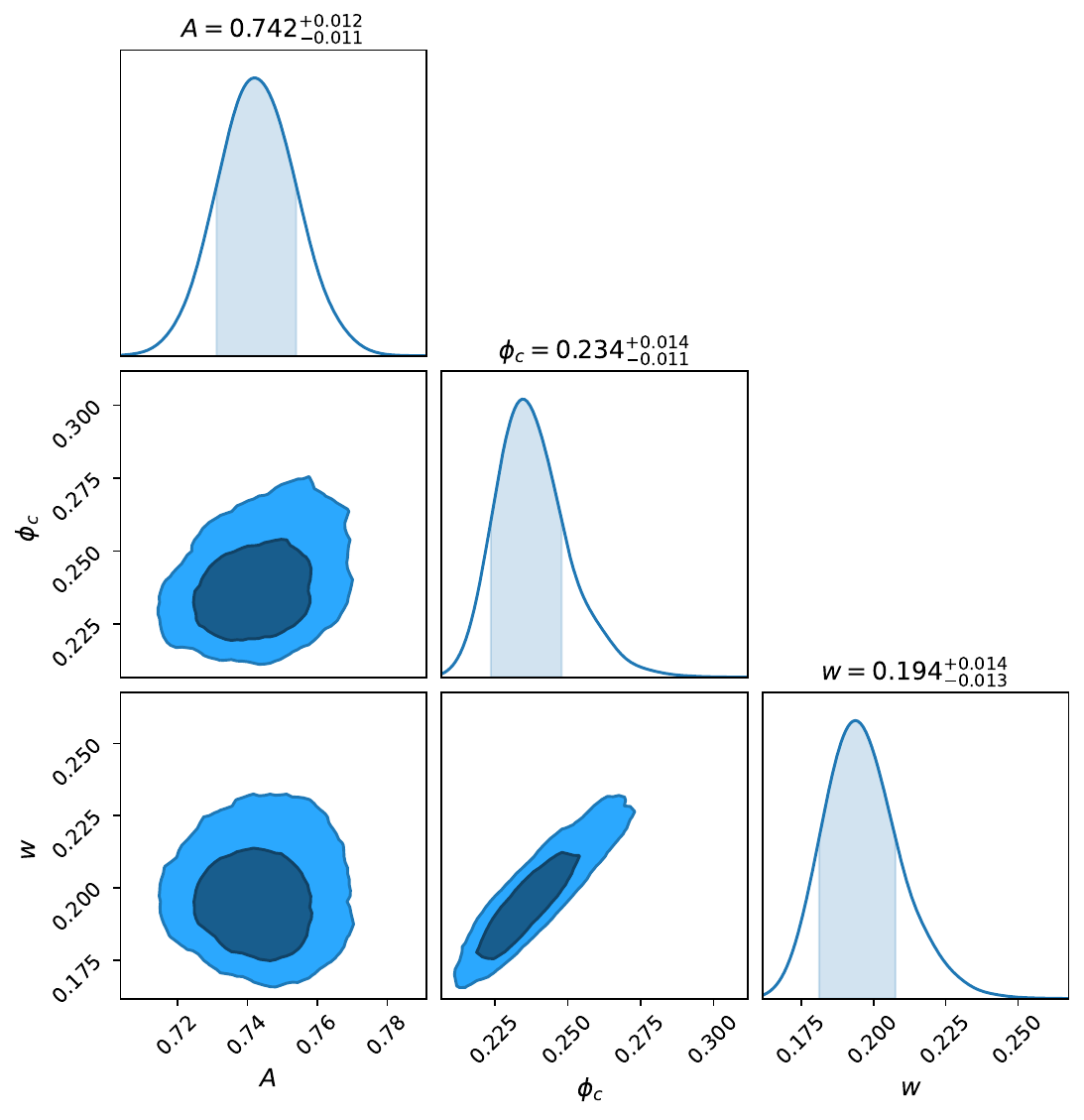}\hfill
    \includegraphics[width=0.32\linewidth]{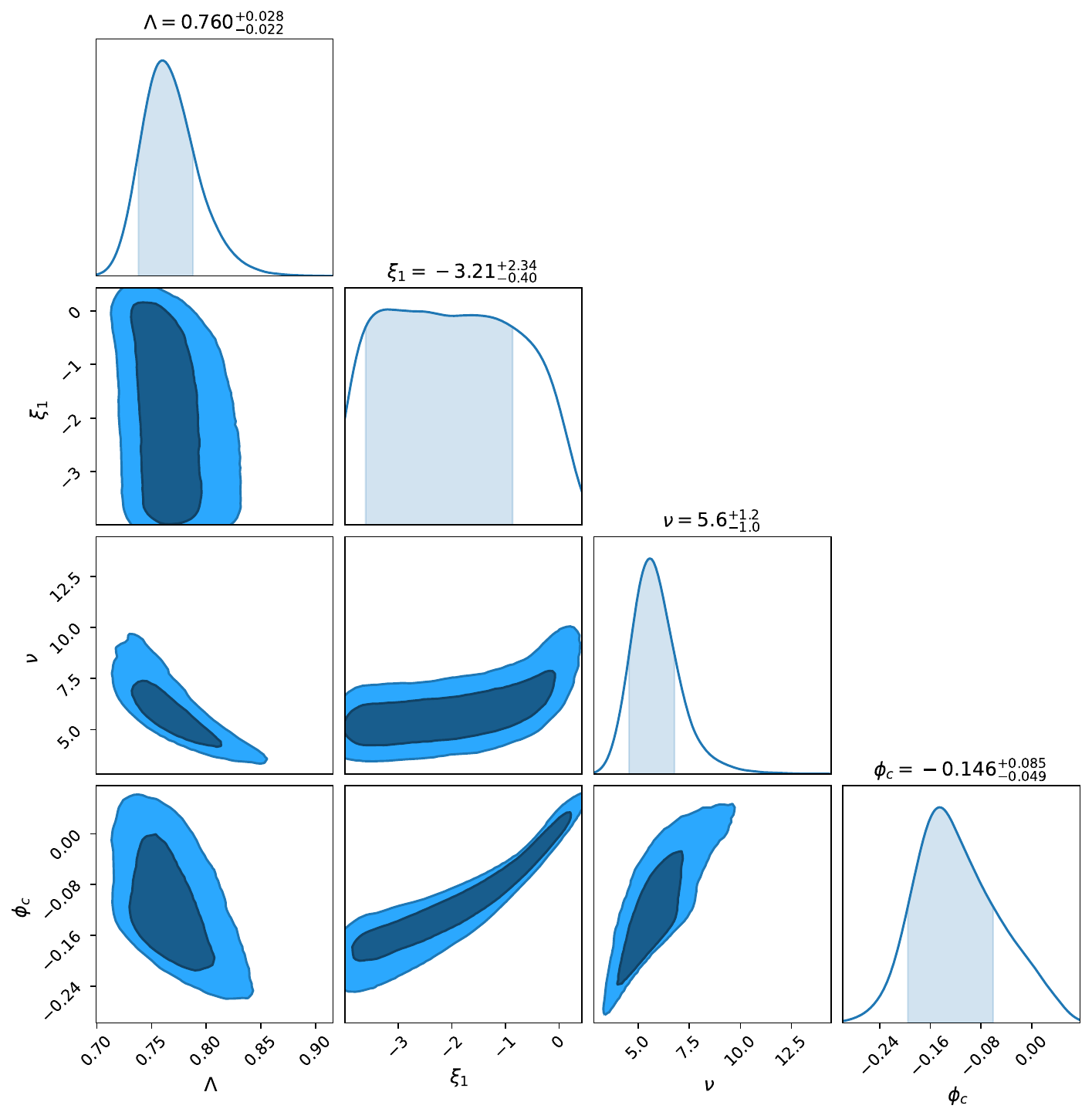}\\[-2pt]
    \parbox[t]{0.32\linewidth}{\centering (d) Inverse power law}\hfill
    \parbox[t]{0.32\linewidth}{\centering (e) Gaussian bump}\hfill
    \parbox[t]{0.32\linewidth}{\centering (f) shifted-$\tanh$}\\

    \caption{
    Posterior distributions for six analytic scalar-field potentials fitted to the \texttt{CPL} reconstructed target potential in \emph{potential space}
    (\cref{subsec:bayes}).
    The constraints are inferred using nested sampling applied to the mock dataset $D$ constructed from $V_{\rm tar}(\phi)$.
    Contours show the $68\%$ and $95\%$ credible regions, while the 1D panels show the corresponding marginalized posteriors.
    For the \texttt{CPL} case, the fit is performed on the single-valued branch of the reconstructed target used for the potential-space comparison
    (\cref{sec:method}).
    }
    \label{fig:corner_cpl_all}
\end{figure*}

The corresponding posterior constraints on the model parameters are summarised in \cref{fig:corner_cpl_all}.
For the exponential potential (\cref{fig:corner_cpl_all}a), we obtain tight constraints on all parameters, $V_0 = -0.468^{+0.025}_{-0.028}$, $\lambda = 9.5^{+1.7}_{-1.6}$, and $V_1 = 0.770^{+0.030}_{-0.025}$, with strong correlations among them, indicating a well-determined exponential realization of the restricted \texttt{CPL} branch.

For the hilltop quartic potential (\cref{fig:corner_cpl_all}b), the inferred constraints are $V_0 = 0.738^{+0.013}_{-0.012}$, $m^2 = 13.00^{+0.84}_{-1.99}$, $\lambda_4 = 0.001^{+0.007}_{-0.000}$, and $\phi_{\rm c} = 0.29^{+0.006}_{-0.006}$.
Although the curvature scale is constrained, the quartic coupling is consistent with negligible higher-order corrections within the allowed prior range, suggesting that the reconstruction does not require a strongly stabilising quartic term over the field interval probed by $D$.

For the PNGB potential (\cref{fig:corner_cpl_all}c), we obtain $A = 0.37^{+0.006}_{-0.005}$, $f = 0.145^{+0.011}_{-0.009}$, and $\phi_{\rm c} = 0.238^{+0.015}_{-0.012}$, with visible correlations among the parameters. 
The relatively narrow uncertainties indicate that the PNGB shape can reproduce the restricted \texttt{CPL} target reasonably well in potential space, under the adopted noise model.

For the inverse power-law potential (\cref{fig:corner_cpl_all}d), we find $V_0 = 2.39^{+0.26}_{-1.31}$, $\alpha = 4.11^{+0.36}_{-0.55}$, $\phi_{\rm c} = 0.225^{+0.005}_{-0.004}$, and $\varepsilon = 1.267^{+0.075}_{-0.179}$.
The broader posteriors and the residual structure in \cref{fig:band_cpl} indicate that this functional form struggles to reproduce the \texttt{CPL} target in the fitted range.

For the Gaussian bump potential (\cref{fig:corner_cpl_all}e), we obtain $A = 0.742^{+0.012}_{-0.011}$, $\phi_{\rm c} = 0.234^{+0.014}_{-0.011}$, and $w = 0.194^{+0.014}_{-0.013}$.
Although the parameters are reasonably constrained, the fit-band comparison indicates that an additional localised feature is not strongly required by the
\texttt{CPL} target under the adopted noise model.

Finally, for the shifted-$\tanh$ potential (\cref{fig:corner_cpl_all}f), the inferred parameters are $\Lambda = 0.764^{+0.028}_{-0.022}$, $\xi_1 = -3.21^{+2.34}_{-0.40}$, $\nu = 5.6^{+1.2}_{-1.0}$, with $\phi_{\rm c}=-0.146^{+0.085}_{-0.049}$.
Compared to the exponential case, the posteriors show more pronounced degeneracies (reflecting the additional flexibility of this form), while the fit-band comparison shows that it nevertheless provides a competitive representation of the restricted \texttt{CPL} branch in potential space.

% --- Multi-panel figure without subcaption/subfigure ---

% ============================
%   CPL: Corner plots (merged)
% ============================

%%============================================================================================================%%
\subsubsection{\texorpdfstring{$\tanh$ reconstruction}{Hyperbolic tangent reconstruction}}
\label{subsubsec:tanh_potspace}
%%============================================================================================================%%

We now repeat the same potential-space comparison for the sign-switching $\tanh$ benchmark~\cite{DeFelice:2020cpt,Akarsu:2022typ,Akarsu:2024qsi,Akarsu:2024eoo}. 
In this case the reconstructed target $V_{\rm tar}(\phi)$ is smooth and single-valued over the fitted field interval (cf.\ \cref{fig:tanh_de_profile2}), so the Bayesian analysis is performed directly on the full reconstructed branch without the additional branch restriction required in the \texttt{CPL} case.

\begin{figure*}[!t]
    \centering

    % Row 1
    \includegraphics[width=0.46\linewidth]{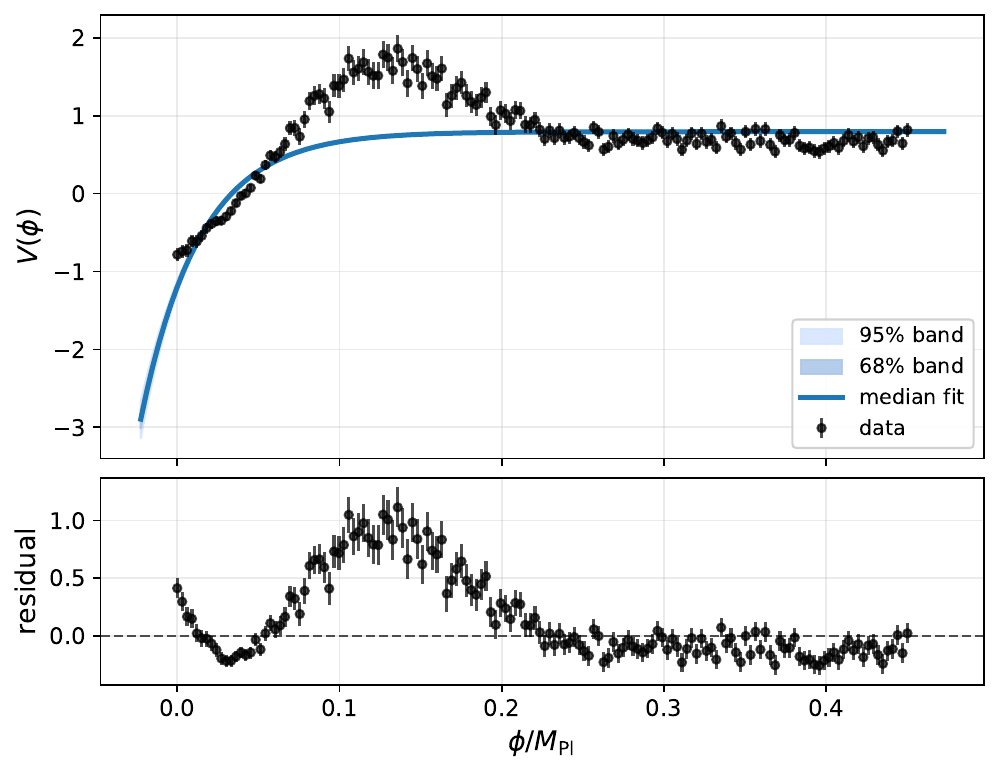}\hfill
    \includegraphics[width=0.46\linewidth]{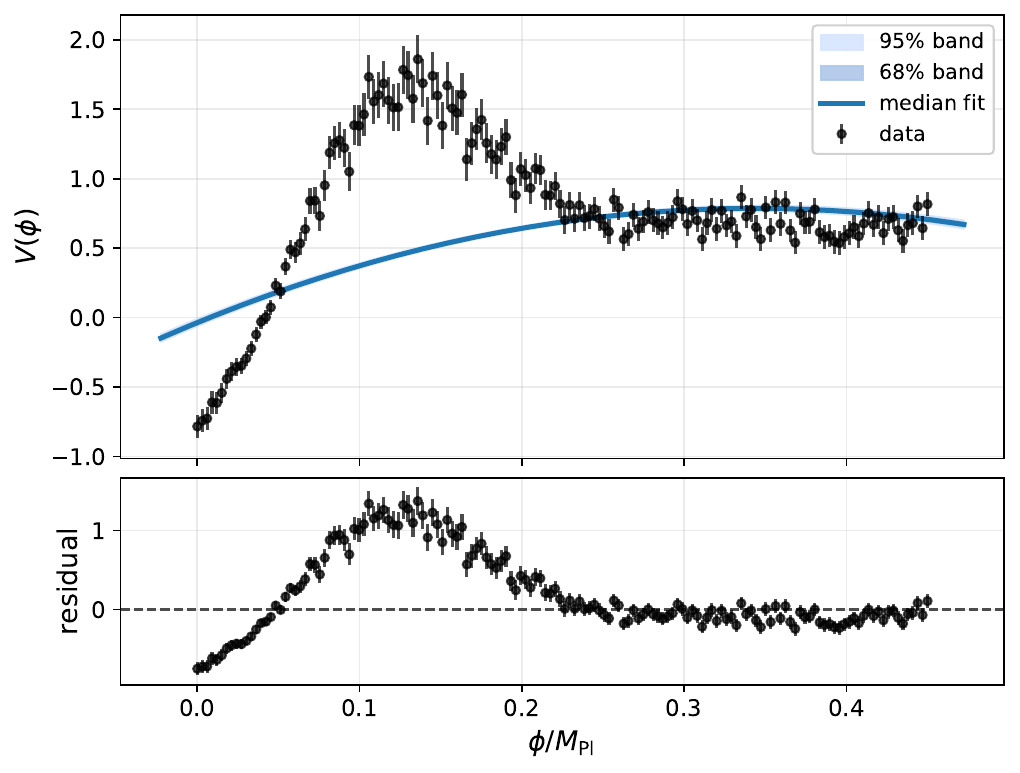}\\[-2pt]
    \parbox[t]{0.46\linewidth}{\centering (a) Exponential}\hfill
    \parbox[t]{0.46\linewidth}{\centering (b) Hilltop quartic}\\[10pt]

    % Row 2
    \includegraphics[width=0.46\linewidth]{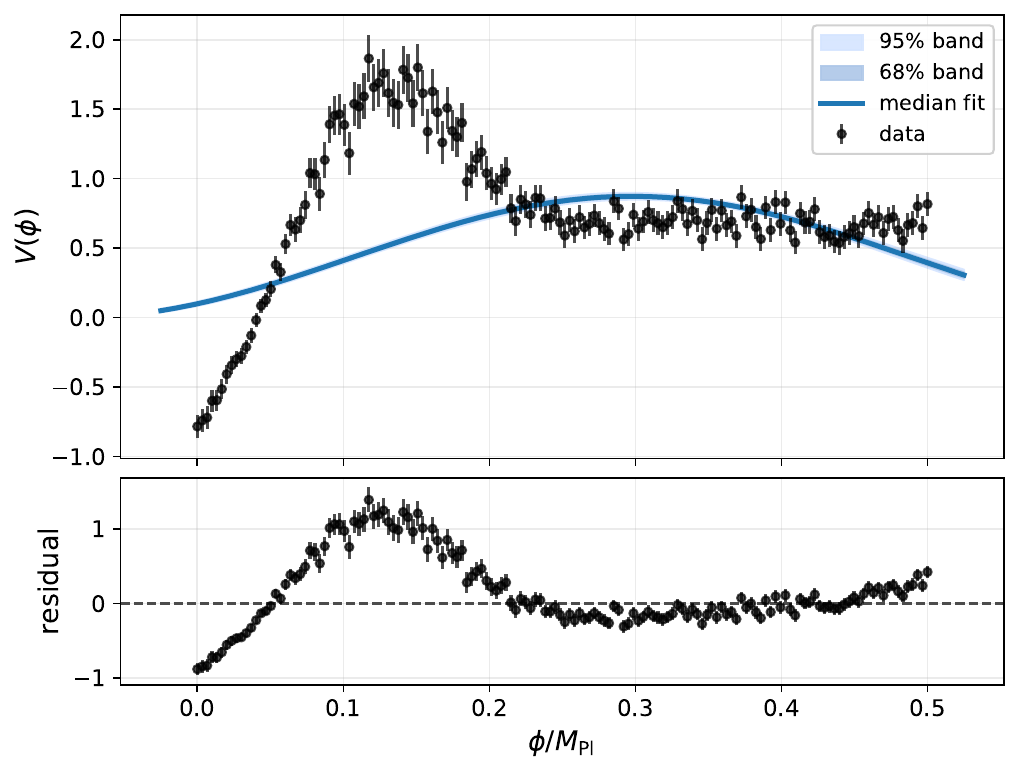}\hfill
    \includegraphics[width=0.46\linewidth]{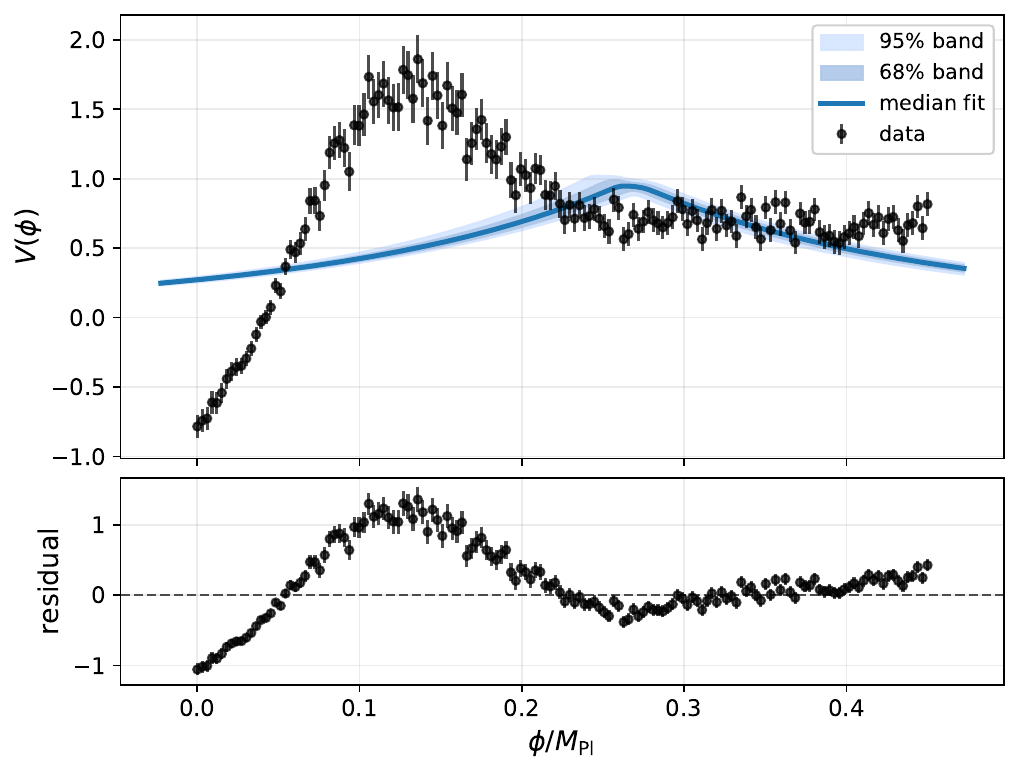}\\[-2pt]
    \parbox[t]{0.46\linewidth}{\centering (c) PNGB}\hfill
    \parbox[t]{0.46\linewidth}{\centering (d) Inverse power law}\\[10pt]

    % Row 3
    \includegraphics[width=0.46\linewidth]{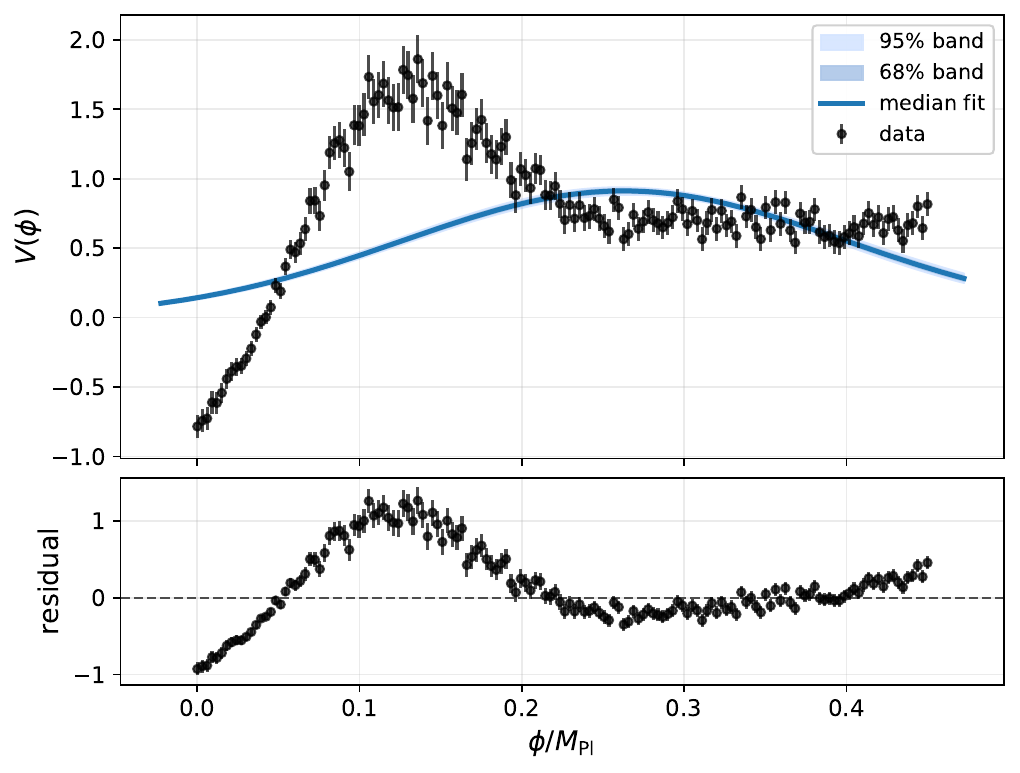}\hfill
    \includegraphics[width=0.46\linewidth]{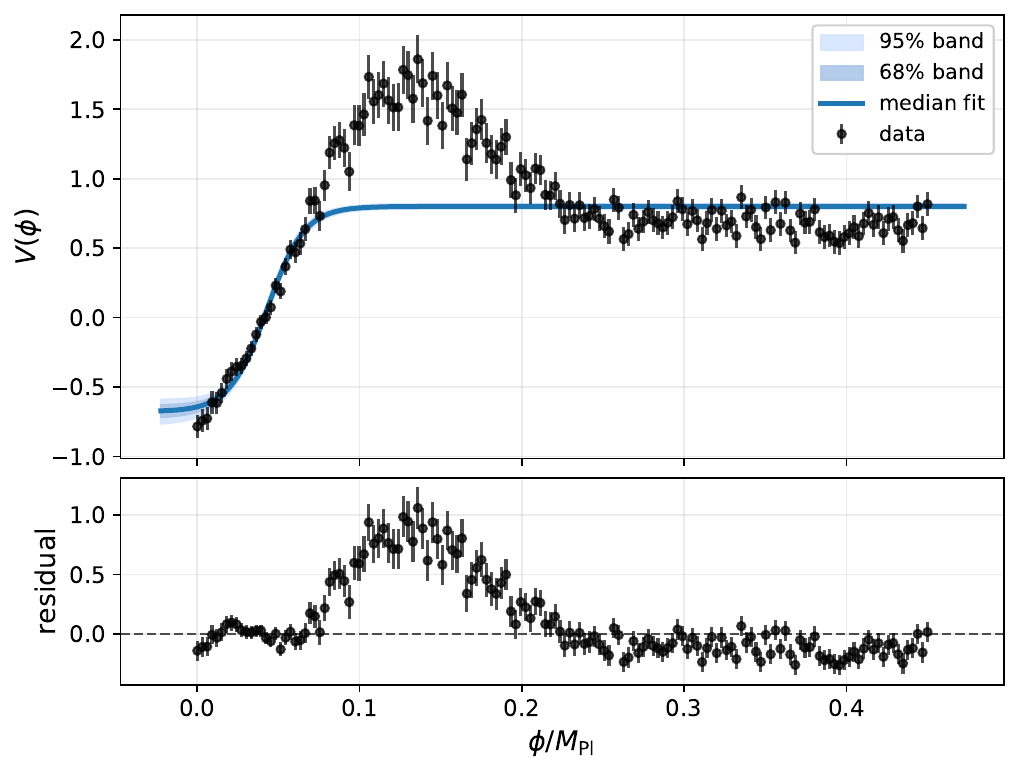}\\[-2pt]
    \parbox[t]{0.46\linewidth}{\centering (e) Gaussian bump}\hfill
    \parbox[t]{0.46\linewidth}{\centering (f) shifted-$\tanh$}\\

\caption{
Posterior-predictive fit bands for the $\tanh$ target in \emph{potential space}, shown for six representative analytic potentials:
(a) Exponential, (b) Hilltop quartic, (c) PNGB, (d) Inverse power law, (e) Gaussian bump, and (f) Shifted-$\tanh$.
In each panel, the solid curve denotes the median reconstructed $V(\phi)$ and the shaded regions show the corresponding $68\%$ and $95\%$ credible bands.
Black points with error bars are the mock potential-space dataset $D$ constructed from $V_{\rm tar}(\phi)$ (\cref{subsec:bayes}), and the lower strips show the corresponding residuals.
}

    \label{fig:fit_band_tanh}
\end{figure*}

\cref{fig:fit_band_tanh} shows the posterior-predictive median $V_{\rm med}(\phi)$ and the associated $68\%$ and $95\%$ credible bands for each analytic potential family potential family fitted to the $\tanh$ mock data set $D=\{(\phi_i,V_i^{\rm(mock)},\sigma_i)\}$ constructed from $V_{\rm tar}(\phi)$ (\cref{subsec:bayes}).
A characteristic feature of the $\tanh$ target is a localized structure in field space around $0.05\lesssim \phi/M_{\rm Pl}\lesssim 0.12$, visible in the mock points and consistent with the detailed shape of the reconstructed target potential (cf.\ \cref{fig:tanh_de_profile2}, top-left).
None of the tested analytic families reproduces this feature perfectly; instead, most models trade off fitting this localized structure against matching the low- and high-$\phi$ behavior. In practice, the exponential and shifted-$\tanh$ families provide the best global agreement away from the localized feature, with the shifted-$\tanh$ model yielding the most accurate overall match across the fitted range. 
Consistently, the residuals in \cref{fig:fit_band_tanh} are typically larger than in the \texttt{CPL} branch fit, reflecting the sharper structure present in the $\tanh$ target under the same noise model adopted.

\begin{figure*}[!t]
    \centering

    % Row 1
    \includegraphics[width=0.32\linewidth]{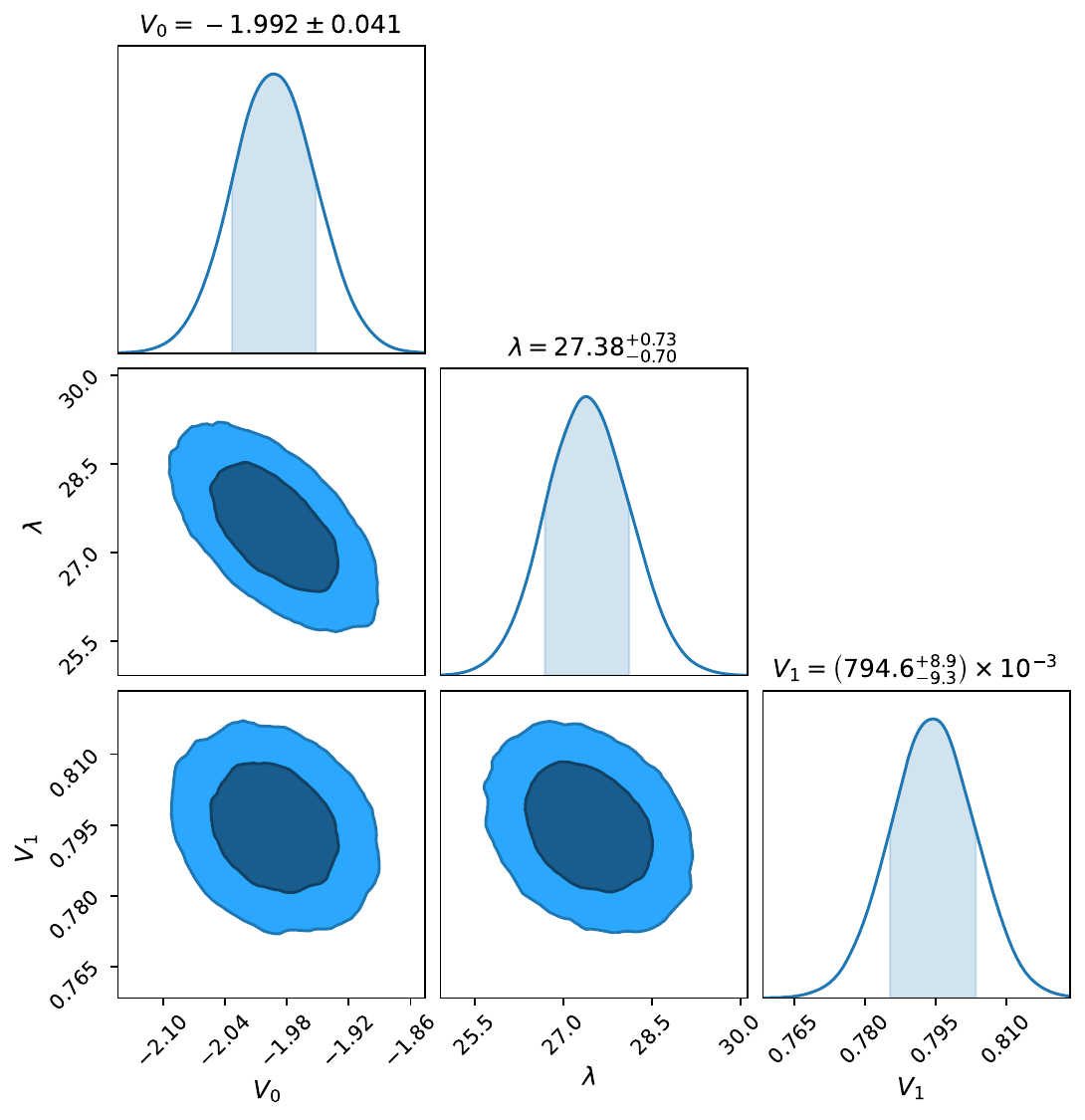}\hfill
    \includegraphics[width=0.32\linewidth]{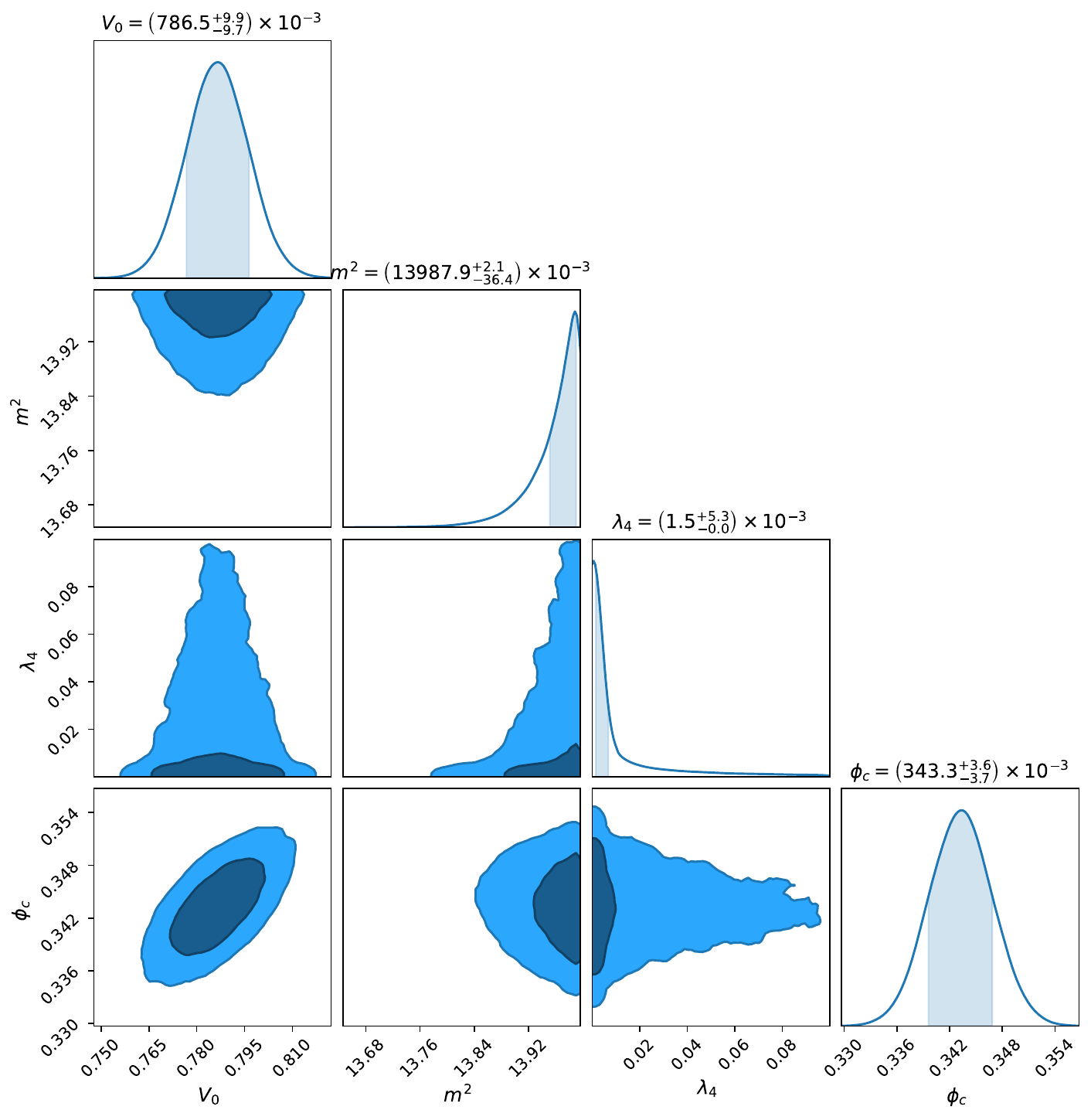}\hfill
    \includegraphics[width=0.32\linewidth]{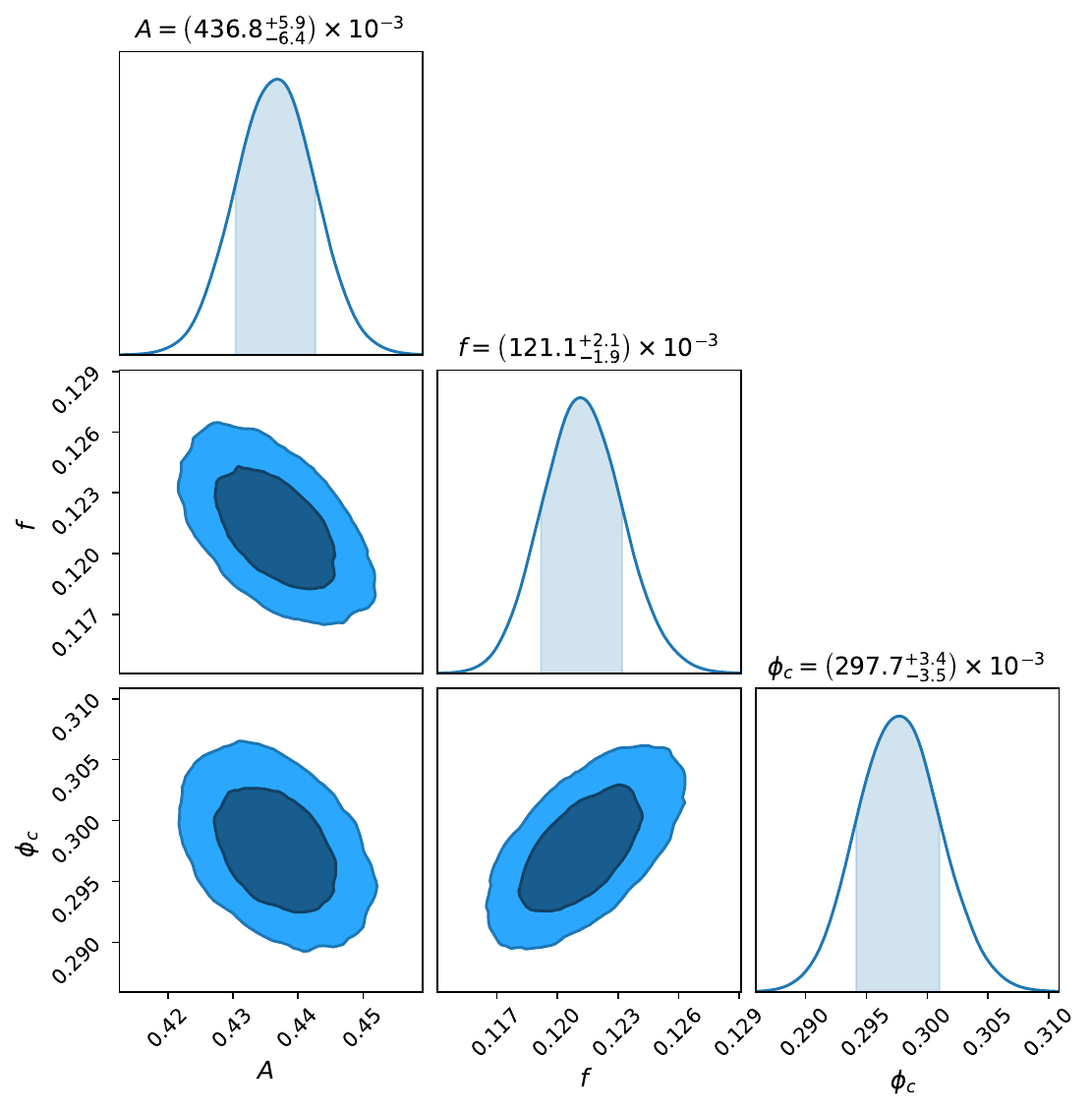}\\[-2pt]
    \parbox[t]{0.32\linewidth}{\centering (a) Exponential}\hfill
    \parbox[t]{0.32\linewidth}{\centering (b) Hilltop quartic}\hfill
    \parbox[t]{0.32\linewidth}{\centering (c) PNGB}\\[8pt]

    % Row 2
    \includegraphics[width=0.32\linewidth]{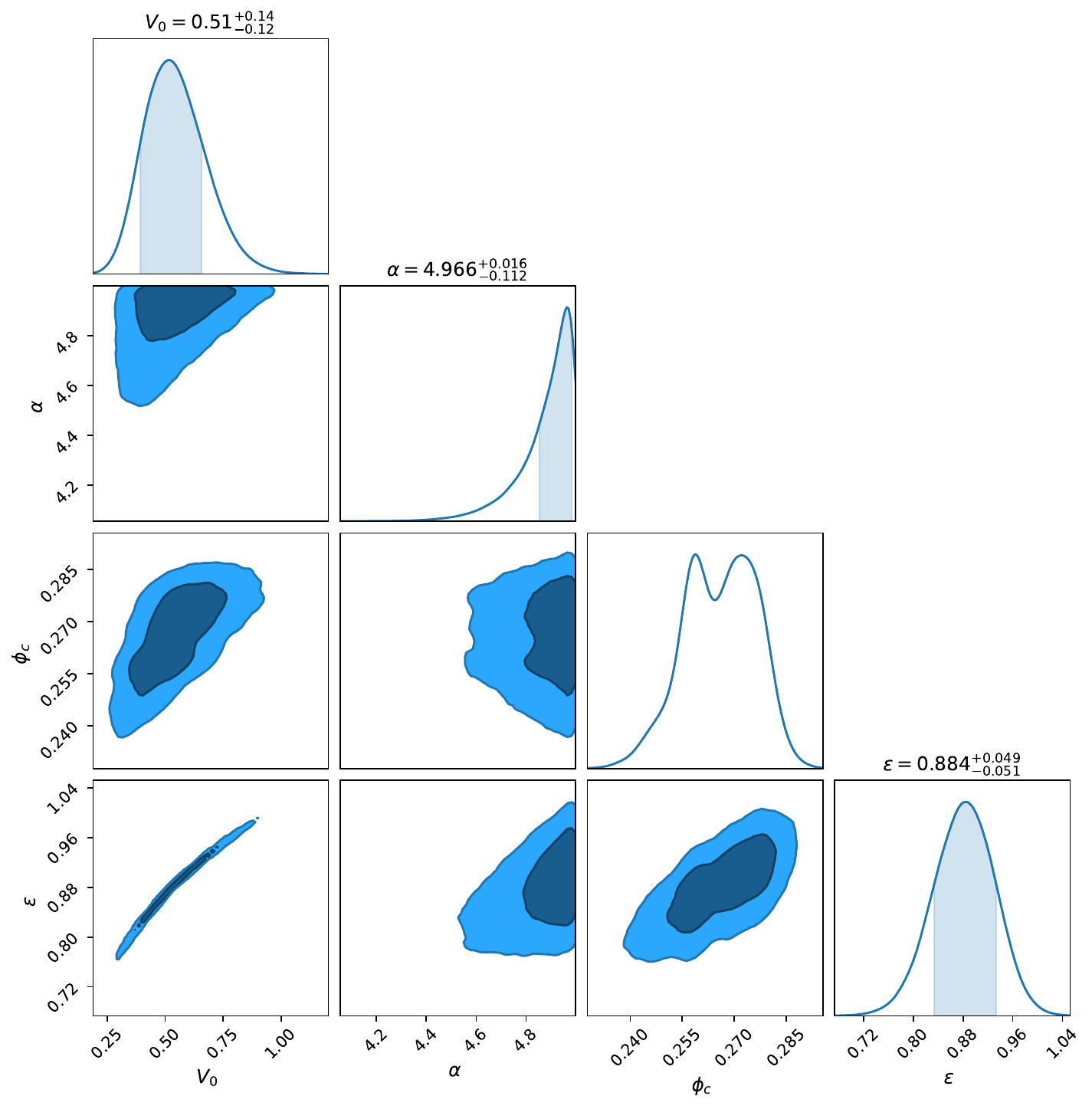}\hfill
    \includegraphics[width=0.32\linewidth]{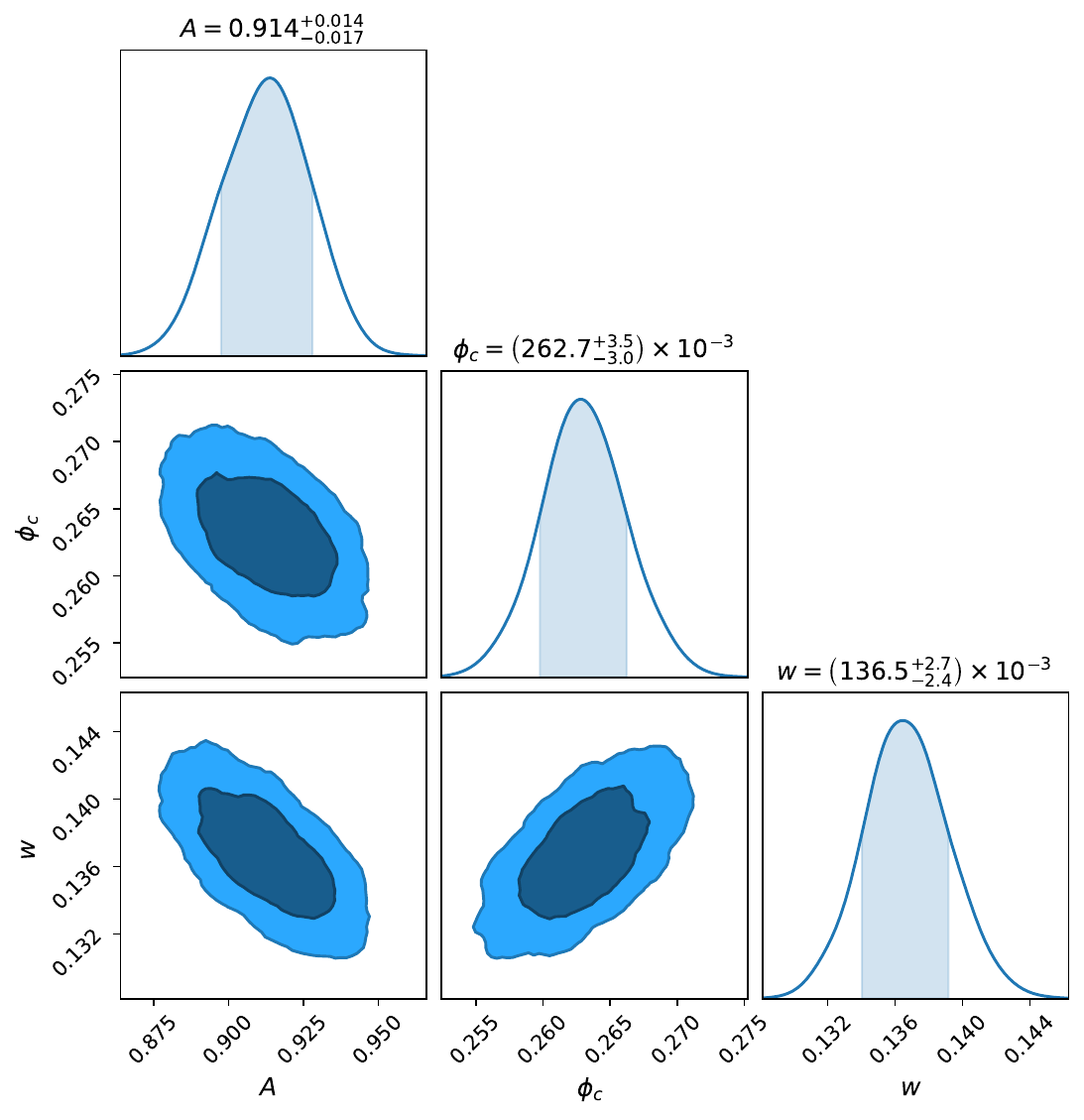}\hfill
    \includegraphics[width=0.32\linewidth]{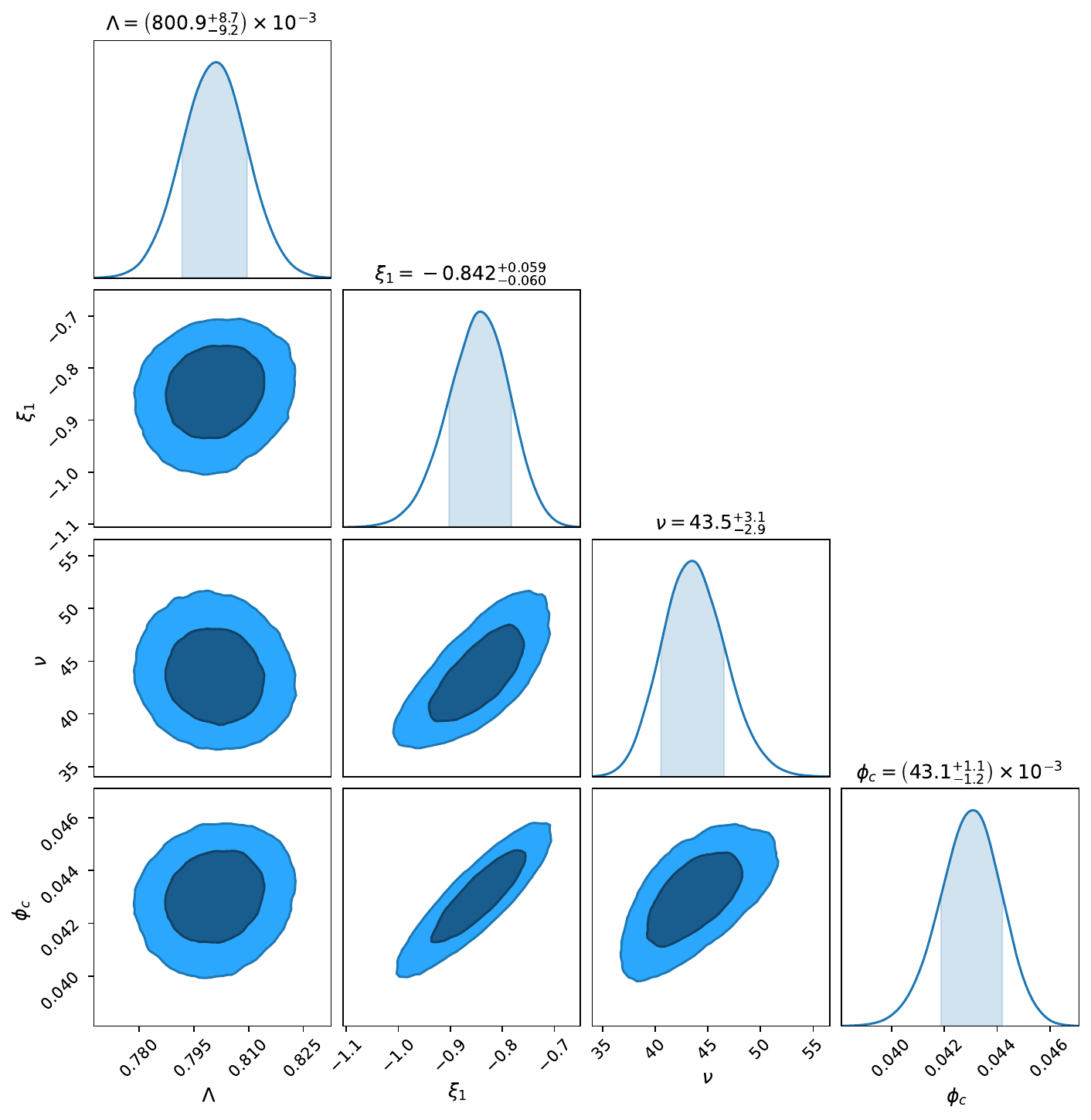}\\[-2pt]
    \parbox[t]{0.32\linewidth}{\centering (d) Inverse power law}\hfill
    \parbox[t]{0.32\linewidth}{\centering (e) Gaussian bump}\hfill
    \parbox[t]{0.32\linewidth}{\centering (f) shifted-$\tanh$}\\

    \caption{
    Posterior distributions for six analytic scalar-field potentials fitted to the sign-switching $\tanh$ reconstructed target potential in \emph{potential space}
    (\cref{subsec:bayes}).
    The constraints are obtained with nested sampling using the mock dataset $D$ constructed from the reconstructed $V_{\rm tar}(\phi)$ for the $\tanh$ history.
    Contours show the $68\%$ and $95\%$ credible regions, and the 1D panels show the corresponding marginalized posteriors.
    }
    \label{fig:corner_tanh_all}
\end{figure*}

The corresponding posterior constraints on model parameters are summarized in \cref{fig:corner_tanh_all}.
For the exponential potential (\cref{fig:corner_tanh_all}a) we obtain well-constrained posteriors, $V_0=-1.99\pm 0.041$, $\lambda=27.38^{+0.73}_{-0.70}$, and $V_1=0.794^{+0.008}_{-0.009}$, with clear parameter correlations.
For the hilltop quartic potential (\cref{fig:corner_tanh_all}b), $V_0$ and $\phi_{\rm c}$ are well constrained ($V_0=0.786^{+0.010}_{-0.009}$, $\phi_{\rm c}=0.343^{+0.004}_{-0.003}$), while $m^2$ and $\lambda_4$ exhibit one-sided posteriors, indicating that the preferred region lies near a limiting regime of the prior volume; this behavior is consistent with the comparatively poorer fit quality seen in \cref{fig:fit_band_tanh}. 
For the PNGB potential (\cref{fig:corner_tanh_all}c), the parameters are tightly constrained, $A = 0.436^{+0.005}_{-0.006}$, $f = 0.121^{+0.0021}_{-0.0019}$, and $\phi_{\rm c} = 0.297^{+0.0034}_{-0.0035}$, with visible correlations among them; nevertheless, the fit-band comparison shows that a good local parameter determination does not automatically imply a globally superior description of the $\tanh$ target under the adopted likelihood.

For the inverse power-law potential (\cref{fig:corner_tanh_all}d), we find mixed behavior: $V_0$ and $\varepsilon$ are reasonably constrained ($V_0=0.51^{+0.14}_{-0.12}$, $\varepsilon=0.884^{+0.049}_{-0.051}$), while $\alpha$ is one-sided ($\alpha=4.966^{+0.016}_{-0.112}$) and $\phi_{\rm c}$ shows bimodality, reflecting parameter degeneracies and a general poor match to the target.
For the Gaussian bump model (\cref{fig:corner_tanh_all}e), the parameters are comparatively well constrained ($A=0.914^{+0.014}_{-0.017}$, $\phi_{\rm c}=0.262^{+0.0035}_{-0.0030}$, $w=0.136^{+0.003}_{-0.002}$), but the fit-band figure indicates that this functional form struggles to reproduce the full $\tanh$ target (in particular the low-$\phi$ behavior), in line with its low statistical preference in the evidence ranking.

Finally, the shifted-$\tanh$ potential (\cref{fig:corner_tanh_all}f) yields tightly constrained posteriors, $\Lambda=0.800^{+0.0087}_{-0.0092}$, $\xi_1=-0.842^{+0.059}_{-0.060}$, $\nu=43.5^{+3.1}_{-2.9}$, and $\phi_{\rm c}=0.043^{+0.0011}_{-0.0012}$, with strong correlations among $(\xi_1,\nu,\phi_{\rm c})$.
Together with the fit-band behavior in \cref{fig:fit_band_tanh}, this supports the conclusion that, within the model set explored here, the shifted-$\tanh$ family provides the most natural analytic representation of the $\tanh$ target potential in field space.

% --- Fit-band panels without subcaption/subfigure ---

% ============================
%   tanh: Corner plots (merged)
% ============================

\begin{table}[!t]
\centering
\footnotesize
\renewcommand{\arraystretch}{1.18}
\setlength{\tabcolsep}{4pt}
\begin{tabular}{lccccc}
\hline\hline
\multicolumn{6}{c}{\textbf{CPL target}} \\
\hline
\textbf{Model} & $N_{\rm par}$ & $\log\mathcal{Z}$ & $\sigma_{\log\mathcal{Z}}$ & $\Delta\log\mathcal{Z}$ & $N_{\rm like}$ \\
\hline
\textbf{Exponential}      & 3 & 165.23    & 0.0860 & 0.00  & 281158  \\
Shifted-$\tanh$           & 4 & 165.10    & 0.0490 & 0.13  & 3556431 \\
Hilltop quartic           & 4 & 165.05    & 0.0700 & 0.18  & 508858  \\
PNGB                      & 3 & 164.28    & 0.0640 & 0.95  & 339684  \\
Gaussian bump             & 3 & 163.12    & 0.1880 & 2.11  & 47077   \\
Inverse power law         & 4 & 147.95    & 0.0540 & 17.28 & 5154918 \\
\hline
\multicolumn{6}{c}{\textbf{$\tanh$ target}} \\
\hline
\textbf{Model} & $N_{\rm par}$ & $\log\mathcal{Z}$ & $\sigma_{\log\mathcal{Z}}$ & $\Delta\log\mathcal{Z}$ & $N_{\rm like}$ \\
\hline
\textbf{Shifted-$\tanh$}  & 4 & -278.39    & 0.0479 & 0.00    & 1991526 \\
Exponential               & 3 & -422.65    & 0.0829 & 144.26  & 403430  \\
PNGB                      & 3 & -1125.579  & 0.0990 & 847.19  & 305416  \\
Hilltop quartic           & 4 & -1129.626  & 0.0970 & 851.24  & 791565  \\
Gaussian bump             & 3 & -1260.427  & 0.2120 & 982.04  & 44395   \\
Inverse power law         & 4 & -1859.8309 & 0.0670 & 1581.44 & 364935  \\
\hline\hline
\end{tabular}
\caption{
Bayesian evidence comparison of analytic scalar-field potentials fitted in \emph{potential space} to the mock datasets $D$
constructed from the \texttt{CPL} and $\tanh$ target potentials $V_{\rm tar}(\phi)$ (\cref{subsec:bayes}).
For each target we also report $\Delta\log\mathcal{Z}\equiv \log\mathcal{Z}_{\rm best}-\log\mathcal{Z}$, so that smaller values indicate stronger support relative to the best model.
We additionally list $N_{\rm like}$, the total number of likelihood evaluations used by the nested sampler, only as a computational diagnostic of sampling effort; it is not itself an inferential statistic and does not enter the model ranking.
}
\label{tab:model_comparison_combined}
\end{table}

\cref{tab:model_comparison_combined} summarises the Bayesian evidence values for the analytic potentials fitted in \emph{potential space} to the mock datasets $D$ constructed from the \texttt{CPL} and $\tanh$ target potentials $V_{\rm tar}(\phi)$ (\cref{subsec:bayes}).
The evidence $\log\mathcal{Z}$ quantifies the \emph{global} performance of each model by integrating the likelihood over the prior volume and therefore balances goodness of fit against model complexity, automatically penalising unnecessary flexibility. In this sense, the evidence provides a robust model-selection criterion beyond simple $\chi^2$-type comparisons. Since the absolute normalisation of $\log\mathcal{Z}$ depends on the adopted noise model in potential space, the meaningful diagnostic is the \emph{relative} evidence ranking within each target, interpreted conditionally on the baseline noise prescription and prior volume adopted here. For this reason, we also report $\Delta\log\mathcal{Z}\equiv \log\mathcal{Z}_{\rm best}-\log\mathcal{Z}$. We list $N_{\rm like}$ only as a computational diagnostic, i.e.\ the total number of likelihood evaluations performed by the nested sampler for each model; it is included to document the computational effort and to show that the evidence ranking is not an artefact of some models having been explored much less thoroughly than others. Within this baseline setup, very large evidence separations may be regarded as decisive, whereas close rankings should be interpreted more cautiously as indicating a competitive subset rather than a uniquely isolated model.

For the \texttt{CPL} target, the exponential potential has the highest evidence in the baseline analysis, $\log\mathcal{Z}=165.23\pm0.086$ ($N_{\rm par}=3$). The shifted-$\tanh$ and hilltop quartic models follow with only slightly smaller evidence values, $\log\mathcal{Z}=165.10\pm0.049$ and $\log\mathcal{Z}=165.05\pm0.070$, corresponding to $\Delta\log\mathcal{Z}=0.13$ and $0.18$, respectively. We therefore interpret these three models as a statistically competitive leading set within the adopted potential-space setup, rather than a sharply separated hierarchy, with the exponential potential being only the nominally best-ranked model. The PNGB potential gives a lower evidence, $\log\mathcal{Z}=164.28\pm0.064$ ($\Delta\log\mathcal{Z}=0.95$), while the Gaussian bump performs worse, $\log\mathcal{Z}=163.12\pm0.188$ ($\Delta\log\mathcal{Z}=2.11$), indicating that an additional localised feature is not statistically required by the \texttt{CPL} target under the adopted likelihood. The inverse power-law potential is strongly disfavoured, $\log\mathcal{Z}=147.95\pm0.054$ ($\Delta\log\mathcal{Z}=17.28$), consistent with its visibly poorer fit-band performance.

For the sign-switching $\tanh$ target, the hierarchy is qualitatively different. The shifted-$\tanh$ potential is overwhelmingly preferred, with $\log\mathcal{Z}=-278.39\pm0.0479$ ($N_{\rm par}=4$), while the exponential potential is the next best model but is already strongly disfavoured, with $\log\mathcal{Z}=-422.65\pm0.0829$ and hence $\Delta\log\mathcal{Z}=144.26$. The PNGB and hilltop quartic potentials follow with much lower evidences, and the Gaussian bump and inverse power-law potentials yield the smallest evidences. Thus, within the tested set, the shifted-$\tanh$ family provides by far the most natural analytic representation of the reconstructed $\tanh$ target in field space, whereas the exponential family is the most favoured representation of the restricted \texttt{CPL} branch. The ranking is therefore not driven by differences in sampling effort (cf.\ $N_{\rm like}$ in \cref{tab:model_comparison_combined}), but reflects a genuine functional mismatch or agreement between the analytic templates and the reconstructed targets.

\section{Conclusions}\label{sec:conclude}

We have developed a unified background-level framework that maps a prescribed late-time dark-energy density history $\rho_{\rm de}(z)$ into an effective scalar-field description on a spatially flat FLRW background, yielding the associated pressure $p_{\rm de}(z)$, kinetic contribution $K(z)$, field trajectory $\phi(z)$, and reconstructed potential $V(\phi)$. A key feature of the approach is that it is formulated directly in terms of $\rho_{\rm de}(z)$ and $\dd\rho_{\rm de}/\dd z$ (equivalently $\tilde\Omega_{\rm de}(z)$ and $\dd\tilde\Omega_{\rm de}/\dd z$), and therefore remains well defined even in sign-changing scenarios where the derived equation-of-state parameter $w_{\rm de}=p_{\rm de}/\rho_{\rm de}$ becomes ill-defined at a zero crossing. In such cases, the physically meaningful discriminator is instead the NEC combination
$\rho_{\rm de}+p_{\rm de}=\tfrac13(1+z)\,\dd\rho_{\rm de}/\dd z$, whose sign controls the kinetic sector and fixes the admissible single-field branch when a minimally coupled realization exists~\cite{Akarsu:2025gwi,Akarsu:2026anp,Gokcen:2026pkq}.

This viewpoint makes the classification of regimes transparent once $\rho_{\rm de}$ is allowed to take either sign. The usual $w_{\rm de}=-1$ line---often referred to as the phantom-divide line, a notion that is physically meaningful only as long as the sign of the dark-energy density remains fixed---is not a universal separator of physical behavior; what matters is the sign of $\rho_{\rm de}+p_{\rm de}$ (equivalently the sign of $(1+w_{\rm de})\rho_{\rm de}$ when $w_{\rm de}$ is defined)~\cite{Akarsu:2025gwi,Akarsu:2026anp,Gokcen:2026pkq}. In the notation of~\cref{tab:pq-branches}, this yields a clean four-way taxonomy ($p/n$-quintessence and $p/n$-phantom) that connects directly to the scalar-field identity $\rho_\phi+p_\phi=\epsilon\dot\phi^{\,2}$. The associated single-field sign-consistency condition, Eq.~\eqref{eq:sign_consistency}, provides an immediate diagnostic: if $(1+z)\,\dd\tilde\Omega_{\rm de}/\dd z$ changes sign over the reconstruction range, then no single minimally coupled real scalar with fixed kinetic signature can realize the full history, and the mapping must instead be interpreted as an effective one-dimensional representation of an extended sector (e.g.\ a quintom-like or non-canonical completion).

We applied the formalism to three benchmark histories chosen to span both standard phenomenological practice and late-time transition scenarios motivated by
sign-switching vacuum-energy ideas. Specifically, we considered: (i) the widely used \texttt{CPL} history~\cite{Chevallier:2000qy,Linder:2002et} as a canonical benchmark of smooth dynamical dark
energy; (ii) a sign-switching $\tanh$ history~\cite{Akarsu:2022typ,Akarsu:2024qsi,Akarsu:2024eoo} that realizes a smooth mirror AdS$\rightarrow$dS transition in $\rho_{\rm de}$ (a $\Lambda_{\rm s}$CDM-like
background history); and (iii) a shifted-$\tanh$ (emergent) profile~\cite{DeFelice:2020cpt} that remains positive definite and approaches $\rho_{\rm de}\to 0^{+}$ at high redshift,
providing a controlled comparison case without a density sign change. For the \texttt{CPL} case, $\rho_{\rm de}>0$ by construction but the evolution crosses the NEC boundary $\rho_{\rm de}+p_{\rm de}=0$ (equivalently $w_{\rm de}=-1$), which forces $K(z)=\tfrac12(\rho_{\rm de}+p_{\rm de})$ to change sign and renders $\phi(z)$ non-monotonic. The resulting $V(\phi)$ is consequently multivalued, and the \texttt{CPL} history cannot be realized globally by a single minimally coupled scalar field with fixed $\epsilon=\pm1$. By contrast, both $\tanh$-based transition histories satisfy $\dd\tilde\Omega_{\rm de}/\dd z<0$ over the reconstruction range, implying $\rho_{\rm de}+p_{\rm de}<0$ throughout and selecting a consistent single-field realization on the phantom branch ($\epsilon=-1$) at the background level. In the sign-switching $\tanh$ case this corresponds to a continuous evolution from $p$-phantom to $n$-phantom across $\rho_{\rm de}=0$, while the shifted-$\tanh$ profile remains on the $p$-phantom branch and approaches $\rho_{\rm de}\to 0^{+}$ at high redshift. For  cases admitting a single-field realization, we also verified the internal consistency of the reconstruction by checking the Klein--Gordon residual (\cref{app:KGcheck}).

Treating the reconstructed $V_{\rm tar}(\phi)$ as a target, we then performed Bayesian model comparison directly in \emph{potential space} across a representative set of analytic potential families. The chosen set was designed to span the main model-building archetypes for late-time acceleration: scaling/attractor (exponential)~\cite{Ratra:1987rm,Wetterich:1987fm}, axion-like periodic (PNGB)~\cite{Frieman1995,Caldwell:1997ii,Choi2000}, thawing/hilltop (hilltop-quartic)~\cite{Dutta:2008qn,Matos2009,Chiba2009}, tracker-like (inverse power law)~\cite{Peebles:1987ek,Steinhardt:1999nw}, localized-feature (Gaussian bump), and smooth transition/plateau interpolation (shifted-$\tanh$)~\cite{Akarsu:2025dmj}. This step is intentionally distinct from a direct cosmological parameter fit: the likelihood quantifies agreement with the reconstructed target in field space under the adopted noise model, and the Bayesian evidence $\log\mathcal{Z}$ ranks potential families by balancing global fit quality against model complexity. For the \texttt{CPL} benchmark, the potential-space fit must be performed on a single-valued monotonic branch of the reconstruction (in practice the $p$-phantom branch with $K<0$); within this restricted target, the exponential potential has the highest evidence in our baseline analysis, although the shifted-$\tanh$ and hilltop quartic forms remain statistically competitive.
For the sign-switching $\tanh$ target, the hierarchy is markedly different: the shifted-$\tanh$ potential is overwhelmingly preferred, reflecting its close functional match to the reconstructed target, while other families are strongly disfavoured in evidence even when some parameters are locally well constrained. (Repeating the same potential-space procedure for the shifted-$\tanh$ target yields the same qualitative evidence ranking.) In this way, the potential-space comparison provides a quantitative ``theory-space filter'': phenomenological density histories translate into specific field-space requirements that select (or strongly disfavour) broad classes of scalar potentials.

Several caveats delimit the interpretation of these results. First, our reconstruction and model comparison are performed at the homogeneous background level.
Second, whenever a phantom realization is invoked, it should be understood as an effective description with an implied cutoff/UV completion rather than a fundamental ghost degree of freedom. Third, the evidence values are conditional on the adopted potential-space noise model and priors; while the \emph{relative} evidence ranking within a given target is the meaningful discriminator for the choices made here, the absolute normalization of $\log\mathcal{Z}$ is not itself a cosmological observable. Accordingly, close rankings---such as the leading subset found for the \texttt{CPL} target---should be interpreted as baseline-dependent competition within the present setup, whereas the very large evidence gaps found for the sign-switching $\tanh$ target indicate a much more decisive preference.

Looking ahead, the framework can be extended in several directions that directly build on the present results. A first step is to move beyond the benchmark histories and perform a fully data-driven reconstruction of $\rho_{\rm de}(z)$ (or equivalently $E(z)$) from cosmological observations, propagating the resulting uncertainties into $V_{\rm tar}(\phi)$ and the potential-space likelihood. 

A second, complementary extension is a full perturbative treatment of the reconstructed histories. As discussed in~\cref{sec:scalarfield}, at fixed homogeneous $\rho_{\rm de}(z)$, linear perturbations do not modify the background reconstruction of $p_{\rm de}(z)$, $K(z)$, $\phi(z)$ or $V(\phi)$, but they provide additional consistency conditions on the microscopic completion. This also highlights that effective-fluid and scalar-field descriptions that are equivalent at the homogeneous level need not be equivalent at the perturbative level: the minimally coupled, linear-in-$X$ scalar sector considered here has $c_{s,\phi}^2=1$ and, when written in fluid language, generally contains a non-adiabatic pressure contribution, whereas a purely adiabatic fluid closure would impose $\delta p_{\rm de}=c_a^2\delta\rho_{\rm de}$. In particular, density-zero crossings require perturbation variables that remain regular when
$\rho_{\rm de}=0$, while histories that cross the NEC boundary---equivalently, the phantom divide when $\rho_{\rm de}>0$, as in the \texttt{CPL} benchmark considered here---require a specified extended-sector completion before their perturbations are uniquely defined. A full analysis of growth, stability and observational constraints from perturbations is left for future work. For such histories, it is also natural to use the reconstruction as a guide for explicit completions (quintom sectors or non-canonical dynamics) and to ask which completions reproduce the effective one-dimensional target with minimal additional structure.

Finally, forthcoming high-precision data from surveys such as DESI and DES~\cite{DESI:2025zgx,DES:2025sig}, Euclid~\cite{Amendola:2016saw,Euclid:2019clj,Euclid:2021qvm}, and LSST~\cite{LSST:2008ijt,Zhan:2017uwu,LSSTDarkEnergyScience:2018jkl} will substantially sharpen late-time expansion constraints and thereby tighten the mapping between phenomenological reconstructions and field-space requirements. In parallel, it would be particularly interesting to pursue a fully data-driven reconstruction of the scalar field potential directly from observational datasets (e.g. DESI and DES~\cite{DESI:2025zgx,DES:2025sig} or Pantheon+~\cite{Brout:2022vxf}) without assuming a specific parametric form for $\rho_{\rm de}(z)$, providing a more agnostic route toward identifying the dynamics responsible for cosmic acceleration.

%\newpage

\begin{acknowledgments}
S.A.A. acknowledges the support of the DGAPA postdoctoral fellowship program at ICF-UNAM, Mexico. S.A.A. also acknowledges the use of High Performance Computing cluster Pegasus at IUCAA, Pune, India. M.A.Z.  acknowledges the support of the SECIHTI. \"{O}.A. acknowledges the support from the Turkish Academy of Sciences in the scheme of the Outstanding Young Scientist Award  (T\"{U}BA-GEB\.{I}P). J.A.V. acknowledges support from FOSEC SEP-CONACYT Ciencia B\'asica A1-S-21925,  UNAM-DGAPA-PAPIIT IN109126, IN110325 and Cátedra de Investigación Marcos Moshinsky. Special acknowledgments to Ing. Francisco Bustos and Lic. Reyes García who assisted considerably with the High Performance Computing at the ICF-UNAM. This article is based upon work from COST Action CA21136 Addressing observational tensions in cosmology with systematics and fundamental physics (CosmoVerse) supported by COST (European Cooperation in Science and Technology).
\end{acknowledgments}

\appendix
\section{Klein-Gordon consistency check}
\label{app:KGcheck}

\begin{figure*}[ht!]
    \centering

    \begin{minipage}[t]{0.32\linewidth}
        \centering
        \includegraphics[width=\linewidth]{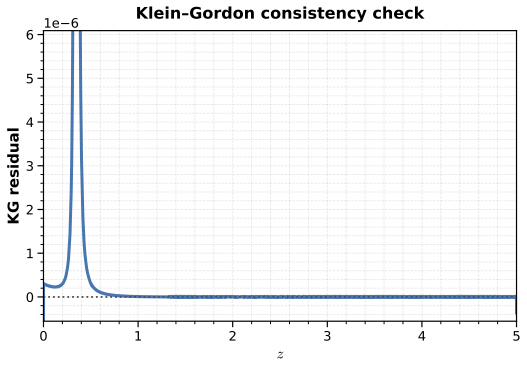}\\[-4pt]
        (a) CPL
    \end{minipage}\hfill
    \begin{minipage}[t]{0.32\linewidth}
        \centering
        \includegraphics[width=\linewidth]{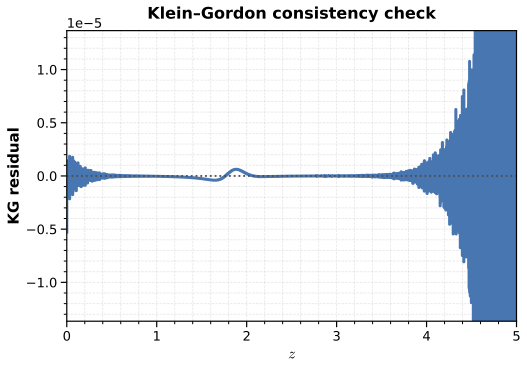}\\[-4pt]
        (b) \(\tanh\)
    \end{minipage}\hfill
    \begin{minipage}[t]{0.32\linewidth}
        \centering
        \includegraphics[width=\linewidth]{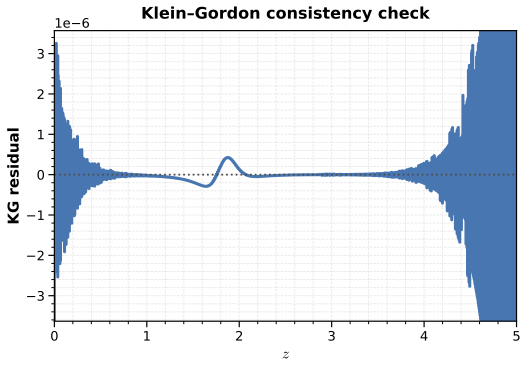}\\[-4pt]
        (c) shifted \(\tanh\)
    \end{minipage}

    \caption{Klein--Gordon consistency residual ${\cal R}(z)$ for (a) CPL, (b) \(\tanh\), and (c) shifted \(\tanh\) reconstructions.}
    \label{fig:KG-check}
\end{figure*}

For cases that admit a single-field realization with fixed kinetic sign, we verify the reconstruction by evaluating the Klein--Gordon equation,~\cref{eq:KG,eq:KG_z}. 
In particular, we define the residual
\begin{equation}
{\cal R}(z)\equiv \epsilon\big(\ddot\phi+3H\dot\phi\big)+\frac{dV}{d\phi},
\end{equation}
and check numerically that ${\cal R}(z)\simeq 0$ along the reconstructed trajectory $\phi(z)$ with $V(\phi)$.
Because ${\cal R}(z)$ involves numerical derivatives and interpolated functions, small excursions near the endpoints of the plotted redshift interval should be interpreted as expected edge effects rather than as physical failures of the reconstruction.
\cref{fig:KG-check} shows ${\cal R}(z)$ for the three benchmark profiles considered in this work.
For the $\tanh$ and shifted-$\tanh$ cases (which satisfy the single-field sign-consistency condition with $\epsilon=-1$ over the plotted range), the residual remains close to zero over the physically relevant interval, confirming the internal consistency of the background-level reconstruction.
For the CPL benchmark, a localized deviation is expected around the phantom-divide crossing, reflecting the fact that the CPL history cannot be described globally by a single minimally coupled scalar field with fixed $\epsilon$.

\newpage

\bibliography{Bibliography}

\end{document}